\documentclass[reqno,12pt]{amsart}
\usepackage{fullpage}
\usepackage{amsfonts}
\usepackage{amssymb}

\numberwithin{equation}{section}
\def\X{{\bf X}}

\def\pr{{\rm pr}}

\def\Esp{{\mathcal E}}
\def\C{{\mathcal C}}
\def\triv{{\rm triv}}
\def\D{{\mathcal D}}
\def\scal{{\rm scal}}

\def\Div{{\rm Div\,}}

\def\Grad{{\rm Grad\,}}
\def\div{{\rm div}}
\def\curl{{\rm curl}}

\def\Rnum{\mathbb{R}}

\def\diag{{\rm diag}}

\def\const{{\rm const.}}
\def\lot{\text{ lower order terms}}

\def\t{{\rm t}}

\def\p{\partial}
\def\smallint{{\textstyle\int}}

\def\til{\tilde}

\def\parder#1#2{\frac{\partial{#1}}{\partial{#2}}}
\def\parderop#1{\partial/\partial{#1}}

\newtheorem{prop}{Proposition}
\newtheorem{thm}{Theorem}
\newtheorem{cor}{Corollary}
\newtheorem{lem}{Lemma}

\def\propref#1{Proposition~\ref{#1}}
\def\thmref#1{Theorem~\ref{#1}}
\def\lemref#1{Lemma~\ref{#1}}
\def\thmrefs#1#2{Theorems~\ref{#1} and~\ref{#2}}
\def\Ref#1{Ref.\cite{#1}}

\def\secref#1{Sec.~\ref{#1}}
\def\Ex#1{{Ex.\ {#1}}}

\def\ie/{i.e.}
\def\eg/{e.g.}
\def\etc/{etc.}

\def\scrpt#1{$\scriptstyle {#1}$}

\begin{document}
\allowdisplaybreaks[3]

\title{Generalization of Noether's theorem in modern form to non-variational partial differential equations}

\author{
Stephen C. Anco
\\\lowercase{\scshape{
Department of Mathematics and Statistics\\
Brock University\\
St. Catharines, ON L\scrpt2S\scrpt3A\scrpt1, Canada}}
}

\begin{abstract}
A general method using multipliers for finding 
the conserved integrals admitted by any given partial differential equation (PDE)
or system of partial differential equations 
is reviewed and further developed in several ways. 
Multipliers are expressions whose (summed) product with a PDE (system) 
yields a local divergence identity which has the physical meaning of 
a continuity equation involving a conserved density and a spatial flux 
for solutions of the PDE (system). 
On spatial domains, 
the integral form of a continuity equation yields a conserved integral. 
When a PDE (system) is variational, 
multipliers are known to correspond to 
infinitesimal symmetries of the variational principle,
and the local divergence identity relating a multiplier to a conserved integral 
is the same as the variational identity used in Noether's theorem 
for connecting conserved integrals to invariance of a variational principle. 
From this viewpoint, the general multiplier method is shown to constitute 
a modern form of Noether's theorem in which the variational principle is not directly used.
When a PDE (system) is non-variational, 
multipliers are shown to be an adjoint counterpart to infinitesimal symmetries, 
and the local divergence identity that relates a multiplier to a conserved integral 
is shown to be an adjoint generalization of the variational identity that underlies Noether's theorem. 
Two main results are established for a general class of PDE systems 
having a solved-form for leading derivatives,
which encompasses all typical PDEs of physical interest, 
such as dynamical systems that model diffusion-reaction-convection, wave propagation, fluid flow, gas dynamics and plasma dynamics, continuum mechanics, 
as well as non-dynamical (static equilibrium) systems. 
First, 
all non-trivial conserved integrals are shown to arise from non-trivial 
multipliers in a one-to-one manner, taking into account certain equivalence freedoms.
Second, a simple scaling formula based on dimensional analysis is derived to obtain 
the conserved density and the spatial flux in any conserved integral, 
just using the corresponding multiplier and the given PDE (system). 
Furthermore, 
a general class of multipliers that captures physically important 
conserved integrals such as mass, momentum, energy, angular momentum
is identified. 
The derivations use a few basic tools from variational calculus, 
for which a concrete self-contained formulation is provided. 
\end{abstract}

\maketitle

\section{Introduction and overview}
\label{intro}

In the study of partial differential equations (PDEs),
conserved integrals and local continuity equations have many important uses. 
They yield fundamental conserved quantities and constants of motion, 
which along with symmetries are an intrinsic coordinate-free aspect of the structure 
of a PDE system. 
They also yield potentials and nonlocally-related systems. 
They provide conserved norms and estimates, 
which are central to the analysis of solutions. 
They detect if a PDE system admits an invertible transformation into a target class of PDE systems
(\eg/, nonlinear to linear, or linear variable coefficient to constant coefficient). 
They typically indicate if a PDE system has integrability structure. 
They allow checking the accuracy of numerical solution methods 
and also give rise to good discretizations (\eg/, conserving energy or momentum).

For a dynamical PDE system in one spatial dimension, 
a local continuity equation is a total divergence expression 
\begin{equation}
D_t T + D_x X =0
\end{equation}
vanishing on the solution space of the system,
where $T$ is a conserved density and $X$ is a spatial flux. 
(Here $D_t$ and $D_x$ are total derivatives with respect to time and space coordinates.)
Every local continuity equation physically represents a conservation law 
for the quantity $T$. 
The conservation law can be formulated by integrating the local continuity equation over any spatial domain $\Omega\subseteq\Rnum$, 
yielding 
\begin{equation}\label{conslaw-onedim}
\frac{d}{dt}\int_\Omega T dx = -X\Big|_{\p\Omega} . 
\end{equation}
This shows that the rate of change of the integral of the conserved density $T$ 
on the domain $\Omega$ is balanced by the net outward flux 
through the domain endpoints $\p\Omega$. 

In two and three spatial dimensions, 
local continuity equations have the more general total divergence form 
\begin{equation}\label{continuityeqn}
D_t T + \Div\vec X=0 . 
\end{equation}
The corresponding physical conservation law is given by 
\begin{equation}\label{conslaw-multidim}
\frac{d}{dt}\int_\Omega T dV = -\oint_{\p\Omega}\vec X\cdot\vec\nu dA
\end{equation}
where $\Omega$ is a spatial domain 
and $\vec\nu$ is the outward unit normal of the domain boundary. 
This conservation law shows that the net outward flux of $\vec X$ integrated over $\p\Omega$ 
balances the rate of change of the integral of the conserved density $T$ on $\Omega$. 

Another type of conservation law in two and three spatial dimensions 
can be formulated on the boundary of a spatial domain $\Omega$, 
\begin{equation}\label{conslaw-boundary}
\frac{d}{dt}\oint_{\p\Omega} \vec T\cdot\vec\nu dA = 0
\end{equation}
holding on the solution space of a PDE system. 
This boundary conservation law corresponds to a local continuity equation \eqref{continuityeqn}
in which the conserved density is a total spatial divergence, $T=\Div\vec T$,
and the flux is a total spatial curl, $\vec X = \Div\mathbf\Gamma$,
where $\mathbf\Gamma$ is an antisymmetric tensor. 
Its physical meaning is that the net flux of $\vec T$ over $\p\Omega$ 
is a constant of the motion for the PDE system. 

When hydrodynamical PDE systems for fluid/gas flow are considered, 
a more physically useful formulation of conservation laws is given by 
considering moving spatial domains $\Omega(t)$, 
or moving spatial boundaries $\p\Omega(t)$, 
that are transported by the flow of the fluid/gas. 

For a moving domain, 
a physical conservation law has the form
\begin{equation}\label{conslaw-movingdomain}
\frac{d}{dt}\int_{\Omega(t)} T dV = -\oint_{\p\Omega(t)}(\vec X-T\vec u)\cdot\vec\nu dA
\end{equation}
where $\vec u$ is the fluid/gas velocity, 
and $\vec X-T\vec u=\vec{\mathcal X}$ is the moving flux.
The local continuity equation \eqref{continuityeqn} 
is then equivalent to a transport equation 
\begin{equation}\label{transporteqn}
(D_t +\vec u\cdot D_x)T = -(\nabla\cdot\vec u)T -\Div\vec{\mathcal X}
\end{equation}
for the conserved density $T$,
with $D_t +\vec u\cdot D_x$ being the material (advective) derivative,
and $\nabla\cdot\vec u$ being the expansion or contraction factor of 
an infinitesimal moving volume of the fluid/gas. 
If the net moving flux over the domain boundary $\p\Omega(t)$ vanishes, 
then the integral of the conserved density $T$ on the moving domain $\Omega(t)$
is a constant of motion. 

For a moving boundary, 
the physical form of a conservation law is given by 
\begin{equation}\label{conslaw-movingboundary}
\frac{d}{dt}\oint_{\p\Omega(t)}\vec T\cdot\vec\nu dA =0
\end{equation}
which shows the net flux of $\vec T$ integrated over $\p\Omega(t)$ 
is a constant of motion. 
In the corresponding transport equation \eqref{transporteqn},
the conserved density is a total spatial divergence, 
$T=\Div\vec T$,
and the moving flux is a total spatial curl, 
$\vec{\mathcal X} = \Div\mathbf\Gamma$,
where $\mathbf\Gamma$ is an antisymmetric tensor.

A related type of conservation law in two and three spatial dimensions 
arises from the total spatial divergence of a flux vector that is not a total spatial curl,
\begin{equation}
\Div\vec X=0, 
\quad 
\vec X \neq \Div\mathbf\Gamma
\end{equation}
holding on the solution space of a PDE system. 
This yields a physical conservation law 
on any spatial domain $\Omega$ enclosed by 
an inner boundary $\p_-\Omega$ and an outer boundary $\p_+\Omega$. 
The conservation law shows that the net outward flux across each boundary 
is the same, 
\begin{equation}\label{div-conslaw}
\oint_{\p_-\Omega}\vec X\cdot\vec\nu_- dA = \oint_{\p_+\Omega}\vec X\cdot\vec\nu_+ dA
\end{equation}
where $\vec\nu_\mp$ is the outward unit normal of the respective boundaries. 

The most well-known method \cite{Olv,1stbook,2ndbook}
for finding conservation laws is Noether's theorem, 
which is applicable only to PDE systems that possess a variational principle. 
Noether's theorem shows that the infinitesimal symmetries of the variational principle 
yield conserved integrals \eqref{conslaw-onedim}, \eqref{conslaw-multidim}, \eqref{div-conslaw} of the PDE system, 
including conserved boundary integrals \eqref{conslaw-boundary}
when the PDE system satisfies a differential identity. 
In the case of PDE systems that possess a generalized Cauchy-Kovalevskaya form \cite{Mar-Alo,Olv}, 
all conserved integrals \eqref{conslaw-onedim}, \eqref{conslaw-multidim}, \eqref{div-conslaw} arise from Noether's theorem. 
This direct connection between conserved integrals and symmetries 
is especially useful because, 
typically, the symmetries of a given PDE system have a direct physical meaning 
related to basic properties of the system, 
while, computationally, 
all infinitesimal symmetries of a given PDE system can be found in a systematic way by solving a linear system of determining equations. 

Over the past few decades, 
a modern formulation of Noether's theorem has been developed 
in which the components of a variational symmetry are expressed 
as the components of a multiplier whose product with a given variational PDE system 
yields a total divergence that reduces on the space of solutions of the PDE system 
to a local continuity equation. 
The main advantage of this reformulation is that 
multipliers can be sought for any given PDE system, 
regardless of whether it possesses a variational principle or not. 
In general, multipliers are simply the natural PDE counterpart of 
integration factors for ordinary differential equations \cite{1stbook}, 
and for any given PDE system, 
a linear system of determining equations can be formulated \cite{Olv,2ndbook}
to yield all multipliers. 
As a consequence, local continuity equations can be derived 
without any restriction required on the nature of the PDE system. 
Moreover, for PDE systems that possess a generalized Cauchy-Kovalevskaya form, 
all conserved integrals \eqref{conslaw-onedim}, \eqref{conslaw-multidim}, \eqref{div-conslaw} arise from multipliers \cite{Mar-Alo,Olv}. 
A review of the history of Noether's theorem and of the multiplier method for finding conservation laws can be found in \Ref{Kos-Sch}. 

In recent years, 
the multiplier method has been cast into the form of a generalization of Noether's theorem 
which is applicable to PDE systems without a variational principle. 
The generalization \cite{AncBlu97,AncBlu02a,AncBlu02b,Anc03} 
is based on the structure of the determining system for multipliers,
which turns out to be an augmented, adjoint version of the determining equations for infinitesimal symmetries. 
In particular, 
multipliers can be viewed as an adjoint generalization of variational symmetries,
and most significantly, the determining system for multipliers 
can be solved by the use of the same standard procedure that is used 
for solving the determining equations for symmetries \cite{Olv,1stbook,2ndbook}. 
Moreover, the physical conservation law determined by a multiplier 
can be constructed directly from the multiplier and the given PDE system 
by various integration methods \cite{Olv,AncBlu02a,AncBlu02b,Wol02b,Anc03,2ndbook}. 

In this modern generalization, 
the problem of finding all conservation laws for a given PDE system 
thereby becomes a kind of adjoint of the problem of finding all infinitesimal symmetries. 
As a consequence, for any PDE system, 
there is no need to use special methods or anstazes 
(\eg/, \cite{Mor,KarMah,Ibr07,Ibr11,PooHer})
for determining its conservation laws, 
just as there is no necessity to use special methods or ansatzes for finding its symmetries.

The present paper is intended to review and extend these recent developments, 
with an emphasis on applications to PDE systems arising in physical models. 
The most natural mathematical framework 
for understanding the methods and the results 
is variational calculus in jet spaces \cite{Olv}. 
This framework will be given a concrete formulation, 
which is useful both for formulating general statements and for doing calculations
for specific PDE systems.  

As a starting point, in \secref{examples}, 
a wide range of examples of local conservation laws and conserved integrals
are presented, 
covering dynamical systems that model convection, diffusion, wave propagation, fluid flow, gas dynamics and plasma dynamics, 
as well as non-dynamical (static equilibrium) systems. 

In \secref{prelim}, 
for general PDE systems, 
the standard formulations of local conservation laws, conserved integrals, 
and symmetries,
as well as some other preliminaries, are stated. 
Additionally, 
local versus global aspects of conservation laws are discussed
and are related to the distinction between trivial and non-trivial conservation laws
and their physical meaning. 
This discussion clears up some confusion in the existing literature. 

In \secref{tools},
some basic modern tools from variational calculus are reviewed 
using a concrete self-contained approach. 
These tools are employed in \secref{characteristiceqns}
to derive the determining equations for multipliers and symmetries, 
based on a characteristic form for conservation laws and symmetry generators. 
An important technical step in this derivation is the introduction of 
a coordinatization for the solution space of PDE systems in jet space,
which involves expressing a given PDE system in a solved form for a set of leading derivatives,
after the system is closed by appending all integrability conditions (if any). 
This coordinatization is applicable to all PDE systems of physical interest, 
including systems that possess differential identities. 
It is used to show that the characteristic form for trivial conservation laws is given by 
trivial multipliers which vanish on the solution space of a PDE system 
when the system has no differential identities. 
This directly leads to an explicit one-to-one correspondence 
between non-trivial conservation laws and non-trivial multipliers,
taking into account the natural equivalence freedoms in conservation laws and multipliers.
An explicit generalization of this correspondence is established 
in the case when a PDE system possesses a differential identity (or set of identities). 
The generalization involves considering gauge multipliers \cite{AncPoh} 
that arise from a conservation law connected with the differential identity. 

These new results significantly broaden the explicit correspondence 
between non-trivial conservation laws and non-trivial multipliers 
previously obtained \cite{Mar-Alo,Olv,AncBlu02a,AncBlu02b} 
only by requiring PDE systems to have a generalized Cauchy-Kovalevskaya form 
(which restricts a system from possessing any differential identities).

Furthermore, as another result, 
a large class of multipliers that captures physically important conserved integrals 
such as mass, momentum, energy, angular momentum 
is identified for general PDE systems 
by examining the numerous examples of conservation laws presented earlier. 

In \secref{Noether}, 
the variational calculus tools are used to state Noether's theorem in a modern form
for variational PDE systems, 
along with the determining equations for variational symmetries. 
The generalization of Noether's theorem in modern form to non-variational PDE systems 
is explained in \secref{mainresults}. 
First, the determining equations for multipliers are shown to be 
an augmented, adjoint counterpart of the determining equations for symmetries. 
More precisely, the multiplier determining system has a natural division into two subsystems \cite{AncBlu97,AncBlu02a,AncBlu02b}. 
One subsystem is the adjoint of the symmetry determining system, 
whose solutions can be viewed as adjoint-symmetries (also known as cosymmetries).
The remaining subsystem comprises equations that are necessary and sufficient for
an adjoint-symmetry to be a multiplier, 
analogously to the conditions required for an infinitesimal symmetry to be a variational symmetry in the case of a variational PDE system. 
Next, the role of a Lagrangian in constructing a conserved integral from a symmetry of a variational principle 
is replaced for non-variational PDE systems by several different constructions:
an explicit integral formula, an explicit algebraic scaling formula,
and a system of determining equations, 
all of which use only a multiplier and the PDE system itself. 
The scaling formula is based on dimensional analysis 
and generalizes a formula previously derived only for PDE systems that admit a scaling symmetry \cite{Anc03}. 

These main results cover both the case of PDE systems without differential identities
and the case of PDE systems with differential identities. 
It is emphasized that this general method for explicitly deriving the conservation laws of PDE systems
reproduces the content of Noether's theorem whenever a PDE system has a variational principle.
(For comparison,
an abstract, cohomological approach to determining conservation laws of PDE systems
can be found in \Ref{KraVin,Ver,Hen}.)

Some concluding remarks, 
including discussion of the geometrical meaning of adjoint-symmetries and multipliers,
are provided in \secref{remarks}.

Several running examples will be used to illustrate 
the main ideas and the main results in every section.

\section{Examples}
\label{examples}

The following seven examples illustrate some basic conserved densities and fluxes \eqref{conslaw-onedim} 
arising in physical PDE systems in one spatial dimension. 

{\bf\Ex{1}}: transport equation
\begin{gather}
u_t=(c(x,u)u)_x
\label{transport-eqn}
\end{gather}
\begin{gather}
T=u 
\text{ is mass density}, 
\quad
X=-c(x,u)u 
\text{ is mass flux \ie/, momentum}.
\end{gather}

{\bf\Ex{2}}: diffusion/heat conduction equation
\begin{gather}
u_t= (k(x,u)u_x)_x
\label{diffusionheat-eqn}
\end{gather}
\begin{gather}
T=u 
\text{ is heat density (temperature)},
\quad
X=-k(x,u)u_x 
\text{ is heat flux}. 
\end{gather}

{\bf\Ex{3}}: telegraph equation
\begin{gather}
u_{tt}+a(t)u_t-(c(x)^2 u_x)_x = 0
\label{telegraph-eqn}
\end{gather}
\begin{gather}
\begin{aligned}
& T=\tfrac{1}{2}\exp(2\smallint a(t)dt)(u_t{}^2+c(x)^2 u_x{}^2)
\text{ is energy density},\\ 
& X=-c(x)^2\exp(2\smallint a(t)dt) u_xu_t
\text{ is energy flux} . 
\end{aligned}
\end{gather}

{\bf\Ex{4}}: nonlinear dispersive wave equation
\begin{gather}
u_t+f(u)u_x+u_{xxx}=0,
\quad
f(u)\neq \const
\label{dispersivewave-eqn}
\end{gather}
\begin{subequations}
\begin{align}
& T=u
\text{ is mass density},
\quad
X=\smallint f(u)du +u_{xx}
\text{ is mass flux \ie/, momentum};
\\
& T=u^2
\text{ is elastic energy density},
\quad
X=2\smallint uf(u)du +2uu_{xx}-u_x{}^2 
\text{ is elastic energy flux};
\\
&\begin{aligned} 
T & =\smallint g(u)du -\tfrac{1}{2}u_x{}^2
\text{ is gradient energy density},
\\
X & =\tfrac{1}{2}(g(u)+u_{xx})^2 +u_xu_t
\text{ is gradient energy flux},
\\&\quad
g(u)=\smallint f(u)du .
\end{aligned} 
\end{align}
\end{subequations}

{\bf\Ex{5}}: compressible viscous fluid equations
\begin{gather}
\begin{aligned}
&\rho_t+( u\rho)_x=0 \\
&\rho(u_t+uu_x) = -p_x+\mu u_{xx}
\end{aligned}
\label{viscousfluid-eqn}
\end{gather}
\begin{subequations}
\begin{align}
& T=\rho
\text{ is mass density},
\quad
X=u\rho
\text{ is mass flux};
\\
& T=\rho u
\text{ is momentum density}, 
\quad
X=p-\mu u_x+Tu
\text{ is momentum flux};
\\
&\begin{aligned} 
T &=\rho(tu-x)
\text{ is Galilean momentum density},
\\
X &=t(p -\mu u_x)+T u
\text{ is Galilean momentum flux}.
\end{aligned}
\end{align}
\end{subequations}

The next two examples  are integrable PDE systems that possess
an infinite hierarchy of higher-order conservation laws. 

{\bf\Ex{6}}: barotropic gas flow/compressible inviscid fluid equations
\begin{gather}
\begin{aligned}
& \rho_t+( u\rho)_x=0 \\
& u_t+uu_x = -p_x/\rho \\
& p=p(\rho) \text{ (barotropic equation of state) }\\
& e=\smallint p/\rho^2 d\rho \text{ (thermodynamic energy) }
\end{aligned}
\label{compressiblefluid-eqn}
\end{gather}
\begin{subequations} 
\begin{align}
& T=\rho(\tfrac{1}{2}u^2+e)
\text{ is energy density},
\quad
X=(p+T)u
\text{ is energy flux};
\\
&\begin{aligned}
T & =\rho_x/(u_x^2-p'\rho_x^2/\rho^2)
\text{ is higher-derivative quantity},
\\
X & =\rho u_x/(u_x^2-p'\rho_x^2/\rho^2)
\text{ is higher-derivative flux}. 
\end{aligned}
\label{higherTXcompressiblefluid}
\end{align}
\end{subequations} 

{\bf\Ex{7}}: breaking wave (Camassa-Holm) equation
\begin{gather}
m_t + 2 u_xm + um_x =0,
\quad
m=u-u_{xx}
\label{breakingwave-eqn}
\end{gather}
\begin{subequations} 
\begin{align}
& T =  m
\text{ is momentum density},
\quad
X = \tfrac{1}{2}(u^2-u_x^2) +um-u_{tx}
\text{ is momentum flux};
\\
& T = \tfrac{1}{2}(u^2+u_x^2)
\text{ is energy density},
\quad
X = u(um-u_{tx})
\text{ is energy flux};
\\
&\begin{aligned}
T & =  \tfrac{1}{2}u(u^2+u_x^2),
\text{ is energy-momentum density},
\\ 
X & = \tfrac{1}{2}(u_{tx}-u(m+\tfrac{1}{2}u)+\tfrac{1}{2}u_x^2)^2 -u_t(uu_x+\tfrac{1}{2}u_t) 
\text{ is energy-momentum flux};
\end{aligned}
\\
& T = m^{1/2} 
\text{ is Hamiltonian Casimir},
\quad
X = 2um^{1/2}
\text{ is Casimir flux};
\\
&\begin{aligned}
T & = m^{-5/2}m_x^2 + 4m^{-1/2} 
\text{ is higher-derivative energy density},
\\
X & = -m^{-5/2}(2m_t+um_x)m_x -4m^{-3/2}u_xm_x -4m^{-1/2}u-8m^{1/2} 
\text{ is higher-derivative flux}. 
\end{aligned}
\label{higherTXbreakingwave}
\end{align}
\end{subequations} 

The following three examples illustrate 
some intrinsically multi-dimensional conservation laws \eqref{conslaw-multidim}
that arise in physical PDE systems in two or more dimensions.

{\bf\Ex{8}}: porous media equation
\begin{gather}
u_t= \nabla\cdot(k(u)\nabla u)
\label{porousmedia-eqn}
\end{gather}
\begin{gather}
\begin{aligned}
T &=\alpha(x)u
\text{ is a general mass-density moment}, 
\\ 
\vec X &=\smallint k(u)du \nabla\alpha(x) - \alpha(x)\nabla\smallint k(u)du
\text{ is flux moment of mass-density}, 
\\&
\Delta\alpha =0
\quad\text{ (arbitrary solution of Laplace equation)}.
\end{aligned}
\end{gather}

{\bf\Ex{9}}: non-dispersive wave equation
\begin{gather}
u_{tt}-c^2\Delta u = f(u)
\label{nondispersivewave-eqn}
\end{gather}
\begin{subequations} 
\begin{align}
&\begin{aligned}
T &=u_t (\mathbf{a}\cdot\vec x)\cdot\nabla u
\text{ is angular momentum density},
\\ 
\vec X &=(\tfrac{1}{2}c^2|\nabla u|^2 -\tfrac{1}{2}u_t^2-\smallint f(u)du)\mathbf{a}\cdot\vec x -c^2((\mathbf{a}\cdot\vec x)\cdot\nabla u)\nabla u 
\text{ is angular momentum flux}, 
\\&
\text{ (arbitrary constant antisymmetric tensor $\mathbf{a}$)}.
\end{aligned}
\\
&\begin{aligned}
T &=\vec b\cdot\vec x(\tfrac{1}{2} u_t^2 +\tfrac{1}{2}c^2|\nabla u|^2 -\smallint f(u)du) +c^2tu_t \vec b\cdot\nabla u
\text{ is boost momentum density},
\\ 
\vec X &=c^2t(\tfrac{1}{2}c^2|\nabla u|^2 -\tfrac{1}{2} u_t^2-\smallint f(u)du)\vec b 
-c^2(\vec b\cdot\vec x u_t +c^2 t\vec b\cdot\nabla u)\nabla u 
\text{ is boost momentum flux}, 
\\&
\text{ (arbitrary constant antisymmetric vector $\vec b$)}.
\end{aligned}
\end{align}
\end{subequations} 

{\bf\Ex{10}}: inviscid (compressible/incompressible) fluid equation
\begin{gather}
\begin{aligned}
& \vec u_t + \vec u\cdot\nabla\vec u= -(1/\rho)\nabla p 
\\
& e=\smallint p/\rho^2 d\rho \text{ (thermodynamic energy) }
\end{aligned}
\label{inviscidfluid-eqn}
\end{gather}
\begin{subequations} 
\begin{align}
&\text{in three dimensions }
\begin{cases}
T= \vec u\cdot(\nabla\times\vec u)
\text{ is local helicity},
\\
\vec{\mathcal X}= \vec X - T\vec u = \tfrac{1}{2}(|\vec u|^2 +(p/\rho) +e) 
\text{ is moving helicity flux};
\end{cases}
\\
&\text{in two dimensions }
\begin{cases}
T= \rho f((\curl\,\vec u)/\rho)
\text{ is local enstrophy}, 
\\
\vec{\mathcal X}= \vec X - T\vec u = 0 
\text{ is moving enstrophy flux},
\\\quad 
\text{(arbitrary function $f$)}. 
\end{cases}
\end{align}
\end{subequations} 

The last four examples illustrate spatial boundary conservation laws \eqref{conslaw-boundary}
and spatial flux conservation laws \eqref{div-conslaw}
for physical PDE systems in three dimensions. 

{\bf\Ex{11}}: electric (displacement) field equation inside matter
\begin{gather}
\begin{aligned}
& \vec D_t = c\nabla\times\vec H \\
& \nabla\cdot\vec D = 4\pi \rho \\
& \vec J=0 \text{ (no currents) } \\
& \rho_t =0 \text{ (static charges) }
\end{aligned}
\end{gather}
\begin{gather}
\vec T= \vec D 
\text{ is flux density of electric field lines.}
\end{gather}

{\bf\Ex{12}}: magnetohydrodynamics (infinite conductivity) equations
\begin{gather}
\begin{aligned}
& \vec u_t + \vec u\cdot\nabla\vec u= (1/\rho)(\vec J\times\vec B -\nabla p) \\
& \vec B_t = \nabla\times(\vec u\times \vec B) \\
& \nabla\times\vec B =4\pi\vec J \\
& \nabla\cdot\vec B = 0
\end{aligned}
\end{gather}
\begin{gather}
\vec T=\vec X= \vec B 
\text{ is flux density of magnetic field lines.}
\end{gather}

{\bf\Ex{13}}: fluid incompressibility equation
\begin{gather}
\nabla\cdot\vec u=0 \\
\vec X= \vec u
\text{ is flux density of streamlines}.
\end{gather}

{\bf\Ex{14}}: charge source equation (in empty space)
\begin{gather}
\nabla\cdot\vec E=0 \\
\vec X= \vec E
\text{ is flux density of electric field lines}.
\end{gather}

\section{Conserved integrals, conservation laws, and symmetries}
\label{prelim}

Throughout, the following notation will be used. 
Let $t$, $x=(x^1,\ldots,x^n)$ be independent variables, $n\geq1$,
and let $u=(u^1,\ldots,u^m)$ be dependent variables, $m\geq1$. 
Partial derivatives of $u$ with respect to $t,x$ are denoted 
$\p u=(u_t,u_{x^1},\ldots,u_{x^n})$,
and $k$th-order partial derivatives are denoted $\p^k u$, $k\geq 2$. 
The coordinate space $J=(t,x,u,\p u,\p^2 u,\ldots)$ is called 
the {\em jet space} associated with the variables $t,x,u$. 
Partial derivatives with respect to these variables are given by 
$\parderop{t}$, 
$\parderop{x}=(\parderop{x^1},\ldots,\parderop{x^n})^\t$, 
$\parderop{u}=(\parderop{u^1},\ldots,\parderop{u^m})^\t$,
with a superscript ``$\t$'' denoting the transpose, 
and similarly for partial derivatives with respect to the derivative variables in $J$. 
Total derivatives with respect to $t,x$, acting by the chain rule, 
are denoted $D=(D_t,D_{x^1},\ldots,D_{x^n})$.
In particular, $Du=\p u$, $D\p u = \p^2 u$, and so on. 
$D^k$ denotes all of the $k$th order total derivatives with respect to $t,x$. 
Spatial divergences are denoted $\Div=D_x\cdot$, 
with a dot denoting the vector dot product. 

Consider an $N$th-order system of $M\geq 1$ PDEs 
\begin{equation}\label{pde}
G=(G^1(t,x,u,\p u,\ldots,\p^N u),\ldots,G^M(t,x,u,\p u,\ldots,\p^N u))=0 . 
\end{equation}
The space of all locally smooth solutions $u(t,x)$ of the system will be denoted $\Esp$. 
This space has an embedding as a subspace in $J$, 
since $u(t,x)\in\Esp$ determines $(t,x,u(t,x),\p u(t,x),\p^2 u(t,x),\ldots)\in J$. 
(In the applied mathematics and physics literature, 
$\Esp$ is commonly identified with the set of equations $G=0$, $DG=0$, $D^2G=0$, $\ldots$ in $J$,
which assumes these equations are locally solvable \cite{Olv}.)

A {\em local conservation law} of a given PDE system \eqref{pde} 
is a local continuity equation 
\begin{equation}\label{conslaw}
(D_t T + D_x\cdot X)|_\Esp =0
\end{equation}
which holds on the whole solution space $\Esp$ of the system, 
where $T(t,x,u,\p u,\ldots,\p^r u)$ is the {\em conserved density}
and $X=(X^1(t,x,u,\p u,\ldots,\p^r u),\ldots,X^n(t,x,u,\p u,\ldots,\p^r u))$ 
is the {\em spatial flux}. 
The pair 
\begin{equation}\label{current}
(T,X)=\Phi 
\end{equation}
is called a {\em conserved current}. 

Every conservation law \eqref{conslaw} can be integrated over 
any given spatial domain $\Omega\subseteq\Rnum^n$ to get
\begin{equation}\label{globalconslaw}
\frac d{dt} \int_{\Omega} T|_\Esp dV = -\oint_{\p\Omega} X|_\Esp\cdot\nu dA
\end{equation}
by the divergence theorem,
where $\p\Omega$ is the boundary of the domain
and $\nu$ denotes the outward pointing unit normal vector. 
This shows that the rate of change of the quantity 
\begin{equation}\label{C}
\C[u]= \int_{\Omega} T|_\Esp dV 
\end{equation}
in the domain is balanced by the net flux escaping through the domain boundary. 
The quantity \eqref{C} is called a {\em conserved integral},
and the relation \eqref{globalconslaw} is called a {\em global conservation law} or {\em global balance equation}. 

Two conservation laws are {\em locally equivalent}
if they give the same global balance equation  \eqref{globalconslaw} up to boundary terms. 
This occurs iff their conserved densities
differ by a total spatial divergence $D_x\cdot\Theta$ 
on the solution space $\Esp$,
and correspondingly, 
their fluxes differ by a total time derivative $-D_t\Theta$ 
modulo a divergence-free vector. 
A conservation law is thereby called {\em locally trivial} if 
\begin{equation}\label{trivconslaw}
T_\triv|_\Esp = D_x\cdot\Theta|_\Esp,
\quad
X_\triv|_\Esp = - D_t\Theta|_\Esp +D_x\cdot\Gamma|_\Esp
\end{equation}
holds for some vector function $\Theta(t,x,u,\p u,\ldots,\p^{r-1} u)$
and some antisymmetric tensor function $\Gamma(t,x,u,\p u,\ldots,\p^{r-1} u)$.
The {\em differential order of a conservation law} is defined to be 
the smallest differential order among all locally equivalent conserved currents.
(It is common in the mathematics literature to define a local conservation law itself
as the equivalence class of locally equivalent conserved currents.)

The global form of a locally trivial conservation law is given by 
\begin{equation}\label{loctriv}
\frac d{dt} \oint_{\p\Omega} \Theta|_\Esp \cdot\nu dA
= \oint_{\p\Omega} D_t\Theta|_\Esp\cdot\nu dA
\end{equation}
since $\oint_{\p\Omega} (D_x\cdot\Gamma)|_\Esp\cdot\nu dA=0$ by Stokes' theorem. 
This integral equation \eqref{loctriv} is just an identity, with no physical content, 
unless the spatial flux integral of $D_t\Theta|_\Esp$ vanishes. 
From the divergence theorem, 
this integral will vanish for all domains $\Omega$ iff
$D_x\cdot D_t\Theta|_\Esp = 0$ holds. 
In such cases, the boundary integral 
\begin{equation}
\int_{\Omega} T|_\Esp dV 
= \oint_{\p\Omega} \Theta|_\Esp \cdot\nu dA 
\end{equation}
will be a constant of motion for solutions of the given PDE system. 
This type of boundary conservation law arises for PDE systems typically 
when the PDEs in the system are related by obeying a differential identity,
as will be discussed further in \secref{characteristiceqns}. 
In all cases when both 
$D_x\cdot\Theta|_\Esp$  and $D_x\cdot D_t\Theta|_\Esp$ do not vanish identically, 
a locally trivial conservation law has no physical content.

For a given PDE system \eqref{pde}, 
the set of all non-trivial conservation laws (up to local equivalence)
forms a vector space on which the symmetries of the system have a natural action 
\cite{Olv,2ndbook}. 

An {\em infinitesimal symmetry} \cite{Olv,1stbook,2ndbook} 
of a given PDE system \eqref{pde} is a generator 
\begin{equation}\label{generator}
\X=\tau\parderop{t} +\xi\parderop{x} +\eta\parderop{u}
\end{equation}
whose prolongation leaves invariant the PDE system, 
\begin{equation}\label{Xsymmcond}
\pr\X(G)|_\Esp =0 
\end{equation}
which holds on the whole solution space $\Esp$ of the system. 
Here $\tau(t,x,u,\p u,\ldots,\p^r u)$, 
$\xi=(\xi^1(t,x,u,\p u,\ldots,\p^r u),\ldots,\xi^n(t,x,u,\p u,\ldots,\p^r u))$,
and $\eta=(\eta^1(t,x,u,\p u,\ldots,\p^r u),\ldots,$ 
$\eta^m(t,x,u,\p u,\ldots,\p^r u))$ 
are called the {\em characteristic functions} in the symmetry generator. 
When acting on the solution space $\Esp$, 
an infinitesimal symmetry generator can be formally exponentiated to produce
a one-parameter group of transformations $\exp(\epsilon\pr\X)$, 
with parameter $\epsilon$,
where the infinitesimal transformation is given by 
\begin{equation}
\begin{aligned}
u(t,x) \rightarrow 
u(t,x) +& \epsilon\big( \eta(t,x,u(t,x),\p u(t,x),\ldots,\p^r u(t,x)) 
\\&\qquad
-u_t(t,x)\tau(t,x,u(t,x),\p u(t,x),\ldots,\p^r u(t,x)) 
\\&\qquad
-u_x(t,x)\cdot\xi(t,x,u(t,x),\p u(t,x),\ldots,\p^r u(t,x)) \big) +O\big(\epsilon^2\big)
\end{aligned}
\end{equation}
for all solutions $u(t,x)$ of the PDE system.

Two infinitesimal symmetries are equivalent if they have the same action 
on the solution space $\Esp$ of a given PDE system. 
An infinitesimal symmetry is thereby called {\em trivial} 
if it leaves all solutions $u(t,x)$ unchanged. 
This occurs iff its characteristic functions satisfy the relation $\hat\eta|_\Esp=0$, 
where
\begin{equation}
\hat\eta = \eta - u_t\tau -u_x\cdot\xi. 
\end{equation}
The corresponding generator \eqref{generator} of a trivial symmetry 
on the solution space $\Esp$ 
is thus given by 
\begin{equation}\label{trivsymm}
\X_\triv = \tau\parderop{t} + \xi\cdot\parderop{x} +(u_t\tau +u_x\cdot\xi)\parderop{u}
\end{equation}
which has the prolongation $\pr\X_\triv=\tau D_t + \xi\cdot D_x$. 
Conversely, any generator of this form \eqref{trivsymm} represents a trivial symmetry. 
The {\em differential order of an infinitesimal symmetry} is defined to be 
the smallest differential order among all equivalent generators. 

In jet space $J$, 
a group of transformations $\exp(\epsilon\pr\X)$ 
in general will not act in a closed form 
on $t,x,u$, and derivatives $\p^k u$ up to a finite order,
except \cite{Olv,2ndbook} for point transformations acting on $(t,x,u)$,
and contact transformations acting on $(t,x,u,u_t,u_x)$. 
Moreover, a contact transformation is a prolonged point transformation 
when the number of dependent variables is $m=1$ \cite{Olv,2ndbook}.
A {\em point symmetry} is defined as a symmetry transformation group on $(t,x,u)$,
whose generator is given by characteristic functions of the form 
\begin{equation}\label{Xpointsymm}
\X=\tau(t,x,u)\parderop{t} +\xi(t,x,u)\parderop{x} +\eta(t,x,u)\parderop{u}
\end{equation}
corresponding to the infinitesimal point transformation 
\begin{equation}\label{pointgroup}
\begin{aligned}
& t\rightarrow t+\epsilon \tau(t,x,u) + O(\epsilon^2),
\\
& x\rightarrow x+\epsilon \xi(t,x,u) + O(\epsilon^2),
\\
& u\rightarrow u+\epsilon \eta(t,x,u) + O(\epsilon^2) . 
\end{aligned}
\end{equation}
Likewise, 
a {\em contact symmetry} is defined as a symmetry transformation group on $(t,x,u,u_t,u_x)$ 
whose generator corresponds to an infinitesimal transformation that preserves 
the contact relations $u_t =\p_t u$, $u_x =\p_x u$. 
The set of all admitted point symmetries and contact symmetries 
for a given PDE system comprises its group of {\em Lie symmetries}. 

Common examples of point symmetries 
admitted by PDE systems arising in physical applications 
are time translations, space translations, and scalings. 
Higher-order symmetries are typically admitted only by integrable PDE systems. 
However, 
it is worth emphasizing that any admitted symmetry can be used 
to obtain a mapping of a given solution $u=f(t,x)$ of a PDE system 
into a one-parameter family of solutions 
$u =\til f(t,x,\epsilon) 
= \big(\exp(\epsilon\pr\hat\X)u\big)|_{u=f(t,x)} 
= \big(u + \epsilon \hat\eta + \tfrac{1}{2}\epsilon^2 \pr\hat\X \hat\eta+ \cdots\big)|_{u=f(t,x)}$ 
where $\hat\X = \X-\X_\triv = \hat\eta\p/\p u$;
and also to find symmetry-invariant solutions $u=f(t,x)$ of a PDE system 
by considering the invariance condition 
$(\hat\X u)|_{u=f(t,x)} = \hat\eta|_{u=f(t,x)} =0$. 
Thus, for these two main purposes, symmetries of any differential order 
are equally useful. 

Similar remarks can be made for conservation laws. 
In physical applications, 
the most common examples of conserved densities admitted by PDE systems 
are mass, momentum, and energy.
These densities are always of low differential order,
whereas higher-order densities are typically admitted only by integrable PDE systems. 
Nevertheless, for the many purposes outlined in \secref{intro}, 
any admitted conservation law of a given PDE system can be useful.

\subsection{Regular PDE systems and computation of symmetry generators, conserved densities and fluxes}

To determine if a current \eqref{current} is conserved for a given PDE system,
and if a generator \eqref{generator} is an infinitesimal symmetry of a given PDE system, 
it is necessary to coordinatize the solution space $\Esp$ of the system in jet space $J$.
This can be accomplished in a general way by the following steps. 
First, for any PDE system \eqref{pde}, 
introduce an index notation for the components of $x$ and $u$:
$x^i$, $i=1,\ldots,n$; and $u^\alpha$, $\alpha=1,\ldots,m$. 
Next, suppose each PDE $G^a=0$, $a=1,\ldots,M$, in the given system 
can be expressed in a solved form 
\begin{equation}\label{pde-solvedform}
G^a = \p_{(\ell_a)} u^{\alpha_a} -g^a
\end{equation}
for some derivative of a single dependent variable $u^{\alpha_a}$,
after a point transformation (change of variables) if necessary, 
such that all other terms in the system contain 
neither this derivative nor its differential consequences, 
namely 
\begin{equation}\label{pde-leadingder}
\begin{gathered}
\p_{(\ell_a)}u^{\alpha_a} \neq \p^k \p_{(\ell_b)}u^{\alpha_b}, 
\quad 
a,b=1,\ldots,M,
\quad
k\geq 1 , 
\\
\parder{g^a}{(\p^k\p_{(\ell_b)} u^{\alpha_b})} =0, 
\quad 
a,b=1,\ldots,M,
\quad
k\geq 0 . 
\end{gathered}
\end{equation}
Such derivatives $\{\p_{(\ell_a)}u^{\alpha_a}\}_{a=1,\ldots, M}$
are called a set of {\em leading derivatives} for the PDE system. 
Last, suppose the given PDE system is closed in the sense that 
it has no integrability conditions and all of its differential consequences 
produce PDEs that have a solved form in terms of differential consequences of the leading derivatives. 
Note that if a PDE system is not closed then it can always be enlarged to get a closed system 
by appending any integrability conditions and differential consequences that involve the introduction of more leading derivatives. 
Then, coordinates for the solution space $\Esp$ of the closed PDE system in $J$ 
are provided by 
the independent variables $t,x^i$, 
the dependent variables $u^\alpha$, 
and all of the non-leading derivatives of $u^\alpha$. 
A closed PDE system \eqref{pde} admitting such a solved form \eqref{pde-solvedform}--\eqref{pde-leadingder}
will be called {\em regular}. 

A more restrictive class of PDE systems is given by Cauchy-Kovalevskaya systems
and their generalizations. 
Recall, a PDE system \eqref{pde} is of Cauchy-Kovalevskaya form \cite{Eva,Olv} 
if the leading derivatives in the solved form of the system 
consist of pure derivatives of $u$ with respect to a single independent variable, 
namely $\p_{(\ell_a)}u^{\alpha_a}=\p_z^{k_a}u^{\alpha_a}$, 
$a=1,\ldots,M$, $z\in \{t,x^i\}$, 
and if their differential order $k_a$ is equal to the differential order $N$ of the system,
namely $k_a=N$, $a=1,\ldots,M$. 
Cauchy-Kovalevskaya systems, 
and their generalizations \cite{Mar-Alo} in which $k_a$ differs from $N$, 
have the feature that they do not possess any differential identities 
and that none of their differential consequences possess differential identities. 
Such PDE systems are usually called {\em normal}. 
Note that, in contrast to normal systems, 
the leading derivatives in a regular PDE system can be, for instance, 
a mixed derivative of all the dependent variables $u^\alpha$ 
or a different derivative of each of the dependent variables $u^\alpha$. 

{\bf Running \Ex{(1)}}:
Generalized Korteweg-de Vries (gKdV) equation 
\begin{equation}\label{gkdv-eqn}
u_t+u^pu_x+u_{xxx}=0,
\quad
p>0 . 
\end{equation}
This is a regular PDE since it has the leading derivative 
$u_t= -u^pu_x-u_{xxx}$. 
It also has a third-order leading derivative 
$u_{xxx}=-u_t-u^pu_x$. 
Both of these solved forms are of generalized Cauchy-Kovalevskaya type. 

{\bf Running \Ex{(2)}}:
Breaking wave equation \cite{DegHonHol}
\begin{equation}\label{b-fam-eqn}
m_t + bu_xm + um_x =0,
\quad
m=u-u_{xx},
\quad
b\neq -1 . 
\end{equation}
This is a regular PDE system since it has the leading derivatives 
$m_t= -bu_xm -um_x$, $u_{xx}=u-m$. 
Equivalently, if $m$ is eliminated through the second PDE, 
this yields a scalar equation $u_t -u_{txx} + (b+1)uu_x = bu_xu_{xx} + uu_{xxx}$
which is a regular PDE with respect to the leading derivative 
\begin{equation}\label{b-fam-u-eqn}
u_{txx} =u_t + (b+1)uu_x -bu_xu_{xx} - uu_{xxx} .
\end{equation}
Neither of these solved forms are of generalized Cauchy-Kovalevskaya type. 
However, the alternative solved forms 
$u_{xxx}=u_x + u^{-1}(bu_x(u-u_{xx})- u_t+u_{txx})$
and $m_x= -u^{-1}(m_t +bu_x m)$, $u_{xx}=u-m$ 
are of generalized Cauchy-Kovalevskaya type. 

{\bf Running \Ex{(3)}}:
Euler equations for constant density, inviscid fluids in two dimensions 
\begin{equation}\label{fluid-eqn}
\begin{gathered}
\nabla\cdot\vec u =0,
\quad
\rho =\const , 
\\
\vec u_t +\vec u\cdot\nabla\vec u = -(1/\rho)\nabla p , 
\\
\Delta p = -\rho (\nabla\vec u)\cdot(\nabla\vec u)^\t . 
\end{gathered}
\end{equation}
In this system, the independent variables are $t$ and $(x,y)$,
and the dependent variables consist of $p$ and $\vec u=(u^1,u^2)$, 
in Cartesian components. 
Leading derivatives are given by writing the PDEs in the solved form 
\begin{equation*}\label{fluid-eqn-solvedform}
\begin{aligned}
u^1_x & =-u^2_y , 
\\
u^1_t &= -(u^1u^1_x + u^2u^1_y +(1/\rho)p_x)\big|{}_{u^1_x=-u^2_y}
= -(u^2u^1_y -u^1u^2_y + +(1/\rho)p_x) , 
\\
u^2_t &= -(u^1u^2_x + u^2u^2_y +(1/\rho)p_y) , 
\\
p_{xx} &= -p_{yy} -\rho((u^1_x)^2 +(u^2_y)^2 +2u^1_yu^2_x)\big|{}_{u^1_x=-u^2_y}
= -p_{yy} -2\rho((u^2_y)^2 +u^1_yu^2_x) . 
\end{aligned}
\end{equation*}
Thus, this system is a regular PDE system, 
but it does not have a generalized Cauchy-Kovalevskaya form. 
A related feature is that the PDEs in the system obey a differential identity
\begin{equation}\label{fluid-eqn-diffid}
\Div( \vec u_t +\vec u\cdot\nabla\vec u +(1/\rho)\nabla p ) 
- (D_t +\vec u\cdot\nabla)(\nabla\cdot\vec u) 
=(1/\rho)\Delta p +(\nabla\vec u)\cdot(\nabla\vec u)^\t . 
\end{equation}
Note that the pressure equation is often not explicitly considered in writing down the Euler equations. 
However, without including the pressure equation, the system would not be closed, 
since the differential identity \eqref{fluid-eqn-diffid} shows that the pressure equation 
arises as an integrability condition of the other equations. 
Correspondingly, the pressure equation does not have a solved form in terms of
differential consequences of the set of derivatives $\{u^1_x,u^1_t,u^2_t\}$. 

{\bf Running \Ex{(4)}}:
Magnetohydrodynamics equations for a compressible, infinite conductivity plasma 
in three dimensions
\begin{equation}\label{mhd-eqn}
\begin{gathered}
p=P(\rho),
\quad
\nabla\times\vec B =4\pi\vec J,
\quad
\nabla\cdot\vec B = 0 , 
\\
\rho_t +\nabla\cdot(\rho\vec u) =0 , 
\\
\vec u_t + \vec u\cdot\nabla\vec u= (1/\rho)(\vec J\times\vec B -\nabla p) , 
\\
\vec B_t = \nabla\times(\vec u\times \vec B) . 
\end{gathered}
\end{equation}
The independent variables in this system are $t$ and $(x,y,z)$,
and the dependent variables consist of $\rho$, 
$\vec u=(u^1,u^2,u^3)$ and $\vec B=(B^1,B^2,B^3)$, 
in Cartesian components. 
This is a regular PDE system, 
where a set of leading derivatives is given by writing the PDEs in the solved form 
\begin{equation*}\label{mhd-eqn-solvedform}
\begin{aligned}
B^1_x &=-B^2_y-B^3_z , 
\\
\rho_t &= \rho(u^1_x+u^2_y+u^3_z) +\rho_x u^1+\rho_y u^2+\rho_z u^3 , 
\\
u^1_t &= -(u^1u^1_x + u^2u^1_y + u^3u^1_z+(1/\rho)(P'(\rho)\rho_x +B^2J^3-B^3J^2)) , 
\\
u^2_t &= -(u^1u^2_x + u^2u^2_y + u^3u^2_z+(1/\rho)(P'(\rho)\rho_y +B^3J^1-B^1J^3)) , 
\\
u^3_t &= -(u^1u^3_x + u^2u^3_y + u^3u^3_z+(1/\rho)(P'(\rho)\rho_z +B^1J^2-B^2J^1)) , 
\\
B^1_t &= u^1B^2_y -u^2B^1_y +u^1B^3_z-u^3B^1_z +u^1_yB^2 -u^2_yB^1 +u^1_zB^3-u^3_zB^1 , 
\\
B^2_t & = (u^2B^3_z -u^3B^2_z +u^2B^1_x-u^1B^2_x +u^2_zB^3 -u^3_zxB^2+u^2_xB^1-u^1_xB^2)\big|{}_{B^1_x=-B^2_y-B^3_z} 
\\
& = -u^2B^2_y-u^3B^2_z -u^1B^2_x +u^2_zB^3 -u^3_zB^2+u^2_xB^1-u^1_xB^2 , 
\\
B^3_t & = (u^3B^1_x -u^1B^3_x +u^3B^2_y-u^2B^3_y +u^3_xB^1 -u^1_xB^3 +u^3_yB^2-u^2_yB^3)\big|{}_{B^1_x=-B^2_y-B^3_z}
\\
& = -u^3B^3_z -u^1B^3_x -u^2B^3_y +u^3_xB^1 -u^1_xB^3 +u^3_yB^2-u^2_yB^3 , 
\end{aligned}
\end{equation*}
with 
\begin{equation*}
4\pi J^1= B^3_y -B^2_z,
\quad
4\pi J^2= B^1_z-B^3_x, 
\quad
4\pi J^3= B^2_x-B^1_y . 
\end{equation*}
These PDEs lack a generalized Cauchy-Kovalevskaya form,
which is related to the feature that they obey a differential identity
\begin{equation}\label{mhd-eqn-diffid}
\Div( \vec B_t -\nabla\times(\vec u\times \vec B) )
=D_t(\nabla\cdot\vec B) .
\end{equation}

As seen from the examples here and in \secref{examples}, 
all typical PDE systems arising in physical applications belong to the class of regular systems. 

For any given regular PDE system, 
the standard approach \cite{Ovs,AndIbr,Ibr,CRC}
to look for symmetries consists of 
solving the invariance condition $\pr\X(G)|_\Esp =0$ to find 
the characteristic functions $\eta$, $\tau$, $\xi$ in the generator $\X$. 
The computations in this approach are reasonable for finding point symmetries, 
but become much more complicated for finding contact symmetries and higher-order symmetries. 

{\bf Running \Ex{(1)}}
Consider the gKdV equation \eqref{gkdv-eqn}. 
Since this is a scalar PDE, 
its Lie symmetries are generated by point transformations and contact transformations,
with the general infinitesimal form 
\begin{equation*}
\X=\tau(t,x,u,u_t,u_x)\parderop{t} +\xi(t,x,u,u_t,u_x)\parderop{x} +\eta(t,x,u,u_t,u_x)\parderop{u} . 
\end{equation*}
Substitution of this generator into the determining condition
$\pr\X(u_t+u^pu_x+u_{xxx})|_\Esp =0$ 
requires prolonging $\X$ to first-order with respect to $t$ 
and third-order with respect to $x$:
\begin{equation*}
\pr\X= \X + \eta^{(t)}\parderop{u_t} + \eta^{(x)}\parderop{u_x} + D_x^2\eta^{(xx)}\parderop{u_{xx}} + D_x^3\eta^{(xxx)}\parderop{u_{xxx}} 
\end{equation*}
where
\begin{equation*}
\begin{aligned}
\eta^{(t)} &= D_t\eta -u_tD_t\tau -u_xD_t\xi , 
\\
\eta^{(x)} &= D_x\eta -u_tD_x\tau -u_xD_x\xi , 
\\
\eta^{(xx)} &= D_x\eta^{(x)} -u_{tx}D_x\tau -u_{xx}D_x\xi , 
\\
\eta^{(xxx)} &= D_x\eta^{(xx)} -u_{txx}D_x\tau -u_{xxx}D_x\xi . 
\end{aligned}
\end{equation*}
This yields
\begin{equation*}
\begin{aligned}
& \big( 
u_x\eta +D_t\eta -u_tD_t\tau -u_xD_t\xi 
+uD_x\eta -(uu_t+3u_{txx})D_x\tau -(uu_x+3u_{xxx})D_x\xi
\\&\qquad
-3u_{tx}D_x^2\tau -3u_{xx}D_x^2\xi+ D_x^3\eta -u_tD_x^3\tau -u_xD_x^3\xi
\big)|_\Esp =0 . 
\end{aligned}
\end{equation*}
There are two steps in solving this determining condition. 
First, since the condition is formulated on the gKdV solution space $\Esp$,
a leading derivative of $u$ (and all of its differential consequences)
needs to be eliminated. 
The most convenient choice is $u_{xxx}=-u_t-u^pu_x$ 
rather than $u_t=-u^pu_x-u_{xxx}$, 
since $\tau$, $\xi$, $\eta$ depend on $u_t$. 
Next, after the total derivatives of $\tau$, $\xi$, $\eta$ are expanded out,
the resulting equation needs to be split with respect to the jet variables
$u_{tt},u_{tx},u_{xx},u_{txx},u_{xxx},u_{txxx},u_{xxxx}$
which do not appear in $\tau$, $\xi$, $\eta$.
Finally, the split equations need to be simplified, 
as some are differential consequences of others. 
After these lengthy computations and simplifications, 
a linear system of 6 determining equations is obtained for $\tau$, $\xi$, $\eta$:
\begin{gather*}
2\tau_{u_t} +u_t\tau_{u_t u_t}  +u_x\xi_{u_t u_t} -\eta_{u_t u_t} =0 , 
\\
2\xi_{u_x} +u_t\tau_{u_x u_x}  +u_x\xi_{u_x u_x} -\eta_{u_x u_x} =0 , 
\\
\tau_{u_x} +\xi_{u_t} +u_t\tau_{u_t u_x}  +u_x\xi_{u_t u_x} -\eta_{u_t u_x} =0 , 
\\
\begin{aligned}
& 3pu_t\eta +2u(u_t\tau_t+u_x\xi_t-\eta_t) 
+3p\big( u_t^2(u_t\tau_{u_t} +u_x\tau_{u_x}) 
\\&\qquad
+ u_tu_x(u_t\xi_{u_t} +u_x\xi_{u_x}) - u_t(u_t\eta_{u_t} +u_x\eta_{u_x}) \big)
=0 , 
\end{aligned}
\\
\begin{aligned}
& pu_x\eta +2u(u_t\tau_x+u_x\xi_x-\eta_x) 
+p\big( u_tu_x(u_t\tau_{u_t} +u_x\tau_{u_x}) 
\\&\qquad
+ u_x^2(u_t\xi_{u_t} +u_x\xi_{u_x}) - u_x(u_t\eta_{u_t} +u_x\eta_{u_x}) \big)
=0 , 
\end{aligned}
\\
\begin{aligned}
& \eta +u(u_t\tau_u+u_x\xi_u-\eta_u) 
+u_t(u_t\tau_{u_t} +u_x\tau_{u_x}) 
\\&\qquad
+ u_x(u_t\xi_{u_t} +u_x\xi_{u_x}) - (u_t\eta_{u_t} +u_x\eta_{u_x}) 
=0 . 
\end{aligned}
\end{gather*}
This system can be solved, with $p$ treated as an unknown, to get 
\begin{equation*}
\begin{gathered}
\tau = \til\tau(t,x,u,u_t,u_x),
\quad
\xi = \til\xi(t,x,u,u_t,u_x),
\\
\eta = u_t(\til\tau -c_1 -3c_3)+u_x(\til\xi -c_2 -c_3 x-c_4 t) -\tfrac{2}{p}c_3 u +c_4,
\quad
c_4 = 0 \text{ if $p\neq1$,}
\end{gathered}
\end{equation*}
which is a linear combination of 
a time translation ($c_1$), a space translation ($c_2$), 
a scaling ($c_3$), and a Galilean boost ($c_4$),
plus a trivial symmetry involving two arbitrary functions $\til\tau(t,x,u,u_t,u_x)$, $\til\xi(t,x,u,u_t,u_x)$. 

Clearly, for finding higher-order symmetries, 
or for dealing with PDE systems that have a high differential order 
or that involve more spatial dimensions, 
the previous standard approach becomes increasingly complicated, 
as the general solution of the symmetry determining condition 
will always contain a trivial symmetry 
involving arbitrary differential functions. 
In particular, 
the resulting linear system of determining equations for finding $\tau$, $\xi$, $\eta$ 
becomes less over-determined and hence more computationally difficult to solve
when going to higher orders. 

The situation for finding conservation laws is quite similar. 
For any given regular PDE system, 
it is possible to look for conservation laws by solving 
the local continuity equation $(D_t T + D_x\cdot X)|_\Esp =0$ 
to find $T$ and $X$. 
This approach is workable when the conserved densities $T$ and fluxes $X$ 
being sought have a low differential order 
and when the number of spatial dimensions is low. 

{\bf Running \Ex{(1)}}
Consider again the gKdV equation \eqref{gkdv-eqn}. 
This is a time evolution PDE of third order in spatial derivatives, 
while the conserved currents in lowest order form 
for mass, energy, and $L^2$ norm 
are of first order in derivatives for the densities
and of second order in derivatives for the fluxes. 
Substitution of functions 
\begin{equation*}
T(t,x,u,u_t,u_x),
\quad
X(t,x,u,u_t,u_x,u_{tt},u_{tx},u_{xx})
\end{equation*}
into the determining condition $(D_t T + D_x X)|_\Esp =0$ yields 
\begin{equation*}
\big( 
T_t + u_tT_u + u_{tx}(T_{u_x}+X_{u_t}) + u_{tt}T_{u_t} 
+ X_x + u_xX_u +u_{xx}X_{u_x} +u_{txx}X_{u_{tx}} +u_{ttx}X_{u_{tt}} +u_{xxx}X_{u_{xx}}
\big)|_\Esp =0 . 
\end{equation*}
The steps in solving this determining condition are similar to 
those used in solving the symmetry determining equation. 
First, a leading derivative of $u$ (and all of its differential consequences)
needs to be eliminated. 
The most convenient choice is $u_{xxx}=-u_t-u^pu_x$ 
rather than $u_t=-u^pu_x-u_{xxx}$, 
since $T$ and $X$ depend on $u_t$. 
Next, the resulting equation needs to be split with respect to the jet variables
$u_{ttx},u_{txx}$, which do not appear in $T$, $X$. 
This splitting immediately leads to a further splitting with respect to $u_{tx},u_{tt}$,
giving a linear system of 5 PDEs for $T$, $X$:
\begin{gather*}
T_{u_t} =0,
\quad
X_{u_{tt}} =0,
\quad
X_{u_{tx}} =0,
\quad
T_{u_x}+X_{u_t} =0, 
\\
T_t + u_tT_u  + X_x + u_xX_u +u_{xx}X_{u_x} -(u_t+u^pu_x)X_{u_{xx}} =0 . 
\end{gather*}
This system can be solved, treating $p$ as an unknown, to obtain 
\begin{equation*}
\begin{aligned}
T = & 
c_1u^2 + c_2u + c_3(\tfrac{1}{(p+1)(p+2)}u^{p+2}-\tfrac{1}{2}u_x{}^2) 
+c_4(xu-\tfrac{1}{2}tu^2)
\\&\qquad
+c_5(t(\tfrac{1}{2}u^2-3u_x{}^2)-xu^2)
+D_x\Theta(t,x,u) , 
\\
X= &
c_1(\tfrac{2}{p+2}u^{p+2}+2uu_{xx}-u_x{}^2) +c_2(\tfrac{1}{p+1}u^{p+1}+u_{xx}) 
\\&\qquad
+c_3(\tfrac{1}{2}(\tfrac{1}{p+1}u^{p+1}+u_{xx})^2 +u_xu_t) 
+c_4(x(\tfrac{1}{2}u^2+u_{xx})-t(\tfrac{1}{3}u^3 +uu_{xx}-\tfrac{1}{2}u_x{}^2)-u_x)
\\&\qquad
+c_5(t(3(\tfrac{1}{3}u^3+u_{xx})^2+6u_tu_x)+x(u_x{}^2-2uu_{xx}-\tfrac{1}{2}u^3)+2uu_x)
-D_t\Theta(t,x,u) , 
\\
& c_4 = 0 \text{ if $p\neq1$},
\quad
c_5 = 0 \text{ if $p\neq2$}
\end{aligned}
\end{equation*}
which yields a linear combination of the densities and the fluxes
representing conserved currents for the $L^2$-norm ($c_1$), 
mass ($c_2$), energy ($c_3$), Galilean momentum ($c_4$), and Galilean energy ($c_5$), 
plus a term involving an arbitrary function $\Theta(t,x,u)$
which represents a locally trivial conserved current. 

However, when going to higher orders or to higher spatial dimensions, 
it becomes increasingly more difficult to solve the local continuity equation 
$(D_t T + D_x\cdot X)|_\Esp =0$, 
as the general solution will contain 
a trivial density term $D_x\cdot\Theta$ in $T$ 
and a trivial flux term $-D_t\Theta +D_x\cdot\Gamma$ in $X$
involving a differential vector function $\Theta$ 
and a differential antisymmetric tensor function $\Gamma$, which are arbitrary. 
In particular, 
the resulting linear system of determining equations for finding $T$ and $X$ 
will be less over-determined and hence more computationally difficult to solve,
compared to the low order case or the one dimensional case. 

These difficulties motivate introducing a characteristic form 
(or canonical representation) for conserved currents 
so that all locally equivalent conserved currents have the same characteristic form,
and likewise for symmetry generators. 
To derive this formulation, 
some tools from variational calculus will be needed.

\section{Tools in variational calculus}
\label{tools}

For working with symmetries and conservation laws of PDE systems,
the natural setting in which to apply variational calculus is 
the space of {\em differential functions} 
defined by locally smooth functions of finitely many variables 
in jet space $J=(t,x,u,\p u,\p^2 u,\ldots)$.

As examples, 
in the nonlinear dispersive wave equation \Ex{4}, 
if the constitutive nonlinearity function $f(u)$ is smooth, 
then the conserved density and flux for mass and energy 
are smooth functions of $u,u_x,u_{xx}$ in $J$,
but if $f(u)$ blows up when $u=0$ then these functions are singular 
at points in $J$ such that $u=0$;
in the barotropic gas flow \Ex{6}, 
the higher-derivative density and flux are singular functions of $\rho,u,\rho_x,u_x$ 
at points in $J$ where $u_x^2=p(\rho)'/\rho$, 
but at all other points these functions are smooth. 

The basic tools that will be needed from variational calculus are 
the Fr\'echet derivative and adjoint derivative, 
the Euler operator,
a homotopy integral,
a total null-divergence identity, 
and a scaling identity. 
Throughout, 
$f(t,x,u,\p u,\ldots,\p^k u)$ denotes a differential function of order $k\geq 0$,
and $v=(v^1(t,x,u,\p u,\p^2 u,\ldots),\ldots,v^m(t,x,u,\p u,\p^2 u,\ldots))$, 
$w(t,x,u,\p u,\p^2 u,\ldots)$ 
denote differential functions of arbitrary finite order. 

The {\em Fr\'echet derivative} of a differential function is 
the linearization of the function as defined by 
\begin{equation}\label{frechet}
\begin{aligned}
\delta_v f &
= \parder{}{\epsilon}f(t,x,u+\epsilon v,\p (u+\epsilon v),\ldots,\p^k(u+\epsilon v))\big|_{\epsilon =0}
\\&
= v\parder{f}{u} + Dv\cdot\parder{f}{(\p u)} +\cdots +D^kv\cdot\parder{f}{(\p^k u)}
\end{aligned}
\end{equation}
which can be viewed as a local directional derivative in jet space,
corresponding to the action of a generator $\hat\X=v\p_u$ in characteristic form, 
$\hat\X(f) = \delta_v f$. 
It is useful also to view the Fr\'echet derivative 
as a linear differential operator acting on $v$. 
Then the relation 
\begin{equation}\label{adjoint}
w\delta_v f - v\delta^*_w f = D\cdot\Psi(v,w;f)
\end{equation}
as obtained using integration by parts 
defines the {\em Fr\'echet adjoint derivative}
\begin{equation}\label{adjfrechet}
\delta^*_w f = 
w\parder{f}{u} -D\cdot\Big(w\parder{f}{(\p u)}\Big) +\cdots
+(-D)^k\cdot\Big(w\parder{f}{(\p^k u)}\Big) 
\end{equation}
which is a linear differential operator acting on $w$. 
The associated current $\Psi(v,w;f)=(\Psi^t,\Psi^x)$ is given by 
\begin{equation}\label{frechetcurrent}
\begin{aligned}
\Psi(v,w;f) & 
= vw \parder{f}{(\p u)}
+ (Dv)\cdot\Big(w\parder{f}{(\p^2 u)}\Big)  -vD\cdot\Big(w\parder{f}{(\p^2 u)}\Big) +\cdots 
\\&\qquad
+\sum_{l=1}^{k} (D^{k-l}v)\cdot\Big((-D)^{l-1}\cdot\Big(w\parder{f}{(\p^k u)}\Big)\Big) .
\end{aligned}
\end{equation}
An alternative notation for the Fr\'echet derivative and its adjoint is 
$\delta_v f =f'(v)$ and $\delta^*_w f = f'{}^*(w)$, 
or sometimes $\delta_v f=D_v f$ and $\delta^*_w f=D_w^* f$. 

The Fr\'echet derivative of a differential function $f$ 
can be inverted to recover $f$ 
by using a line integral along any curve $C$ in $J$,
where the endpoints $\p C$ are given by 
a general point $(t,x,u,\p u,\ldots,\p^k u) \in J$ 
and any chosen point $(t,x,u_0,\p u_0,\ldots,\p^k u_0) \in J$
at which $f$ is non-singular. 
This yields 
\begin{equation}\label{line-integral}
f\big|_{\p C} = 
\int_C \parder{f}{u^\t}du^\t + \parder{f}{(\p u^\t)}\cdot d\p u^\t +\cdots 
+\parder{f}{(\p^k u^\t)}\cdot d\p^k u^\t . 
\end{equation}
If the curve $C$ is chosen so that the contact relations hold, 
$d\p u|_C = \p du|_C,\ldots,d\p^k u|_C = \p^k du|_C$, 
then the line integral becomes a general homotopy integral 
\begin{equation}\label{invfrechet}
f = f\big|_{u=u_0} + \int_{0}^{1} (\delta_v f)\Big|_{v=\p_\lambda u_{(\lambda)},u=u_{(\lambda)}}\;d\lambda, 
\quad
u_{(1)} = u,
\quad
u_{(0)} = u_0
\end{equation}
where $u_{(\lambda)}(t,x)$ is a homotopy curve given by a parametric family of functions. 
If $f$ is non-singular when $u=0$, 
then the homotopy curve can be chosen simply to be a homogeneous line, 
which yields a standard linear-homotopy integral \cite{Olv}
\begin{equation}
f = f\big|_{u=0} + \int_{0}^{1} (\delta_u f)\big|_{u=u_{(\lambda)}}\frac{d\lambda}{\lambda},
\quad
u_{(\lambda)} = \lambda u .
\end{equation}

The {\em Euler operator} $E_u$ is defined in terms of the Fr\'echet derivative 
through the relation 
\begin{equation}\label{frechet-euler}
\delta_v f = vE_u(f) +D\cdot\Upsilon_f(v)
\end{equation}
obtained from integration by parts, 
which gives
\begin{equation}\label{eulerop}
E_u(f) = 
\parder{f}{u} -D\cdot\Big(\parder{f}{(\p u)}\Big) + \cdots 
+ (-D)^k\cdot\Big(\parder{f}{(\p^k u)}\Big)
\end{equation}
where 
\begin{equation}\label{frechet-euler-current}
\begin{aligned}
\Upsilon_f(v) = \Psi(v,1;f)  & = 
v\parder{f}{(\p u)} 
+ Dv\cdot\parder{f}{(\p^2 u)} - vD\cdot\parder{f}{(\p^2 u)} +\cdots 
\\&\qquad
+ \sum_{l=1}^{k}(D^{k-l}v)\cdot\Big((-D)^{l-1}\cdot\parder{f}{(\p^k u)}\Big)  
= \sum_{l=0}^{k-1}(D^lv)\cdot E_{\p^{l+1}u}(f) . 
\end{aligned}
\end{equation}

The Euler-Lagrange relation \eqref{frechet-euler} can be combined with the general homotopy integral \eqref{invfrechet}
to obtain the following useful formula. 

\begin{lem}\label{inveulerlagr-lemma}
\begin{equation}\label{inveulerlagr}
f=\int_0^1 \p_\lambda u_{(\lambda)} E_u(f)\big|_{u=u_{(\lambda)}}\;d\lambda
+ D\cdot F
\end{equation}
is an identity, 
where
\begin{equation}\label{inveulerlagr-current}
F= \int_0^1\Upsilon_f(\p_\lambda u_{(\lambda)})\big|_{u=u_{(\lambda)}}\;d\lambda +F_0
\end{equation}
with $F_0=(F_0^t(t,x),F_0^x(t,x))$ being any current such that $D\cdot F_0=f|_{u=u_0}$. 
\end{lem}

A useful relation is 
\begin{equation}\label{highereuler-current}
\Upsilon_f(v) = 
vE_u^{(1)}(f) + D\cdot(vE_u^{(2)}(f)) +\cdots
+ D^{k-1}\cdot(vE_u^{(k)}(f))
\end{equation}
which arises through repeated integration by parts on the expression \eqref{frechet-euler-current},
where 
\begin{equation}\label{highereulerop}
E_u^{(l)}(f)  = 
\parder{f}{(\p^l u)} -\binom{l+1}{l}D\cdot\Big(\parder{f}{(\p^{l+1} u)}\Big) + \cdots 
+ \binom{k}{l}(-D)^{k-l}\cdot\Big(\parder{f}{(\p^k u)}\Big),
\quad
l=1,\ldots,k
\end{equation}
define the {\em higher Euler operators}. 
Equations \eqref{frechet-euler} and \eqref{highereuler-current} 
then provide an alternative formula 
for the Fr\'echet derivative
\begin{equation}\label{frechet-euler-rel}
\delta_v f = vE_u(f) + D\cdot(vE_u^{(1)}(f)) + \cdots + D^k\cdot(vE_u^{(k)}(f))
\end{equation}
which leads to a similar formula for the Fr\'echet adjoint derivative
\begin{equation}\label{adjfrechet-euler-rel}
\delta_w^* f = wE_u(f) -(Dw)\cdot E_u^{(1)}(f) + \cdots + (-D)^kw\cdot E_u^{(k)}(f)
\end{equation}
after integration by parts. 
Explicit coordinate formulas for all of the Euler operators 
are stated in \Ref{Olv}; 
coordinate formulas for the Fr\'echet derivative and its adjoint,
as well as the associated divergence, are shown in \Ref{AncKar}. 

The Euler operators \eqref{eulerop} and \eqref{highereulerop} 
have the following important properties. 

\begin{lem}\label{eulerop-lemma}
(i) $E_u(fg) = \delta_g^* f + \delta_f^* g$ is a product rule. 
(ii) $E_u(f)=0$ holds identically iff $f=D\cdot F$
for some differential current function $F=(F^t,F^x$). 
(iii) $E_u^{(1)}(D\cdot F) = E_u(F^t,F^x) = (E_u(F^t),E_u(F^x))$ 
and 
$E_u^{(l+1)}(D\cdot F) = (E_u^{(l)},E_u^{(l)})\odot (F^t,F^x)$, $l\geq 1$, 
are descent rules,
where $\odot$ denotes the symmetric tensor product. 
\end{lem}

The proof of (i) is an immediate consequence of the ordinary product rule 
applied to each partial derivative term in $E_u(fg)$. 
To prove the first part of (ii), 
if $f=D\cdot F$ then 
$\delta_v f = D\cdot\delta_v F$ combined with the Euler-Lagrange relation \eqref{frechet-euler}
yields $vE_u(f)= D\cdot(\delta_v F-\Upsilon_f(v))$. 
Since $v$ is an arbitrary differential function, this implies $E_u(f)=0$
(and $\Upsilon_f(v) = \delta_v F$ modulo a divergence-free term). 
Conversely, for the second part of (ii), 
if $E_u(f)=0$ then the general homotopy integral \eqref{inveulerlagr} 
shows $f=D\cdot F$ holds,
with $F$ given by the formula \eqref{inveulerlagr-current}. 
The proof of (iii) starts from the property $\delta_v(D\cdot F)= D\cdot\delta_v F$. 
Next, the Fr\'echet derivative relation \eqref{frechet-euler-rel} is applied separately to 
$f=D\cdot F$ and $f=F$. 
This yields $D\cdot(vE_u^{(1)}(D\cdot F))= D\cdot(vE_u(F))$, 
$D^2\cdot(vE_u^{(2)}(D\cdot F))= D\cdot(vD\cdot E_u^{(1)}(F))$, and so on. 
The expressions for $E_u^{(1)}(D\cdot F)$, $E_u^{(2)}(D\cdot F)$, and so on 
are then obtained by recursively expanding out each Euler operator in components 
$E_u^{(1)}= E_u^{(t,x)}= (E_u^{(t)}, E_u^{(x)})$
and $E_u^{(l+1)}= E_u^{(l,t,x)}= (E_u^{(l,t)}, E_u^{(l,x)})$, $l\geq 1$, 
followed by symmetrizing over these components together with the components of $F=(F^t,F^x)$. 
This completes the proof of \lemref{eulerop-lemma}. 

A {\em null-divergence} is a total divergence $D\cdot\Phi=0$ 
vanishing identically in jet space, 
where $\Phi=(\Phi^t,\Phi^x)$ is a differential current function. 
Similarly to Poincar\'e's lemma,
which shows that ordinary divergence-free vectors in $\Rnum^n$
can be expressed as curls, 
null-divergences are total curls in jet space. 

\begin{lem}\label{null-div-lemma}
If a differential current function 
$\Phi=(\Phi^t(t,x,u,\p u,\ldots,\p^k u),\Phi^x(t,x,u,\p u,$ $\ldots,\p^k u))$ 
has a null-divergence,
\begin{equation}\label{total-null-div}
D\cdot\Phi = D_t\Phi^t +D_x\cdot\Phi^x =0 
\text{ in $J$,}
\end{equation}
then it is equal to a total curl
\begin{equation}\label{total-curl}
\Phi=D\cdot\Psi = (D_x\cdot\Theta,-D_t\Theta+D_x\cdot\Gamma)
\text{ in $J$} 
\end{equation}
with
\begin{equation}\label{curl-terms}
\Psi=\begin{pmatrix} 0 & \Theta\\ -\Theta & \Gamma\end{pmatrix}
\end{equation}
holding for some differential vector function 
$\Theta(t,x,u,\p u,\ldots,\p^{k-1} u)$ 
and some differential antisymmetric tensor function 
$\Gamma(t,x,u,\p u,\ldots,\p^{k-1} u)$,
both of which can be expressed in terms of $\Phi^t,\Phi^x$. 
\end{lem}

The proof begins by taking the Fr\'echet derivative of the null-divergence
to get $D\cdot\delta_v\Phi = 0$. 
A descent argument will be used to solve this equation. 
Let the terms in $\delta_v\Phi=(\delta_v\Phi^t,\delta_v\Phi^x)$
containing highest derivatives $\p^k v$ be denoted 
$(T^{(k)}\p^k v,X^{(k)}\p^k v)$,
where the coefficients $T^{(k)}$ and $X^{(k)}$ of each term 
are given by a differential scalar function and a differential vector function in $J$. 
Then the highest derivative terms $\p^{k+1}v$ 
in the equation $D\cdot\delta_v\Phi = 0$ consist of 
$T^{(k)}\p_t\p^k v +X^{(k)}\cdot\p_x\p^k v$.
The coefficients of $\p^{k+1}v$ in this expression must vanish,
which can be shown to give 
$T^{(k)}\p^k v = \theta^{(k-1)}\cdot\p_x\p^{k-1}v$
and 
$X^{(k)}\p^k v = -\theta^{(k-1)}\p_t\p^{k-1}v + \gamma^{(k-1)}\cdot\p_x\p^{k-1}v$, 
where $\theta^{(k-1)}$ is some differential vector function, 
and $\gamma^{(k-1)}$ is some differential antisymmetric tensor function. 
Integration by parts on these expressions yields
\begin{gather*}
T^{(k)}\p^k v = D_x\cdot(\theta^{(k-1)}\p^{k-1}v)+\lot, 
\\
X^{(k)}\p^k v = -D_t(\theta^{(k-1)}\p^{k-1}v) + D_x\cdot(\gamma^{(k-1)}\p^{k-1}v)+\lot, 
\end{gather*}
and hence
\begin{equation*}
(T^{(k)}\p^k v,X^{(k)}\p^k v) = D\cdot\Psi^{(k-1)}(v) +\lot
\end{equation*}
where
\begin{equation*}
\Psi^{(k-1)}(v)=
\begin{pmatrix} 0 & \Theta^{(k-1)}(v)\\ -\Theta^{(k-1)}(v) & \Gamma^{(k-1)}(v)\end{pmatrix}
\end{equation*}
with $\Theta^{(k-1)}(v) = \theta^{(k-1)}\p^{k-1}v$ 
and $\Gamma^{(k-1)}(v) = \gamma^{(k-1)}\p^{k-1}$.
This shows that the highest derivative terms in $\delta_v\Phi$ 
have the form of a total curl, modulo lower order terms. 
Subtraction of this curl $D\cdot\Psi^{(k-1)}(v)$ from $\delta_v\Phi$ 
will now eliminate all terms containing $\p^k v$, so that 
\begin{equation*}
\delta_v\Phi-D\cdot\Psi^{(k)}(v) = (T^{(k-1)}\p^{k-1}v,X^{(k)}\p^{k-1}v) +\lot
\end{equation*}
where the coefficients $T^{(k-1)}$ and $X^{(k-1)}$ of the $\p^{k-1}v$ terms 
are again a differential scalar function and a differential vector function in $J$. 
Since total curls have a vanishing total divergence, 
the highest derivative terms remaining in the null-divergence equation 
$0=D\cdot\delta_v\Phi$ are given by 
$T^{(k-1)}\p_t\p^{k-1}v +X^{(k-1)}\cdot\p_x\p^{k-1}v$,
which has the same form as the expression obtained at highest order. 
This completes the first step in the descent argument. 
Continuing to lower orders, 
the descent argument will terminate at the equation 
$T^{(0)}\p_tv +X^{(0)}\cdot\p_xv = 0$,
which yields $T^{(0)}=0$ and $X^{(0)}=0$. 
As a result, 
the solution of the null-divergence equation $D\cdot\delta_v\Phi = 0$ 
is given by $\delta_v\Phi= \sum_{l=1}^{k} D\cdot\Psi^{(l-1)}(v)$.  

The final step in the proof is simply to apply the general homotopy integral \eqref{invfrechet}
to the Fr\'echet derivative $\delta_v\Phi= \sum_{l=1}^{k} D\cdot\Psi^{(l-1)}(v)$, 
which gives 
\begin{equation*}
\Phi -\Phi\big|_{u=u_0} = 
\int_{0}^{1} \big(\sum_{l=1}^{k} D\cdot\Psi^{(l-1)}(\p_\lambda u_{(\lambda)})\big)\big|_{u=u_{(\lambda)}} \;d\lambda . 
\end{equation*}
This shows $\Phi=D\cdot\Psi$ is a total curl, 
where 
\begin{equation*}
\Psi = \Psi_0 + \int_{0}^{1} \big(\sum_{l=1}^{k} \Psi^{(l-1)}(\p_\lambda u_{(\lambda)})\big)\big|_{u=u_{(\lambda)}}\;d\lambda
\end{equation*}
has the form \eqref{curl-terms},
with $D\cdot\Psi_0$ being an ordinary curl determined by Poincare's lemma 
applied to the vanishing divergence $D\cdot(\Phi|_{u=u_0})=0$.
This completes the proof of \lemref{null-div-lemma}. 

{\em Scaling transformations} are a one-parameter Lie group whose action is given by 
\begin{equation}\label{scaling}
t\rightarrow \lambda^a t,
\quad
x^i\rightarrow \lambda^{b_{(i)}} x^i,
\quad
u^\alpha\rightarrow \lambda^{c_{(\alpha)}} u^\alpha,
\quad
\lambda \neq 0
\end{equation}
prolonged to jet space,
where the constants $a,b_{(i)},c_{(\alpha)}$ are the scaling weights of $t,x^i,u^\alpha$. 
Note the generator of these transformations is $\X_\scal = \tau \p_t + \xi\p_x + \eta \p_u$ 
where 
\begin{equation}
\tau = at,
\quad
\xi =(b_{(1)}x^1,\ldots,b_{(n)}x^n),
\quad
\eta = (c_{(1)} u^1,\ldots,c_{(m)} u^m) . 
\end{equation}
In characteristic form, 
the scaling generator is $\hat\X_\scal = P_\scal \p_u$
with $P_\scal = \eta -u_t\tau - u_x\cdot\xi$. 
Now consider a differential function $f$ that is homogeneous under the action of the scaling transformation \eqref{scaling}, 
such that $f\rightarrow \lambda^s f$. 
Then the infinitesimal action is given by 
$\hat\X_\scal(f) = \delta_{P_\scal} f = s f -\tau D_t f -\xi\cdot D_xf$. 
A useful identity comes from integrating this expression by parts 
and combining it with the Euler-Lagrange relation \eqref{frechet-euler},
yielding 
\begin{equation}\label{inveulerlagr-scaling}
\omega f= P_\scal E_u(f) +D_t F^t +D_x\cdot F^x, 
\quad
\omega= s +D_t\tau +D_x\cdot\xi = s +a + \sum_{i=1}^{n} b_{(i)}
\end{equation}
where
\begin{equation}\label{inveulerlagr-scalingcurrent}
F^t = f\tau  + \Upsilon^t_f(P_\scal), 
\quad
F^x = f\xi  + \Upsilon^x_f(P_\scal)
\end{equation}
with $\Upsilon_f=(\Upsilon^t_f,\Upsilon^x_f)$ given by expression \eqref{frechet-euler-current}.
Note here $\omega$ is equal to the scaling weight of the integral quantity 
$\int_{t_0}^{t_1}\int_\Omega f\;dV\;dt$, 
as defined on any given spatial domain $\Omega\subseteq\Rnum^n$
and any time interval $[t_0,t_1]\subseteq\Rnum$. 

Finally, 
for subsequent developments, 
the following technical result 
(which is a straightforward application of Hadamard's lemma \cite{Nes} 
to the setting of jet space)
will be useful.

\begin{lem}\label{hadamard}
If a differential function $f(t,x,u,\p u,\ldots,\p^k u)$ vanishes 
on the solution space $\Esp$ of a given regular PDE system \eqref{pde}, 
then 
\begin{equation}\label{fRrel}
f= R_f(G)
\end{equation}
holds identically,
where 
\begin{equation}\label{Rop}
R_f = R_f^{(0)} + R_f^{(1)}\cdot D + \cdots + R_f^{(k-N)}\cdot D^{k-N} 
\end{equation}
is a linear differential operator, depending on $f$, 
with coefficients given by differential functions 
$R_f^{(0)}$, $R_f^{(1)}$, $\ldots$, $R_f^{(k-N)}$ that are non-singular 
when evaluated on $\Esp$. 
The operator $R_f|_\Esp$ is canonically determined by the function $f$ 
if the PDE system has no differential identities. 
Otherwise, if the PDE system satisfies a differential identity
\begin{equation}\label{diffid}
\D(G)=\D_1G^1+\cdots+ \D_MG^M=0
\end{equation}
with $\D_1,\ldots,\D_M$ being linear differential operators 
whose coefficients are non-singular differential functions 
when evaluated on $\Esp$, 
then the operator $R_f|_\Esp$ is canonically determined only modulo $\chi\D$,
where $\chi$ is an arbitrary differential function. 
\end{lem}

The proof relies heavily on the coordinatization property \eqref{pde-solvedform} 
that characterizes a PDE system being regular. 
For a regular PDE system $G=0$ of order $N\geq 1$, 
consider its prolongation to order $k\geq 1$, 
$\pr G=(G,DG,\ldots,D^k G)=0$,
which has differential order $k+N$. 
Let $(\zeta^1-g^1(Z),\zeta^2-g^2(Z),\ldots)$ 
be the solved-form derivative expressions for the PDEs in $\pr G$, 
where $\zeta=(\zeta^1,\zeta^2,\ldots)$ $\in J$ 
denotes the leading derivatives of $u^\alpha$ chosen for the prolonged system,
and $Z=(Z^1,Z^2,\ldots)$ $\in J$ 
denotes the coordinates for the prolonged solution space $\Esp\subset J$ of the system.  
Note that $\pr G=0$ represents $\Esp$ as a set of surfaces 
$\zeta^1=g^1(Z)$, $\zeta^2=g^2(Z)$ ,$\ldots$ in $J$.  
Then we have $f(\zeta,Z)|_\Esp = f(g(Z),Z)=0$. 
We now use the standard line integral identity
\begin{equation*}
f(\zeta,Z)= \int_{g(Z)}^{\zeta} \p_y f(y,Z)\cdot dy 
= \int_0^1 (\zeta-g(Z))\cdot\p_\zeta f(s\zeta+(1-s)g(Z),Z)\;ds .
\end{equation*}
This shows that $f(\zeta,Z)= F(\zeta,Z)\cdot(\zeta-g(Z))$, 
with $F(\zeta,Z)= \int_0^1 \p_\zeta f(s\zeta+(1-s)g(Z),Z)\;ds$ being a vector function. 
Note $F(\zeta,Z)|_\Esp = F(g(Z),Z) =\p_\zeta f(g(Z),Z)$ 
is non-singular since $f$ is a differential function. 
Hence we obtain $F(\zeta,Z)\cdot(\zeta-g(Z))= R_f(G)$ 
where $R_f$ is a linear differential operator 
whose coefficients $F^1(\zeta,Z)$, $F^2(\zeta,Z)$, $\ldots$ 
are non-singular when evaluated on $\Esp$. 
Furthermore, 
the expression for $F(\zeta,Z)$ shows that it is canonically determined by $f$, 
unless the PDE system satisfies a differential identity,
whereby $0=\D(G)=h(Z)\cdot(\zeta-g(Z))$ 
holds identically for some vector function $h(Z)$. 
In this case, $R_f(G)$ is well-defined only modulo $\chi\D(G)=0$,
where $\chi$ is any differential function. 
This completes the proof of \lemref{hadamard}.

\section{Characteristic forms and determining equations for\\ conservation laws and symmetries}
\label{characteristiceqns}

Consider an infinitesimal symmetry \eqref{generator} of a regular PDE system \eqref{pde}. 
When acting on the solution space $\Esp$ of the PDE system in jet space $J$, 
the symmetry generator is equivalent to a generator given by 
\begin{equation}\label{symmchar}
\hat\X=\X-\X_\triv = P\parderop{u}, 
\quad 
P =\eta-u_t\tau -u_x\cdot\xi 
\end{equation}
under which $u$ is infinitesimally transformed 
while $t,x$ are invariant. 
This generator \eqref{symmchar} defines the {\em characteristic form} 
(or canonical representation) for the infinitesimal symmetry. 
The symmetry invariance \eqref{Xsymmcond} of the PDE system 
can then be expressed by 
\begin{equation}\label{symmdeteq}
\pr\hat\X(G)|_\Esp = 0 
\end{equation}
holding on the whole solution space $\Esp$ of the given system. 
Note that the action of $\pr\hat\X$ is the same as a Fr\'echet derivative \eqref{frechet},
and hence an equivalent, modern formulation \cite{Olv,2ndbook} of this invariance \eqref{symmdeteq} 
is given by the {\em symmetry determining equation}
\begin{equation}\label{modernsymmdeteq}
(\delta_P G)|_\Esp = 0 . 
\end{equation}

This formulation of infinitesimal symmetries has several advantages compared to
the standard formulation shown in \secref{prelim}. 
Firstly, 
a symmetry is trivial iff its characteristic function $P$ vanishes on $\Esp$. 
Also, the differential order of a symmetry is simply given by 
the differential order of $P|_\Esp$. 
Secondly, 
the symmetry determining equation \eqref{modernsymmdeteq} can be set up
without doing any prolongations of the generator \eqref{symmchar}, 
as only total differentiation is needed. 
Thirdly, 
when contact symmetries or higher-order symmetries are sought, 
the generator can be formulated simply as 
\begin{equation}\label{symmP}
\hat\X = P(t,x,u,\p u,\ldots,\p^r u)\p_u
\end{equation}
with the symmetry determining equation then being a linear PDE 
for the characteristic function $P$. 
This formulation \eqref{symmP} eliminates arbitrary functions 
depending on all of the variables $t,x,u,\p u,\ldots,\p^r u$ in the solution for $P$. 

Now consider a conservation law \eqref{conslaw} of a regular PDE system \eqref{pde}. 
The starting point to obtain an equivalent characteristic form of the conservation law 
is provided by equations \eqref{fRrel} and \eqref{Rop} in \lemref{hadamard}.
These equations show that the conservation law can be expressed 
as a divergence identity 
\begin{equation}\label{conslawoffE}
D_t T +D_x\cdot X = R_\Phi(G)
= R_\Phi^{(0)}G^\t + R_\Phi^{(1)}\cdot DG^\t + \cdots + R_\Phi^{(r+1-N)}\cdot D^{r+1-N}G^\t
\end{equation}
which is obtained by moving off solutions of the PDE system,
where $u(t,x)$ is an arbitrary (sufficiently smooth) function. 
Here $r$ is the differential order of the conserved current $\Phi=(T,X)$,
and $N$ is the differential order of the PDE system. 
The next step is to integrate by parts on the righthand side 
in the divergence identity \eqref{conslawoffE}, 
yielding
\begin{equation}\label{chareqn}
D_t\til T +D_x\cdot\til X= GQ
\end{equation}
with 
\begin{equation}\label{equivTX}
\begin{aligned}
(\til T, \til X) = 
& (T,X) + R_\Phi^{(1)}G^\t + R_\Phi^{(2)}\cdot DG^\t -(D\cdot R_\Phi^{(2)})G^\t 
\\&\qquad
+\cdots 
+ \sum_{l=0}^{r-N}\big((-D)^{l}\cdot R_\Phi^{(r+1-N)}\big)\cdot D^{r-N-l}G^\t
\end{aligned}
\end{equation}
and
\begin{equation}\label{Q}
Q^\t = (Q_1,\ldots,Q_M)
= R_\Phi^{(0)} - D\cdot R_\Phi^{(1)} +\cdots +(-D)^{r+1-N}\cdot R_\Phi^{(r+1-N)} . 
\end{equation}
On the solution space $\Esp$, 
note that $(\til T,\til X)|_\Esp = (T,X)|_\Esp$ reduces to 
the conserved density and the flux in the given conservation law 
$(D_t T +D_x\cdot X)|_\Esp=0$, 
and hence 
\begin{equation}
(D_t\til T +D_x\cdot\til X)|_\Esp= 0
\end{equation}
is a locally equivalent conservation law.
The identity \eqref{chareqn} is called the {\em characteristic equation}
for the conservation law, 
and the set of functions \eqref{Q} is called the {\em multiplier}. 
Explicit coordinate formulas for the density $\til T$ and the flux $\til X$ in terms of $T$ and $X$ 
are shown in \Ref{AncKar}. 

When a regular PDE system is expressed in 
a solved form \eqref{pde-solvedform}--\eqref{pde-leadingder} 
for a set of leading derivatives,
note that these leading derivatives (and their differential consequences) 
can be eliminated from the expression for a conserved current $\Phi=(T,X)$ 
without loss of generality,
since this only changes the conserved current by the addition of a locally trivial current. 
Then it is straightforward to derive explicit expressions for the coefficient functions in the operator $R_\Phi$ by applying the chain rule to $D_t T$ and $D_{x^i} X^i$
with the use of subleading derivatives defined by the relations 
$\p_t\p_{(\ell_a/t)} u^{\alpha_a} = \p_{x^i}\p_{(\ell_a/x^i)} u^{\alpha_a}
= \p_{(\ell_a)} u^{\alpha_a}$. 
This leads to an explicit Euler-Lagrange expression 
\begin{equation}\label{QfromTX}
Q^\t= \big( 
E_{\p_{(\ell_1/t)} u^{\alpha_1}}(T) +\sum_{i=1}^{n}E_{\p_{(\ell_1/x^i)} u^{\alpha_1}}(X^i),\ldots,
E_{\p_{(\ell_M/t)} u^{\alpha_M}}(T) +\sum_{i=1}^{n}E_{\p_{(\ell_M/x^i)} u^{\alpha_M}}(X^i)
\big)
\end{equation}
for the components of the multiplier \eqref{Q}, 
where $\p_{(\ell_a/t)} u^{\alpha_a}$ and $\p_{(\ell_a/x^i)} u^{\alpha_a}$ 
denote the subleading derivatives. 
As a result, the multiplier components \eqref{QfromTX} 
can contain leading derivatives $\p_{(\ell_a)} u^{\alpha_a}$ 
(and their differential consequences) at most polynomially.

Also note that, as asserted by \lemref{hadamard}, 
if a regular PDE system has no differential identities \eqref{diffid},
then the operator $R_\Phi|_\Esp$ will be canonically determined by the expression for $\Phi=(T,X)$. 
This implies the relation 
\begin{equation}
Q^\t|_\Esp = (Q_1,\ldots,Q_M)|_\Esp
= E_G(D_t\til T+D_x\cdot\til X)|_\Esp 
= (E_{G^1}(D_tT+D_x\cdot X),\ldots,E_{G^M}(D_tT+D_x\cdot X))|_\Esp
\end{equation}
for the multiplier \eqref{Q}. 

In general, for a given regular PDE system \eqref{pde}, 
a set of functions 
\begin{equation}\label{generalQ}
Q= (Q_1(t,x,u,\p u,\p^2 u,\ldots\p^r u),\ldots,Q_M(t,x,u,\p u,\p^2 u,\ldots\p^r u))^\t
\end{equation}
will be a multiplier iff each function is non-singular on the PDE solution space $\Esp$ 
and their summed product with the expressions $G=(G^1,\ldots,G^M)$ for the PDEs 
has the form of a total space-time divergence. 

The characteristic equation \eqref{chareqn} establishes that, up to local equivalence, 
all non-trivial conservation laws for any regular PDE system arise from multipliers. 
A determining condition to find all multipliers comes from \lemref{eulerop-lemma}
applied to the characteristic equation \eqref{chareqn}, yielding
\begin{equation}\label{multrdeteq}
0=E_{u}(GQ)=\delta_Q^* G +\delta_G^* Q . 
\end{equation}
This condition, which is required to hold identically in jet space, 
is necessary and sufficient for $Q$ to be a multiplier. 
For each solution $Q$, 
a corresponding conserved current that satisfies the characteristic equation \eqref{chareqn}
can be obtained from the expression $f=GQ$ by using \lemref{inveulerlagr-lemma}. 
This yields 
\begin{equation}
\til\Phi= \int_0^1\Upsilon_{GQ}(\p_\lambda u_{(\lambda)})\big|_{u=u_{(\lambda)}}\;d\lambda
\end{equation}
whose multiplier \eqref{Q} is $Q$. 
An explicit formula for this conserved current is stated next. 

\begin{lem}\label{TXfromQ}
For a regular PDE system \eqref{pde}, 
each multiplier \eqref{generalQ} yields a conserved current \eqref{chareqn} 
which is explicitly given by a homotopy integral 
\begin{align}
&
\til T = 
\int_0^1 \Big( 
\sum_{l=0}^{k-1} \p_\lambda\p^{l}u_{(\lambda)}\cdot\big( E_{\p^{l}\p_t u}(GQ) \big)\big|_{u=u_{(\lambda)}}
\Big)d\lambda + D_x\cdot\Theta, 
\label{TfromQ}\\
&
\til X = 
\int_0^1 \Big( 
\sum_{l=0}^{k-1} \p_\lambda\p^{l}u_{(\lambda)}\cdot\big( E_{\p^{l}\p_x u}(GQ) \big)\big|_{u=u_{(\lambda)}}
\Big)d\lambda - D_t\cdot\Theta + D_x\cdot\Gamma
\label{XfromQ}
\end{align}
along a homotopy curve $u_{(\lambda)}(t,x)$, 
with $u_{(1)}=u$ and $u_{(0)}=u_0$ such that $(GQ)|_{u=u_0}$ is non-singular. 
Here $k=\max(r,N)$. 
\end{lem}

Note the conserved current formula \eqref{TfromQ}--\eqref{XfromQ} can be simplified 
by evaluating it on the solution space $\Esp$ of the given regular PDE system.
Modulo a locally trivial current, this yields
\begin{equation}\label{PhifromQ}
\til\Phi|_\Esp = 
\int_0^1 \sum_{j=1}^{k}\Big( 
\p_\lambda \p^{j-1}u_{(\lambda)}\cdot\Big( 
\sum_{l=j}^{k}(-D)^{l-j}\cdot\Big(\parder{G}{(\p^l u)}Q\Big)\Big|_{u=u_{(\lambda)}} 
\Big) \Big) d\lambda 
\end{equation}
where the curve $u_{(\lambda)}(t,x)$ is now in the solution space $\Esp$.

\subsection{Correspondence between conservation laws and multipliers}

As shown by the following key result, 
multipliers provide a unique characteristic form (or canonical representation) 
for locally equivalent conservation laws, 
in analogy to the characteristic form \eqref{symmchar} for symmetries, 
if a regular PDE system has no differential identities. 
A generalization holding for regular PDE systems with differential identities
will be stated later. 

\begin{prop}\label{correspondence-no-diffids}
For any regular PDE system \eqref{pde} that has no differential identities, 
a conserved current is locally trivial \eqref{trivconslaw} 
iff its corresponding multiplier \eqref{Q} vanishes 
when evaluated on the solution space of the system. 
\end{prop}

The proof has two parts. 
For the ``only if part'', 
suppose a conserved current is locally trivial \eqref{trivconslaw}.
By \lemref{hadamard}, 
the conserved density and the flux will have the respective forms
$T=D_x\cdot\Theta + \hat T(G)$
and 
$X=-D_t\Theta +D_x\cdot\Gamma+ \hat X(G)$
for some linear differential operators $\hat T$ and $\hat X$
whose coefficients are differential functions that are non-singular 
when evaluated on $\Esp$. 
For this conserved current $\Phi=(T,X)$, 
consider the divergence identity \eqref{conslawoffE},
where $R_\Phi(G)= D_t\hat T(G) +D_x\cdot\hat X(G)$. 
As the PDE system is assumed to have no differential identities, 
then the homotopy integral formula for the operator $R_\Phi$ 
from the proof of \lemref{hadamard}
shows that integration by parts applied to $R_\Phi(G)$ 
yields $\til T=T -\hat T(G)$, $\til X=X-\hat X(G)$ 
in the characteristic equation \eqref{chareqn}--\eqref{equivTX},
and hence $GQ=D_t\til T +D_x\cdot \til X =0$. 

It is now straightforward to determine $Q$ from the equation $GQ=0$.
In the case when $G$ comprises a single PDE (\ie/, $M=1$),
then $Q=0$ is immediate. 
In the case when $G$ contains more than one PDE (\ie/, $M>1$),
the equation $GQ=0$ can be solved by linear algebra as follows. 

First express each PDE $G^a=0$, $a=1,\ldots, M$,
in the solved form \eqref{pde-solvedform}--\eqref{pde-leadingder}
for a leading derivative $\p_{(\ell_a)} u^{\alpha_a}$.
Then take the Fr\'echet derivative of $GQ=0$,
which yields 
\begin{equation*}
(\delta_vG)Q + G(\delta_vQ)=0 . 
\end{equation*}
To solve this Fr\'echet derivative equation, 
consider the terms involving $\p^k\p_{(\ell_a)}v^{\alpha_a}$
and let $w=(\p_{(\ell_1)}v^{\alpha_1},\ldots,\p_{(\ell_M)}v^{\alpha_M})$
for ease of notation. 
It is easy to see the expression $\delta_vG$ 
contains only one term of this form, 
which is simply given by $w$ itself,
as a consequence of the solved form of the PDEs $G=(G^1,\ldots,G^M)$. 
The expression $\delta_vQ$ contains a sum of terms involving derivatives of $w$,
which will have the form $\sum_{k=0}^{r} Q^{(k)}\p^kw^\t$,
where $r$ is the differential order of the highest derivatives of 
the variables $\p_{(\ell_a)} u^{\alpha_a}$ in $Q$,
and where the coefficients $Q^{(k)}$ are differential $M\times M$ matrix functions in $J$.
Hence, all of the terms involving $\p^k\p_{(\ell_a)}v^{\alpha_a}$
in the Fr\'echet derivative equation 
consist of $wQ+\sum_{k=0}^{r} GQ^{(k)}\p^kw^\t=0$. 
Then the coefficients of each jet variable $\p^kw$, $k=0,1,\ldots,r$, 
must vanish separately. 
This immediately yields $Q^{(k)}=0$ for $k=1,\ldots,r$. 
The remaining terms are given by $wQ+GQ^{(0)}w^\t=0$. 
This is a linear homogeneous equation in $w^\t$, 
after the transpose relation $wQ=(wQ)^t=Q^tw^t$ is used,
which gives $(Q^\t +GQ^{(0)})w^t=0$. 
The vanishing of the coefficient of $w^t$ yields $Q^\t =-GQ^{(0)}$,
and hence $Q|_\Esp=0$. 

For the ``if part'', 
suppose a multiplier satisfies $Q|_\Esp=0$. 
Then, \lemref{hadamard} can be applied to get $Q= \hat Q(G)$,
where $\hat Q$ is some linear differential operator 
whose coefficients are differential functions that are non-singular 
when evaluated on $\Esp$. 
The characteristic equation \eqref{chareqn} must now be solved to 
determine the corresponding conserved density $\til T$ and flux $\til X$. 
This will be done in two main steps. 

For the first step, 
a descent argument will be given to solve the Fr\'echet derivative equation 
\begin{equation*}
D\cdot(\delta_v\til\Phi)=(\delta_v G)Q+G(\delta_vQ)
\end{equation*}
for $\delta_v\til\Phi=(\delta_v\til T,\delta_v\til X)$,
similarly to the proof of \lemref{null-div-lemma}.
Let $F(v) = (\delta_v G)Q+G(\delta_vQ)$, with $Q= \hat Q(G)$. 
The terms in $F(v)$ containing highest derivatives of $v$ 
will be denoted $F^{(k)}\p^k v$, 
where $k$ is the larger of the differential orders of $Q$ and $G$,
and where the coefficients $F^{(k)}$ are differential functions in $J$
such that $F^{(k)}|_\Esp=0$ since $F(v)|_\Esp = ((\delta_v G)\hat Q(G))|_\Esp = (\delta_v G)|_\Esp \hat Q(0)=0$. 
Note the differential order of $\delta_v\til\Phi$ then can be assumed to be $k-1$. 
Next, let $\Upsilon(v)=(\delta_v\til T,\delta_v\til X)$, 
and denote the terms containing highest derivatives of $v$ 
in $\Upsilon(v)$ as $\til T^{(k-1)}\p^{k-1}v$ and $\til X^{(k-1)}\p^{k-1}v$, respectively,
where the coefficients $\til T^{(k-1)}$ and $\til X^{(k-1)}$ 
are given by a set of differential scalar functions and a set of differential vector functions in $J$. 
In this notation, the Fr\'echet derivative equation becomes
\begin{equation*}
D\cdot\Upsilon(v) = F(v) . 
\end{equation*}
Now the highest derivative terms $\p^k v$ in this equation
are given by 
\begin{equation*}
\til T^{(k-1)}\p_t\p^{k-1}v +\til X^{(k-1)}\cdot\p_x\p^{k-1}v = F^{(k)}\p^k v . 
\end{equation*}
Expand out $F^{(k)}\p^k v= F^{(k-1,t)}\p_t\p^{k-1}v + F^{(k-1,x)}\cdot\p_x\p^{k-1}v$,
and collect the terms $\p_t\p^{k-1}v$ and $\p_x\p^{k-1}v$ in the equation, 
giving
\begin{equation*}
(\til T^{(k-1)}-F^{(k-1,t)})\p_t\p^{k-1}v +(\til X^{(k-1)}-F^{(k-1,x)})\cdot\p_x\p^{k-1}v
=0 . 
\end{equation*}
The same analysis used in the proof of \lemref{null-div-lemma} then yields
\begin{equation*}
\begin{aligned}
(\til T^{(k-1)}\p^{k-1}v,\til X^{(k-1)}\p^{k-1}v) & = 
(F^{(k-1,t)}\p^{k-1}v,F^{(k-1,x)}\p^{k-1}v) 
\\&\qquad
+ D\cdot\Psi^{(k-2)}(v) +\lot
\end{aligned}
\end{equation*}
where
\begin{equation*}
\Psi^{(k-2)}(v)=
\begin{pmatrix} 0 & \Theta^{(k-2)}(v)\\ -\Theta^{(k-2)}(v) & \Gamma^{(k-2)}(v)\end{pmatrix}
\end{equation*}
with $\Theta^{(k-2)}(v) = \theta^{(k-2)}\p^{k-2}v$ 
and $\Gamma^{(k-2)}(v) = \gamma^{(k-2)}\p^{k-2}$
being given by some differential vector function $\theta^{(k-2)}$ 
and some differential antisymmetric tensor function $\gamma^{(k-2)}$. 
Hence the highest derivative terms in $\Upsilon(v)$ involving $v$ have the form
\begin{equation*}
\Upsilon(v) = (F^{(k-1,t)}\p^{k-1}v,F^{(k-1,x)}\p^{k-1}v) + D\cdot\Psi^{(k-2)}(v) + \til\Upsilon(v)
\end{equation*}
where $\til\Upsilon(v)$ comprises all remaining terms, 
which contain derivatives of $v$ up to order $\p^{k-2}v$,
and where $D\cdot\Psi^{(k-2)}(v)$ is a total curl, which has a vanishing total divergence. 
Substitution of this expression $\Upsilon(v)$ into the Fr\'echet derivative equation gives
\begin{equation*}
\begin{aligned}
& (\til T^{(k-2)}+D\cdot F^{(k-1,t)})\p_t\p^{k-2}v+(\til X^{(k-2)}+D\cdot F^{(k-1,x)})\cdot\p_x\p^{k-2}v 
\\&\qquad
= F^{(k-1)}\p^{k-1} v +\lot
\end{aligned}
\end{equation*}
where $\til T^{(k-2)}$ and $\til X^{(k-2)}$ are a set of differential scalar functions and a set of differential vector functions
given by the coefficients of the terms $\p^{k-2}v$ in $\til\Upsilon(v)$.
After $F^{(k-1)}\p^{k-1} v= F^{(k-2,t)}\p_t\p^{k-2}v + F^{(k-2,x)}\cdot\p_x\p^{k-2}v$ is expanded out,
the terms containing highest derivatives of $v$ in this equation are given by 
\begin{equation*}
(\til T^{(k-2)}-F^{(k-2,t)}+D\cdot F^{(k-1,t)})\p_t\p^{k-2}v+(\til X^{(k-2)}-F^{(k-2,x)}+D\cdot F^{(k-1,x)})\cdot\p_x\p^{k-2}v = 0
\end{equation*}
which has the same form as the equation solved previously.
This completes the first step in the descent argument. 

Next, continuing to all lower orders, 
the descent argument yields 
\begin{equation*}
\Upsilon(v) = \sum_{l=1}^{k-1}D\cdot\Psi^{(l-1)}(v) + \sum_{l=0}^{k-1}\sum_{j=l}^{k-1} ((-D)^{j-l}\cdot F^{(j,t)},(-D)^{j-l}\cdot F^{(j,x)})\p^{l}v . 
\end{equation*}
Note the terms in the first sum are a total curl,
and the terms in the second sum vanish on $\Esp$ since each $F^{(l)}|_\Esp=0$.

The final step is to apply the general line integral \eqref{line-integral}
to the Fr\'echet derivative $\delta_v\til\Phi=\Upsilon(v)$
evaluated on $\Esp$. 
Since $\Upsilon(v)|_\Esp= \sum_{l=1}^{k-1}D\cdot\Psi^{(l-1)}(v)|_\Esp$,
this gives 
\begin{equation*}
\til\Phi|_\Esp -\til\Phi|_{u=0} = 
\int_{0}^{1} \sum_{l=1}^{k-1}D\cdot\Psi^{(l-1)}(v)\Big|_{u=u_{(\lambda)},v=\p_\lambda u_{(\lambda)}}\;d\lambda
\end{equation*}
where $u_{(\lambda)}(t,x)$ is a homotopy curve in the solution space $\Esp$ of the regular PDE system,
with $u_{(1)}=u(t,x)$ being an arbitrary solution 
and $u_{(0)}=u_0(t,x)$ being any particular solution. 
Thus $\til\Phi|_\Esp -\til\Phi_0 =D\cdot\Psi$ is a total curl, 
where 
\begin{equation*}
\Psi = \int_{0}^{1} \sum_{l=1}^{k-1} \Psi^{(l-1)}(v)\Big|_{u=u_{(\lambda)},v=\p_\lambda u_{(\lambda)}}\;d\lambda
\end{equation*}
has the form \eqref{curl-terms}.
Now, substitution of $\til\Phi|_\Esp =\til\Phi|_{u=u_0} +D\cdot\Psi$
into $D\cdot\til\Phi = GQ$ yields
$0=(D\cdot\til\Phi -GQ)|_\Esp = D\cdot(\til\Phi|_{u=u_0})$.
This immediately establishes that $\til\Phi|_{u=u_0}=D\cdot\Psi_0$ is an ordinary curl, 
by Poincare's lemma.
Thus, 
\begin{equation*}
\til\Phi|_\Esp =D\cdot(\Psi+\Psi_0)
\end{equation*}
is a locally trivial conserved current,
which completes the proof of \propref{correspondence-no-diffids}.

The correspondence stated in \propref{correspondence-no-diffids} no longer holds 
when a PDE system possesses a differential identity \eqref{diffid}. 
In particular, for a given differential identity, 
multiplication by an arbitrary differential function $\chi$, 
followed by integration by parts, 
yields 
\begin{equation}\label{gaugeconslaw}
0=\chi \D(G) = G\D^*(\chi) + D\cdot\Phi(\chi,G)
\end{equation}
where $\Phi(\chi,G)$ is a conserved current that vanishes on the solution space of the PDE system,
\begin{equation}\label{gaugePhi}
\Phi(\chi,G)|_\Esp =\Phi(\chi,0)=0 . 
 \end{equation}
Hence 
\begin{equation}\label{gaugeQ}
Q=\D^*(\chi)
\end{equation}
is a multiplier which determines a locally trivial conserved current. 
This derivation can be reversed, 
showing that the existence of a multiplier \eqref{gaugeQ} is necessary and sufficient 
for a PDE system to possess a differential identity \eqref{diffid}. 

Multipliers of the form \eqref{gaugeQ},
given by a linear differential operator acting on an arbitrary differential function $\chi$,
will be called {\em gauge multipliers} \cite{AncPoh}, 
in analogy with gauge symmetries. 
Note that a gauge multiplier is non-vanishing on the solution space $\Esp$ of the PDE system
whenever the differential identity is non-trivial,
since $\D|_\Esp\neq0$ implies $Q|_\Esp\neq0$ for $\chi\neq0$.
Two multipliers that differ by a gauge multiplier will be called {\em gauge equivalent}.

{\bf Running \Ex{(3)}}:
The Euler equations for constant density, inviscid fluids in two dimensions
comprise an evolution equation for $\vec u=(u^1,u^2)$, 
\begin{equation*}
\vec G = \vec u_t +\vec u\cdot\nabla\vec u +(1/\rho)\nabla p=0, 
\end{equation*}
a spatial equation relating $\vec u$ to $p$, 
\begin{equation*}
G^p = (1/\rho)\Delta p + (\nabla\vec u)\cdot(\nabla\vec u)^\t=0,
\end{equation*}
and a spatial constraint equation on $\vec u$,
\begin{equation*}
G^\div = \nabla\cdot\vec u =0 . 
\end{equation*}
This PDE system obeys a differential identity 
\begin{equation*}
\Div\vec G -D_t G^\div -G^p =0
\end{equation*}
which has the form \eqref{diffid} where 
$\D = \diag(\Div,-1,-D_t)$ and $G=(\vec G,G^p,G^\div)$. 
The corresponding gauge multiplier is given by 
\begin{equation*}
Q=(\vec Q,Q^p,Q^\div)^\t,
\quad
\vec Q = -\Grad\chi,
\quad
Q^p =-\chi,
\quad
Q^\div = D_t\chi
\end{equation*}
where $\chi$ is an arbitrary differential scalar function. 
The characteristic equation yields 
\begin{equation*}
GQ = -(\Grad\chi)\cdot\vec G -\chi G^p + (D_t\chi)G^\div
= D_t(\chi G^\div) +D_x\cdot(-\chi\vec G)
\end{equation*}
which is a locally trivial conservation law,
where $T=\chi G^\div$ is the conserved density 
and $\vec X=-\chi\vec G$ is the spatial flux. 
If $\chi$ is chosen to be a constant,
$\chi=1$,  
then the conserved density becomes a total spatial divergence 
$T=D_x\cdot\vec u$
which produces a boundary conservation law
\begin{equation*}
\frac{d}{dt}\int_\Omega TdV|_\Esp 
=\frac{d}{dt}\oint_{\p\Omega} \vec u\cdot\vec \nu dA|_\Esp = 0
\end{equation*}
on any closed spatial domain $\Omega\in\Rnum^2$, 
since the flux vanishes on the solution space of the system, 
$\vec X|_\Esp = 0$. 
This boundary conservation law represents conservation of streamlines
in the fluid. 

{\bf Running \Ex{(4)}}:
The magnetohydrodynamics equations for a compressible, infinite conductivity plasma in three dimensions
comprise evolution equations for $\rho$, $\vec u=(u^1,u^2,u^3)$ and $\vec B=(B^1,B^2,B^3)$, 
\begin{gather*}
G^\rho = 
\rho_t +\nabla\cdot(\rho\vec u) =0 , 
\\
\vec G^u = \vec u_t + \vec u\cdot\nabla\vec u+(1/\rho)(P'(\rho)\nabla\rho-\vec J\times\vec B)=0,
\quad
4\pi\vec J = \nabla\times\vec B , 
\\
\vec G^B = \vec B_t -\nabla\times(\vec u\times \vec B) =0,
\end{gather*}
and a spatial constraint equation on $\vec B$,
\begin{equation*}
G^\div = \nabla\cdot\vec B =0 . 
\end{equation*}
This PDE system obeys a differential identity 
\begin{equation*}
\Div( \vec G^B ) - D_t G^\div =0
\end{equation*}
which has the form \eqref{diffid} where 
$\D = \diag(0,0,\Div,-D_t)$ and $G=(G^\rho,\vec G^u,\vec G^B,G^\div)$. 
The corresponding gauge multiplier is given by 
\begin{equation*}
Q=(Q^\rho,\vec Q^u,\vec Q^B,Q^\div)^\t,
\quad
Q^\rho =0,
\quad
\vec Q^u =0,
\quad
\vec Q^B = -\Grad\chi,
\quad
Q^\div = D_t\chi
\end{equation*}
where $\chi$ is an arbitrary differential scalar function. 
The characteristic equation yields 
\begin{equation*}
GQ = -(\Grad\chi)\cdot\vec G^B + (D_t\chi)G^\div
= D_t(\chi G^\div) +D_x\cdot(-\chi\vec G^B)
\end{equation*}
which is a locally trivial conservation law,
where $T=\chi G^\div$ is the conserved density 
and $\vec X=-\chi\vec G^B$ is the spatial flux. 
If $\chi$ is chosen to be a constant, $\chi=1$, 
then the conserved density becomes a total spatial divergence 
$T=D_x\cdot\vec B$
which produces a boundary conservation law
\begin{equation*}
\frac{d}{dt}\int_\Omega TdV|_\Esp 
= \frac{d}{dt}\oint_{\p\Omega} \vec B\cdot\vec \nu dA|_\Esp = 0
\end{equation*}
on any closed spatial domain $\Omega\in\Rnum^3$, 
since the flux vanishes on the solution space of the system, 
$\vec X|_\Esp = 0$. 
This boundary conservation law represents conservation of magnetic flux 
in the plasma. 

The following natural generalization of \propref{correspondence-no-diffids} will now be established. 

\begin{prop}\label{correspondence-diffids}
For any regular PDE system \eqref{pde} that possesses a differential identity \eqref{diffid},
a conserved current is locally trivial \eqref{trivconslaw} 
iff its corresponding multiplier \eqref{Q} evaluated on the solution space of the system
is equal to a gauge multiplier \eqref{gaugeQ} for some differential function $\chi$. 
\end{prop}

The same steps used in proof for \propref{correspondence-no-diffids} go through with only two changes. 
For the ``if part'', 
suppose a multiplier satisfies $Q|_\Esp = \D^*(\chi)$,
which implies $Q=\hat Q(G) +\D^*(\chi)$ by \lemref{hadamard},
where $\hat Q$ is some linear differential operator 
whose coefficients are differential functions that are non-singular 
when evaluated on $\Esp$. 
Then the conservation law identity \eqref{gaugeconslaw} 
combined with the characteristic equation \eqref{chareqn}
yields
\begin{equation*}
G\hat Q(G) = G(Q-\D^*(\chi)) = D_t(\til T+\Phi^t(\chi,G)) +D_x\cdot(\til X+\Phi^x(\chi,G)) . 
\end{equation*}
This equation can be solved by the same steps 
used in proving the ``if part'' of \propref{correspondence-no-diffids},
thus showing that $\til\Phi+\Phi(\chi,G)$ is a locally trivial current.
Since $\Phi(\chi,G)$ itself is a locally trivial current, 
the conservation law given by $\til\Phi$ is therefore locally trivial \eqref{trivconslaw}. 
For the ``only if'' part, 
suppose a conserved current is locally trivial \eqref{trivconslaw},
so then, by \lemref{hadamard},  
the conserved density and the spatial flux will have the respective forms
$T=D_x\cdot\Theta + \hat T(G)$
and 
$X=-D_t\Theta +D_x\cdot\Gamma+ \hat X(G)$
for some linear differential operators $\hat T$ and $\hat X$
whose coefficients are differential functions that are non-singular 
when evaluated on $\Esp$. 
As the PDE system is assumed to satisfy a differential identity \eqref{diffid},
the divergence identity \eqref{conslawoffE} will be unique only up to the addition of 
a multiple of this differential identity, $\chi\D(G)=0$. 
This implies from the homotopy integral formula for the operator $R_\Phi$ 
that the characteristic equation \eqref{chareqn}--\eqref{equivTX} holds with 
$\til T=T-\hat T(G)-\Phi^t(\chi,G)$, $\til X=X-\hat X(G)-\Phi^x(\chi,G)$, 
and $GQ=G\D^*(\chi)$. 
The equation $G(Q-\D^*(\chi))=0$ can be solved by the same steps 
used in proving the ``only if part'' of \propref{correspondence-no-diffids},
thereby showing $(Q-\D^*(\chi))|_\Esp=0$, 
so $Q|_\Esp$ is equal to $\D^*(\chi)|_\Esp$. 
This completes the proof of \propref{correspondence-diffids}. 

The characterization of locally trivial conservation laws 
in \propref{correspondence-no-diffids} and \propref{correspondence-diffids} 
establishes an important general correspondence result 
which underlies the usefulness of multipliers.

For a given regular PDE system, 
the set of multipliers forms a vector space 
on which the symmetries of the system have a natural action 
\cite{Anc16,AncKar}. 
A multiplier is called {\em trivial} if yields a locally trivial conservation law,
and two multipliers are said to be {\em equivalent} if they differ by a trivial multiplier.
When the PDE system has no differential identities, 
then a multiplier $Q$ is trivial iff it vanishes on the solution space, $Q|_\Esp=0$, 
whereas when the PDE system possesses a differential identity \eqref{diffid},
a multiplier $Q$ is trivial iff it equals a gauge multiplier \eqref{gaugeQ} 
on the solution space, $Q|_\Esp = \D^*(\chi)$. 
A set of multipliers is linearly independent if no linear combination of the multipliers is trivial. 
Likewise, a set of conservation laws is linearly independent if no linear combination of the conserved currents is locally trivial. 

\begin{thm}\label{correspondence}
(i) For any regular PDE system \eqref{pde}, 
whether or not it possesses a differential identity, 
there is a one-to-one correspondence between 
its admitted equivalence classes of linearly-independent local conservation laws
and its admitted equivalence classes of linearly-independent multipliers. 
(ii) An explicit formulation of this correspondence is given by 
the homotopy integral formula \eqref{TfromQ}---\eqref{XfromQ}
for conserved currents in terms of multipliers. 
\end{thm}

Infinitesimal symmetries have a well-known action on conserved currents \cite{Olv,2ndbook}. 
This action induces a corresponding action of infinitesimal symmetries on multipliers 
\cite{Anc16,AncKar}, 
and there are several equivalent formulas 
\cite{Kha,AncBlu97,IbrKarMah,BluTemAnc,Ibr07,Ibr11,Anc16,Anc17} 
for the conserved current obtained from the action of a given infinitesimal symmetry 
applied to a given multiplier. 
It is worth noting that this action does not preserve linear independence of equivalence classes. 
For example \cite{Anc16,Anc17}, 
any non-trivial conserved current that does not explicitly contain at least one of the independent variables in a PDE system 
is mapped into a locally trivial current under any translation symmetry.

\subsection{Low-order conservation laws}\label{loworder}

For any given regular PDE system, 
the correspondence between local conservation laws and multipliers 
stated in \thmref{correspondence} gives a straightforward way using the following three steps 
to find all of the non-trivial local conservation laws (up to equivalence) admitted by the PDE system. 
Step 1: solve the determining condition \eqref{multrdeteq} to obtain all multipliers. 
Step 2: find all linearly independent equivalence classes of non-trivial multipliers. 
Step 3: apply the homotopy integral formula \eqref{PhifromQ} to a representative multiplier 
in each equivalence class to obtain a corresponding conserved current. 

In practice, for solving the determining condition \eqref{multrdeteq}, 
it is very useful to know at which differential orders the non-trivial multipliers will be found. 
As seen in the examples in \secref{examples}, 
physically important conservation laws, such as energy and momentum, 
always have a low differential order for the conserved density $T$ and the spatial flux $X$, 
whereas conservation laws having a high differential order are typically connected with integrability. 
A general pattern emerges from these conservation law examples 
when their multipliers are examined. 

In \Ex{1} and \Ex{2}, 
mass conservation 
for the transport equation \eqref{transport-eqn}
and net heat conservation 
for the diffusion/heat conduction equation \eqref{diffusionheat-eqn}
both have $Q=1$
which does not involve $u$ or its derivatives. 

In \Ex{3}, 
energy conservation 
for the telegraph equation \eqref{telegraph-eqn}
has $Q=\exp(2\smallint a(t)dt)u_t$,
while the leading derivative in this equation is $u_{tt}$ or $u_{xx}$. 

In \Ex{4}, 
for the nonlinear dispersive wave equation \eqref{dispersivewave-eqn},
mass conservation, 
$L^2$-norm conservation, 
and energy conservation 
respectively have 
$Q=1$, 
$Q=2u$, 
and 
$Q=g(u)+u_{xx}$.
The leading derivative in this equation is $u_t$ or $u_{xxx}$. 

In \Ex{5}, 
for the viscous fluid equations \eqref{viscousfluid-eqn},
mass conservation has $Q^\t=(1,0)$, 
momentum conservation has $Q^\t=(u,1)$,
and Galilean momentum conservation has $Q^\t=(tu,t)$,
while $\{\rho_t,u_t\}$ is a set of leading derivatives in this system. 

In \Ex{6}, 
energy conservation for 
the barotropic gas flow/compressible inviscid fluid equations \eqref{compressiblefluid-eqn}, 
has $Q^\t=(\tfrac{1}{2}u^2,\rho u)$,
and again $\{\rho_t,u_t\}$ is a set of leading derivatives in this system. 

In \Ex{7}, 
momentum conservation,
energy conservation,
and momentum-energy conservation 
for the breaking wave equation \eqref{breakingwave-eqn}
respectively have 
$Q=1$, $Q=u$, $Q=-(u_{tx}-u(m+\tfrac{1}{2}u)+\tfrac{1}{2}u_x^2)$,
while the Hamiltonian Casimir has $Q=\tfrac{1}{2}(u-u_{xx})^{1/2}$.
The leading derivative in this equation is $u_{txx}$ or $u_{xxx}$. 

In \Ex{8}, 
mass conservation for the porous media equation \eqref{porousmedia-eqn}
has $Q=\alpha(x)$. 

In \Ex{9},
angular momentum conservation and boost momentum conservation 
for the non-dispersive wave equation \eqref{nondispersivewave-eqn}
respectively have 
$Q=(\mathbf{a}\cdot\vec x)\cdot\nabla u$, 
$Q=\vec b\cdot x u_t +c^2 t\vec b\cdot\nabla u$, 
while the leading derivative in this equation is $u_{tt}$ or $\Delta u$. 

In all of these examples, 
each variable $\p^k u$ that appears in the conservation law multiplier 
is related to some leading derivative of $u$ in the PDE system 
by differentiation of this variable $\p^k u$ with respect to $t,x$. 

In contrast, 
the conservation laws for the higher-derivative quantities 
\eqref{higherTXcompressiblefluid} in \Ex{6} 
and \eqref{higherTXbreakingwave} in \Ex{7}
have, respectively,
$Q^\t=\big( (2\rho_xu_x/(u_x^2-p'\rho_x^2/\rho^2)^2)_x,(u_x^2/(u_x^2-p'\rho_x^2/\rho^2)^2)_x+p'/\rho^2(\rho_x^2/(u_x^2-p'\rho_x^2/\rho^2)^2)_x \big)$
and $Q=\tfrac{5}{2}m^{-7/2}m_x^2 -2m^{-5/2}m_{xx} -2m^{-3/2}$,
which involve variables of higher differential order 
than the leading derivatives. 

An exceptional case is the conservation laws for local helicity 
and local enstrophy in \Ex{10}. 
These conservation laws for the inviscid (compressible/incompressible) fluid equation \eqref{inviscidfluid-eqn} 
have, respectively, 
$\vec Q = 2 \nabla\times\vec u$
which involves a variable with the same differential order 
as the leading derivative $\vec u_t$,
and 
$\vec Q = f''((\curl\,\vec u)/\rho)\nabla\cdot((\nabla\wedge\vec u)/\rho)$
which involves a higher-derivative variable. 
Note, however, if the fluid equation is expressed as a system 
for the velocity $\vec u$ and 
the vorticity vector $\vec\omega =\nabla\times\vec u$ in three dimensions
or the vorticity scalar $\omega =\curl\,\vec u$ in two dimensions,
then the multipliers for helicity and enstrophy conservation 
are given by, respectively,
$Q^\t=(\vec\omega,\vec u)$ and $Q^\t=(\vec 0,f'(\omega/\rho))$
in which the variables are related to the leading derivatives 
$\vec u_t$ and $\vec\omega_t$ by differentiation with respect to $t$. 

This pattern motivates introducing the following general class of multipliers.
A multiplier $Q$ for a regular PDE system \eqref{pde} will be called {\em low-order}
if each jet variable $\p^k u^\alpha$ that appears in $Q|_\Esp$ 
is related to some leading derivative of $u^\alpha$ 
by differentiations with respect to $t,x^i$. 
(Note that, therefore, the differential order $r$ of $Q|_\Esp$
must be strictly less than the differential order $N$ of the PDE system.)
Correspondingly, a conservation law is said to be of {\em low-order}
if its multiplier is low-order when evaluated on the solution space of the PDE system. 

For a given regular PDE system, 
the explicit form for low-order conservation laws can be determined from the form for low-order multipliers
by inverting the relation \eqref{Q} which defines a multiplier in terms of a conserved current.

{\bf Running \Ex{(1)}}
The gKdV equation \eqref{gkdv-eqn} 
is a time evolution PDE whose leading derivative is $u_t$ or $u_{xxx}$. 
Its low-order conservation laws $(D_t T+ D_x X)|_\Esp=0$ 
are given by multipliers that have the form
\begin{equation*}
Q(t,x,u,u_x,u_{xx})
\end{equation*}
since, in the jet space $J=(t,x,u,u_t,u_x,u_{tt},u_{tx},u_{xx},\ldots)$, 
the only variables that can be differentiated with respect to $t$ or $x$
to obtain a leading derivative are $u,u_x,u_{xx}$. 
To derive the corresponding form for low-order conserved currents $\Phi=(T,X)$, 
the first step is to expand out $D_t T$ and $D_x X$ 
starting from general expressions for $T$ and $X$
in which a leading derivative $u_t$ or $u_{xxx}$ has been eliminated 
along with all of its differential consequences. 
If $u_t$ is chosen, then the starting expressions will be 
$T(t,x,u,u_x,u_{xx},\ldots)$ and $X(t,x,u,u_x,u_{xx},\ldots)$,
which gives
\begin{equation*}
D_t T = T_t +u_tT_u + u_{tx}T_{u_x} + u_{txx}T_{u_{xx}} + \cdots,
\quad
D_x X = X_x +u_xX_u + u_{xx}X_{u_x} + u_{xxx}X_{u_{xx}} + \cdots . 
\end{equation*}
The second step is to obtain the operator $R_\Phi$ from the terms in the divergence expression 
$D_t T +D_x X$ containing $u_t$ (and its differential consequences). 
This yields 
\begin{equation*}
D_t T +D_x X= (T_u + T_{u_x}D_x + T_{u_{xx}}D_x^2 + \cdots)u_t +T_t + X_x +u_xX_u + u_{xx}X_{u_x} + u_{xxx}X_{u_{xx}} + \cdots
\end{equation*}
and hence
\begin{equation*}
R_\Phi = T_u + T_{u_x}D_x + T_{u_{xx}}D_x^2 + \cdots
\end{equation*}
since $u_t=G-u^pu_x-u_{xxx}$ is the solved form for the PDE expression. 
Then the main steps are, first, to equate $Q$ with the expression $E_{G}(R_\Phi(G))$
and, next, to use the resulting equation together with the characteristic equation 
$D_t\til T+ D_x\til X=GQ$ 
to determine the dependence of $T$ and $X$ on all jet variables that do not appear in $Q$. 
This gives, first, 
\begin{equation*}
R_\Phi(G) = T_uG + T_{u_x}D_xG + T_{u_{xx}}D_x^2G + \cdots 
= \delta_G T = GE_u(T) + D_x\Upsilon^x(G)
\end{equation*}
by using the Euler-Lagrange relation \eqref{frechet-euler},
which yields the equation 
\begin{equation*}
Q(t,x,u,u_x,u_{xx})= E_{G}(R_\Phi(G)) = E_G(\delta_G T) = E_u(T) . 
\end{equation*}
Comparison of the differential order of both sides of this equation directly determines 
\begin{equation*}
T= \til T(t,x,u,u_x) +D_x \Theta(t,x,u,u_x,\ldots) . 
\end{equation*}
This implies $\Upsilon^x(G)= \til T_{u_x} G$. 
Next, the characteristic equation then yields
\begin{equation*}
GQ  = D_t T + D_x(X -\til T_{u_x}G) = D_t\til T +D_x\til X 
\end{equation*}
which gives
\begin{equation*}
D_x(X +D_t\Theta) = D_x(\til X +G\til T_{u_x}) 
= -\til T_t +(u^pu_x+u_{xxx})\til T_u +(u^pu_x+u_{xxx})_x\til T_{u_x} . 
\end{equation*}
Comparison of both sides of this equation now determines 
\begin{equation*}
X= \til X(t,x,u,u_t,u_x,u_{xx}) -D_t \Theta(t,x,u,u_x,\ldots) - G\til T_{u_x}(t,x,u,u_x) . 
\end{equation*}
The same result can be shown to hold if $u_{xxx}$ is chosen as the leading derivative instead of $u_t$.
Hence, all low-order conserved currents have the general form 
\begin{equation*}
\Phi|_\Esp=(\til T(t,x,u,u_x),\til X(t,x,u,u_t,u_x,u_{xx}))
\end{equation*}
modulo locally trivial conserved currents. 

{\bf Running \Ex{(2)}}
The breaking wave equation \eqref{b-fam-u-eqn}
is a regular PDE whose leading derivative is $u_{txx}$ or $u_{xxx}$. 
All low-order conservation laws of this PDE are given by multipliers that have the second-order form 
\begin{equation}\label{bfam-lowordQ}
Q(t,x,u,u_t,u_x,u_{tx},u_{xx})  
\end{equation}
where $u_{tt}$ is excluded because it cannot be differentiated to obtain a leading derivative $u_{txx}$ or $u_{xxx}$. 
The corresponding form for low-order conserved currents $\Phi=(T,X)$ 
is derived by starting from general expressions for $T$ and $X$ 
in which a leading derivative $u_{txx}$ or $u_{xxx}$ has been eliminated 
along with all of its differential consequences. 
It is simplest to use the pure derivative $u_{xxx}$,
which implies $T$ and $X$ are functions only of 
$t$, $x$, $u$, $u_x$, $u_{xx}$, and their $t$-derivatives. 
Then the terms in the divergence expression $D_t T +D_x X$ 
containing the leading derivative $u_{xxx}$ (and its differential consequences) 
are given by 
\begin{equation*}
\begin{aligned}
D_t T +D_x X & = (X_{u_{xx}} + X_{u_{txx}}D_t + X_{u_{ttxx}}D_t{}^2 + \cdots)u_{xxx} 
\\&\qquad
+T_t +X_x +u_tT_u +u_xX_u + u_{tt}T_{u_t} + u_{tx}(T_{u_x}+X_{u_t}) + u_{xx}X_{u_x} 
\\&\qquad
+ u_{ttt}T_{u_{tt}} + u_{ttx}(T_{u_{tx}} +X_{u_{tt}}) + u_{txx}(T_{u_{xx}} + X_{u_{tx}}) + \cdots . 
\end{aligned}
\end{equation*}
This expression yields the operator
\begin{equation*}
R_\Phi = (X_{u_{xx}} + X_{u_{txx}}D_t + X_{u_{ttxx}}D_t{}^2 + \cdots)u^{-1}
\end{equation*}
since $u_{xxx}=-u^{-1}(G+bu_x(u_{xx}-u)-u_{txx} +u_t)+u_x$ is the solved form for the PDE expression.
Now, the main steps consist of, 
first, equating $Q$ with the expression $E_{G}(R_\Phi(G))$
and, next, using the characteristic equation $D_t\til T+ D_x\til X=GQ$ 
to determine the dependence of $T$ and $X$ on all jet variables that do not appear in $Q$. 
The first step gives
\begin{equation*}
R_\Phi(G) = (X_{u_{xx}} + X_{u_{txx}}D_t + X_{u_{ttxx}}D_t{}^2 + \cdots)(-u^{-1}G)
= -u^{-1}G E_{u_{xx}}(X) - D_t\Upsilon^t(u^{-1}G)
\end{equation*}
after using the relation \eqref{frechet-euler},
which yields the equation 
\begin{equation*}
Q(t,x,u,u_t,u_x,u_{tx},u_{xx})  = E_{G}(R_\Phi(G)) = E_G(-u^{-1}G E_{u_{xx}}(X)) = -u^{-1}E_{u_{xx}}(X) . 
\end{equation*}
Comparison of the differential order of both sides of this equation directly determines 
\begin{equation*}
X=\til X(t,x,u,u_t,u_x,u_{tx},u_{xx}) -D_t\Theta(t,x,u,u_t,u_x,u_{tt},u_{tx},u_{xx},\ldots) 
\end{equation*}
which implies $\Upsilon^t(u^{-1}G)=0$. 
Then, for the next step, 
the characteristic equation yields
\begin{equation*}
GQ = D_t T +D_x X = D_t\til T +D_x\til X 
\end{equation*}
giving 
\begin{equation*}
D_t(T-D_x\Theta)= D_t\til T
= -\til X_x -u_x\til X_u -u_{tx}\til X_{u_t} -u_{xx}\til X_{u_x} -u_{txx}\til X_{u_{tx}} . 
\end{equation*}
Comparison of both sides of this equation now determines 
\begin{equation*}
T= \til T(t,x,u,u_x,u_{xx}) +D_x\Theta(t,x,u,u_t,u_x,u_{tt},u_{tx},u_{xx},\ldots) . 
\end{equation*}
Hence, all low-order conserved currents have the general form 
\begin{equation*}
\Phi|_\Esp=(\til T(t,x,u,u_x,u_{xx}),\til X(t,x,u,u_t,u_x,u_{tx},u_{xx}))
\end{equation*}
modulo locally trivial conserved currents.

\section{Variational symmetries and Noether's theorem in modern form}
\label{Noether}

A PDE system \eqref{pde} is {\em globally variational} if it is given by 
the critical points of a variational principle defined on some spatial domain $\Omega\subseteq\Rnum^n$ and some time interval $[t_0,t_1]\subseteq\Rnum$. 
In typical applications, 
this will involve specifying a function space for $u(t,x)$ with $x\in \Omega$
and also posing boundary conditions on $u(t,x)$ for $x\in\p\Omega$. 
Noether's theorem is usually formulated in this context,
where it shows that every transformation group leaving invariant the variational principle 
yields a corresponding conserved integral \eqref{globalconslaw} 
for solutions of the PDE system with $u(t,x)$ belonging to the specified function space. 

However, for the purpose of obtaining local conservation laws \eqref{conslaw}, 
a global variational principle is not necessary, 
and a PDE system instead needs to have just a local variational principle. 

A PDE system \eqref{pde} is {\em locally variational} if it is given by 
the Euler-Lagrange equations 
\begin{equation}\label{ELpde}
0= G=E_u(L)^\t
\end{equation} 
for some differential function $L(t,x,u,\p u,\ldots,\p^k u)$, 
called a {\em Lagrangian}. 
Note that, as shown by \lemref{eulerop-lemma},  
a Lagrangian is unique up to addition of an arbitrary total divergence. 
In particular, $L$ and $\til L= L+D_t\Psi^t +D_x\cdot\Psi^x$ have the same Euler-Lagrange equations, 
for any differential scalar function $\Psi^t$ and differential vector function $\Psi^x$.

There is a well-known condition 
for a given PDE system to be locally variational \cite{Olv,2ndbook}.

\begin{lem}\label{eulerlagrange}
$G=E_u(L)^\t$ holds for some Lagrangian $L(t,x,u,\p u,\ldots,\p^k u)$ 
iff 
\begin{equation}\label{selfadj}
\delta_v G^\t= \delta_v^* G^\t
\end{equation}
holds for all differential functions $v(t,x)$. 
\end{lem}

The ``only if'' part of the proof has two steps. 
First, $\delta_v E_u(L) = E_u(\delta_v L)$ can be directly verified to hold, 
due to $v$ having no dependence on $u$ and derivatives of $u$. 
Next, the Euler-Lagrange relation \eqref{frechet-euler} combined with \lemref{eulerop-lemma}
yields  
$E_u(\delta_v L) = E_u(vE_u(L)) = \delta_v^* E_u(L)$, 
after again using the fact that $v$ has no dependence on $u$ and derivatives of $u$. 
Hence, $\delta_v E_u(L) = \delta_v^* E_u(L)$, 
which completes this part of the proof. 

The ``if'' part of the proof proceeds by first inverting the relation $G^\t=E_u(L)$
through applying \lemref{inveulerlagr-lemma} to $f=L$. 
This yields
$L= \til L +D\cdot F$,
with $\til L=\int_0^1 \p_\lambda u_{(\lambda)} G^\t\big|_{u=u_{(\lambda)}}d\lambda$. 
Then the remaining steps consist of showing that $E_u(L)=E_u(\til L)=G^\t$ 
holds for this Lagrangian when $\delta_v G^\t= \delta_v^* G^\t$. 
First, the Fr\'echet derivative of $\til L$ gives 
$\delta_v\til L = \int_0^1 \big( \p_\lambda v_{(\lambda)} G^\t\big|_{u=u_{(\lambda)}} + \p_\lambda u_{(\lambda)} \delta_{v_{(\lambda)}} G^\t\big|_{u=u_{(\lambda)}} \big)d\lambda$
where $v_{(\lambda)}= \delta_v u_{(\lambda)}$. 
Next, substitute $\delta_{v_{(\lambda)}} G^\t= \delta_{v_{(\lambda)}}^* G^\t$ 
and use the Fr\'echet derivative relation \eqref{adjoint}, 
which yields
\begin{equation*}
\begin{aligned}
\delta_v\til L & 
=\int_0^1 \big( \p_\lambda v_{(\lambda)} G^\t\big|_{u=u_{(\lambda)}} + v_{(\lambda)}\p_\lambda G^\t\big|_{u=u_{(\lambda)}} -D\cdot\Psi(\p_\lambda u_{(\lambda)},v_{(\lambda)};G^\t)\big|_{u=u_{(\lambda)}} \big)d\lambda
\\&
=vG^\t  -v_0G^\t\big|_{u=u_0} 
-D\cdot\int_0^1 \Psi(\p_\lambda u_{(\lambda)},v_{(\lambda)};G^\t)\big|_{u=u_{(\lambda)}}\;d\lambda 
\end{aligned}
\end{equation*}
where $\Psi$ is given by expression \eqref{frechetcurrent}.
Finally, apply $E_v$ to $\delta_v\til L$ to get $E_v(\delta_v\til L) = G^\t$,
and use the identity $E_v(\delta_v\til L) = E_v(vE_u(\til L)) = E_u(\til L)$ 
which follows from the Euler-Lagrange relation \eqref{frechet-euler}. 
This yields $E_u(\til L) = G^\t$, 
which completes the proof. 

The condition \eqref{selfadj} for a PDE system $G=0$ to be locally variational 
states that the linearization of $G^\t$ must be self-adjoint. 
From the relations \eqref{frechet} and \eqref{adjfrechet-euler-rel}, 
or equivalently \eqref{adjfrechet} and \eqref{frechet-euler-rel}, 
this condition splits with respect to $v,\p v,\ldots,\p^k v$ 
into a linear overdetermined system of equations on $G$:
\begin{equation}\label{helmholtz}
\parder{G}{(\p^k u)} = (-1)^k \big(E_u^{(k)}(G)\big)^\t,
\quad
k=0,1,\ldots,N
\end{equation}
where $N$ is the differential order of the PDE system $G=0$. 
These equations are called the {\em Helmholtz conditions}. 
Note the appearance of the transpose implies that the Helmholtz conditions cannot hold 
if $u$ and $G$ have a different number of components. 
Also, the expression \eqref{highereulerop} for the higher Euler operators $E_u^{(k)}$
shows that the Helmholtz condition for $k=N$ reduces to the equation 
\begin{equation}
(1-(-1)^N)\Big(\parder{G}{(\p^N u)} + \Big(\parder{G}{(\p^N u)}\Big)^\t\Big) =0
\end{equation}
which cannot hold if $N$ is odd. 
Consequently, a necessary condition for a PDE system to be locally variational is that 
its differential order $N$ must be even 
and the number $M$ of PDEs must be the same as the number $m$ of dependent variables. 

When a PDE system satisfies the Helmholtz conditions \eqref{helmholtz}, 
a Lagrangian $L$ for the system can be recovered from the expressions $G=(G^1,\ldots,G^M)$ 
by the general homotopy integral formula 
\begin{equation}\label{LfromG}
L= \int_0^1 \p_\lambda u_{(\lambda)} G^\t\big|_{u=u_{(\lambda)}}d\lambda
\end{equation}
(as shown in the proof of \lemref{eulerlagrange}). 
A total divergence can be added to this Lagrangian to obtain an equivalent Lagrangian 
that has the lowest possible differential order, 
which is $N/2$. 

{\bf Running \Ex{(1)}}
The gKdV equation \eqref{gkdv-eqn} is an odd-order PDE. 
Hence, it cannot be locally variational as it stands. 
To verify there is no local variational principle, 
note $G=G^\t=u_t+u^pu_x+u_{xxx}$ gives
\begin{equation*}
\delta_v G^\t = v_t + u^pv_x + pu^{p-1}u_x v + v_{xxx},
\quad
\delta_v^* G^\t = -v_t -u^pv_x -v_{xxx}
\end{equation*}
and hence $\delta_v G^\t - \delta_v^* G^\t = 2v_t + 2u^pv_x + pu^{p-1}u_x v +2 v_{xxx} \neq 0$
whereby $G^\t$ fails to have a self-adjoint linearization. 
Equivalently, the Helmholtz conditions are not satisfied:
\begin{equation*}
\begin{aligned}
(k=0)\quad & 
\parder{G}{u} = pu^{p-1}u_x \neq E_u(G) =0, 
\\
(k=1)\quad & 
\parder{G}{u_t} =1 \neq -E_u^{(t)}(G) = -1,
\quad
\parder{G}{u_x} =u^p \neq -E_u^{(x)}(G) = -u^p, 
\\
(k=2)\quad &
\parder{G}{u_{xx}} = 0= E_u^{(x,x)}(G) =-D_x(1), 
\\
(k=3)\quad &
\parder{G}{u_{xxx}} =1 \neq -E_u^{(x,x,x)}(G) =-1 . 
\end{aligned}
\end{equation*}
However, if a potential variable $w$ is introduced by putting $u=w_x$,
then the PDE becomes $w_{tx}+w_x^pw_{xx}+w_{xxxx}=0$ which has even order. 
Repetition of the previous steps, with $G=G^\t=w_{tx}+w_x^pw_{xx}+w_{xxxx}$, 
now gives
\begin{equation*}
\delta_v G^\t = v_{tx} + w_x^pv_{xx} + pw_x^{p-1}w_{xx} v_x + v_{xxxx} = \delta_v^* G^\t
\end{equation*}
and 
\begin{equation*}
\begin{aligned}
(k=0)\quad & 
\parder{G}{w} =0 = E_w(G), 
\\
(k=1)\quad & 
\parder{G}{w_t} =0 = -E_w^{(t)}(G) = -D_x(1),
\\&
\parder{G}{w_x} = pw_x^{p-1}w_{xx} = -E_w^{(x)}(G) = -pw_x^{p-1}w_{xx} +2D_x(w_x^p) +D_t(1) +D_x^3(1), 
\\
(k=2)\quad &
\parder{G}{w_{tx}} =1 = E_w^{(t,x)}(G), 
\\
(k=3)\quad &
\parder{G}{w_{xxx}} =0 = -E_w^{(x,x,,x)}(G) = D_x(1) , 
\\
(k=4)\quad &
\parder{G}{w_{xxxx}} =1 = E_w^{(x,x,x,x)}(G) .
\end{aligned}
\end{equation*}
Hence, the potential gKdV equation is locally variational.
A Lagrangian is given by the homotopy integral 
\begin{equation*}
L=\int_0^1 w (\lambda w_{tx}+ \lambda^{p+1}w_x^pw_{xx}+\lambda w_{xxxx})\;d\lambda 
= \tfrac{1}{2}ww_{tx}+\tfrac{1}{p+2}ww_x^pw_{xx}+\tfrac{1}{2}ww_{xxxx}
\end{equation*}
using $w_{(\lambda)}=\lambda w$. 
The addition of a total divergence $D_t\Psi^t+D_x\Psi^x$ given by 
\begin{equation*}
\Psi^t = -\tfrac{1}{2}ww_x,
\quad
\Psi^x = -\tfrac{1}{2}(ww_{xxx}-w_xw_{xx}) -\tfrac{1}{(p+1)(p+2)}ww_x^{p+1}
\end{equation*}
yields an equivalent Lagrangian that has minimal differential order, 
\begin{equation*}
\til L = -\tfrac{1}{2}w_xw_t-\tfrac{1}{(p+1)(p+2)}w_x^{p+2}+\tfrac{1}{2}w_{xx}^2 .
\end{equation*}

For a locally variational PDE system, 
a global variational principle on a spatial domain $\Omega$ and a time interval $[t_0,t_1]$
can be defined in terms of a Lagrangian by 
\begin{equation}\label{varprinc}
S[u]= \int_{t_0}^{t_1}\int_{\Omega} ( L(t,x,u,\p u,\ldots,\p^k u) +D_x\cdot\Theta(t,x,u,\p u,\ldots) )\;dV\; dt
\end{equation} 
where the spatial divergence term is chosen to let spatial boundary conditions 
be posed on $u(t,x)$ for $x\in\p\Omega$. 
The critical points of the variational principle \eqref{varprinc} are given by 
the vanishing of the variational derivative of $S[u]$, 
\begin{equation}
\begin{aligned}
0=S'[u] & = \parder{}{\epsilon}S[u+\epsilon v]\big|_{\epsilon=0} 
\\
& = \int_{t_0}^{t_1}\int_{\Omega} vE_u(L)\;dV\; dt 
+ \int_{t_0}^{t_1}\oint_{\p\Omega} ( \delta_v\Theta + \Upsilon_L(v) )\cdot\nu\;dA
\end{aligned}
\end{equation} 
where $v(t,x)$ is an arbitrary differential function that satisfies the same spatial boundary conditions as $u(t,x)$. 
Here $\nu$ denotes the outward unit normal vector on $\p\Omega$, 
and $\Upsilon_L$ is given by the Euler-Lagrange relation \eqref{frechet-euler}.
Provided $\Theta$ is chosen so that the boundary integral vanishes, 
then $S'[u]=0$ yields the PDE system $G=E_u(L)^\t=0$ on the spatial domain $\Omega$.

\subsection{Variational symmetries}

A {\em variational symmetry} \cite{Olv,1stbook} 
of a given variational principle \eqref{varprinc} 
is a generator \eqref{generator} whose prolongation leaves invariant the variational principle. 
This invariance condition has both a global aspect, 
which involves the spatial domain and the spatial boundary conditions, 
and a local aspect, which involves only the Lagrangian. 

For a local variational principle \eqref{ELpde}, 
a {\em variational (divergence) symmetry} \cite{Olv,1stbook} 
is a generator \eqref{generator} whose prolongation satisfies the invariance condition 
\begin{equation}
\pr\X(L) = \tau D_t L +\xi\cdot D_x L + D_t\Psi^t +D_x\cdot\Psi^x
\end{equation}
for some for differential scalar function $\Psi^t$ and differential vector function $\Psi^x$.
This condition can be expressed alternatively as 
\begin{equation}
\pr\X(L) = D_t\til\Psi^t +D_x\cdot\til\Psi^x -(D_t \tau +D_x\cdot\xi)L
\end{equation}
with $\til\Psi^t=\Psi^t +L\tau$ and $\til\Psi^x=\Psi^x +L\xi$, 
where $D_t \tau +D_x\cdot\xi$ represents the infinitesimal conformal change 
in the space-time volume element $dV dt$ under the symmetry generator $\X$. 

A simpler formulation of a variational symmetry is given by 
using the characteristic form \eqref{symmchar} for the symmetry generator. 
Then an infinitesimal symmetry \eqref{symmchar} is a variational symmetry iff 
its prolongation leaves invariant the Lagrangian modulo total a divergence,
\begin{equation}\label{divsymm}
\pr\hat\X(L) = D_t\Psi_P^t +D_x\cdot\Psi_P^x
\end{equation}
for some for differential scalar function $\Psi_P^t$ and differential vector function $\Psi_P^x$
depending on the characteristic function $P$ of the symmetry. 
Note that, since any total divergence is annihilated by the Euler operator $E_u$, 
a variational symmetry preserves the critical points of the Lagrangian $L$. 
As a consequence, 
every variational symmetry is an infinitesimal symmetry of the PDE system $G=E_u(L)=0$. 
The converse is not true in general, 
since (for example) scaling symmetries of Euler-Lagrange equations need not always preserve the Lagrangian. 

There is an equivalent, modern formulation of the variational symmetry condition \eqref{divsymm}
which uses only the Euler-Lagrange equations and not the Lagrangian itself. 

\begin{prop}\label{varsymm-modern}
For any locally variational PDE system \eqref{ELpde}, 
an infinitesimal symmetry in characteristic form 
$\hat\X=P(t,x,u,\p u,\ldots,\p^r u)\p_u$ 
is a variational symmetry iff 
\begin{equation}\label{varsymmdeteq}
\delta_P G^\t= -\delta_G^* P^\t
\end{equation}
holds identically.
\end{prop}

To prove this result, first note that 
$E_u(\pr\hat\X(L))$ vanishes identically iff $\pr\hat\X(L)$ is a total divergence, 
by \lemref{eulerop-lemma}. 
Next, 
$E_u(\pr\hat\X(L)) = E_u(\delta_P L) = E_u(P E_u(L)) = \delta_P^* G^\t + \delta_{G^\t}^* P$
directly follows from the Euler-Lagrange relation \eqref{frechet-euler}
combined with the product rule shown in \lemref{eulerop-lemma} for the Euler operator. 
Finally, $\delta_P^* G^\t = \delta_P G^\t$ holds by \lemref{eulerlagrange},
and $\delta_{G^\t}^* P = \delta_G^* P^\t$ holds as an identity. 
Hence 
$E_u(\pr\hat\X(L)) = \delta_P G^\t + \delta_G^* P^\t$ is an identity. 
This completes the proof. 

An importance consequence of equation \eqref{varsymmdeteq} is that it provides 
a determining condition to find {\em all} variational symmetries 
for a given locally variational PDE system, without the explicit use of a Lagrangian. 
In particular, 
this formulation avoids the need to consider the ``gauge terms'' $D_t\Psi^t +D_x\cdot\Psi^x$ 
which arise in the Lagrangian formulation \eqref{divsymm}.

{\bf Running \Ex{(1)}}
The Lie symmetries of the gKdV equation \eqref{gkdv-eqn} 
consist of a time translation $\hat\X = -u_t\p_u$, 
a space translation $\hat\X = -u_x\p_u$, 
a scaling $\hat\X = -(\tfrac{2}{p} u+3t u_t +xu_x)\p_u$, 
and a Galilean boost $\hat\X = (1-tu_x)\p_u$ if $p\neq1$. 
These symmetries project to corresponding Lie symmetries of the potential gKdV equation 
$w_{tx}+w_x^pw_{xx}+w_{xxxx}=0$ 
through the relation $u=w_x$. 
This yields the generator 
$\hat\X = P\p_w$ with 
\begin{equation}
P=P^\t=(1-\tfrac{2}{p})c_3 w +c_4 -(c_1 +3c_3t)w_t-(c_2 +c_3 x+c_4 t) w_x .
\end{equation}
The variational Lie symmetries can be easily found by checking the condition \eqref{varsymmdeteq}. 
Using $G=G^\t=w_{tx}+w_x^pw_{xx}+w_{xxxx}$, 
a simple computation yields
\begin{equation}
\begin{aligned}
\delta_P G^\t & 
= D_tD_x P +pw_x^{p-1}w_{xx} D_x P + w_x^p D_x^2 P +D_x^4 P 
\\&
= -(c_1 +3c_3 t)D_t G -(c_2 +c_3 x +c_4 t) D_x G -(3+\tfrac{2}{p})c_3 G
\end{aligned}
\end{equation}
and also 
\begin{equation}
\begin{aligned}
\delta_G^* P^\t & 
= G \parder{P}{w} -D_t\Big( G \parder{P}{w_t} \Big) -D_x\Big( G\parder{P}{w_x} \Big) 
\\&
= (5-\tfrac{2}{p})c_3 G  +(c_1 +3c_3t)D_t G + (c_2 +c_3 x+c_4 t)D_x G . 
\end{aligned}
\end{equation}
Hence, 
$0=\delta_P G^\t+\delta_G^* P^\t = (2-\tfrac{4}{p}) c_3 G$
determines $(p-2)c_3=0$. 
This shows that all of the Lie symmetries except the scaling symmetry are variational symmetries
for an arbitrary nonlinearity power $p\neq 0$,
and that the scaling symmetry is a variational symmetry only for the special power $p=2$. 

\subsection{Noether's theorem in modern form}

Variational symmetries have a direct relationship to local conservation laws 
through the variational identity 
\begin{equation}\label{noetherid}
\begin{aligned}
\pr\hat\X(L) & =D_t\Psi_P^t +D_x\cdot\Psi_P^x
\\
& = \delta_P L = P E_u(L) + D\cdot\Upsilon_L(P)
\end{aligned}
\end{equation}
holding due to the Euler-Lagrange relation \eqref{frechet-euler}. 
The identity \eqref{noetherid} yields 
\begin{equation}\label{noetherconslaw}
P E_u(L) =D\cdot\Phi,
\quad
\Phi=(\Psi_P^t-\Upsilon_L^t(P),\Psi_P^x-\Upsilon_L^x(P))
\end{equation}
which is a conservation law in characteristic form for the PDE system given by $E_u(L)=0$. 
When combined with the formula \eqref{PhifromQ} for conserved currents, 
this provides a modern, local form of Noether's theorem, 
which does not explicitly use the Lagrangian. 

\begin{thm}\label{modernNoether}
For any locally variational PDE system $G=E_u(L)^\t=0$, 
variational symmetries $\hat\X=P\p_u$ 
and local conservation laws in characteristic form $D_t\til T+ D_x\cdot\til X=GQ$
have a one-to-one correspondence given by the relation 
\begin{equation}
P=Q^\t . 
\end{equation}
Equivalently, 
this correspondence is given by the homotopy integral 
\begin{equation}
\til\Phi = (\til T,\til X) 
= \int_0^1 \sum_{j=1}^{k}\Big( \p_\lambda \p^{j-1}u_{(\lambda)}\cdot\Big( 
\sum_{l=j}^{k}(-D)^{l-j}\cdot\Big(\parder{(PG^\t)}{(\p^l u)}\Big)\Big|_{u=u_{(\lambda)}} 
\Big) \Big)d\lambda
\end{equation}
modulo a total curl,
along a homotopy curve $u_{(\lambda)}(t,x)$, 
with $u_{(1)}=u$ and $u_{(0)}=u_0$ such that $(GQ)|_{u=u_0}$ is non-singular. 
Here $k=\max(r,N)$. 
\end{thm}

The Noether correspondence stated in \thmref{modernNoether} has a sharper formulation 
using the additional correspondence between multipliers and local conservation laws
provided by \thmref{correspondence}.
This formulation depends on whether a given variational PDE system possesses differential identities or not. 

In particular, when a PDE system satisfies a differential identity \eqref{diffid}, 
there will exist {\em gauge symmetries} 
\begin{equation}\label{gaugeP}
\hat\X= (\D^*(\chi))^\t\p_u
\end{equation}
corresponding to gauge multipliers \eqref{gaugeQ},
where $\D$ is the linear differential operator defining the given differential identity \eqref{diffid},
and $\chi$ is an arbitrary differential function. 
Two symmetries that differ by a gauge symmetry will be called {\em gauge equivalent}.

Recall, for any regular PDE system, 
a symmetry is trivial iff its characteristic function vanishes on the solution space of the PDE system,
and two symmetries are equivalent iff they differ by a trivial symmetry. 

\begin{cor}
(i) If a locally variational, regular PDE system \eqref{ELpde} has no differential identities, 
then there is a one-to-one correspondence between 
its admitted equivalence classes of linearly-independent local conservation laws
and its admitted equivalence classes of linearly-independent variational symmetries. 
(ii) If a locally variational, regular PDE system \eqref{ELpde} satisfies a differential identity,
then its admitted equivalence classes of linearly-independent local conservation laws
are in one-to-one correspondence with 
its admitted equivalence classes of linearly-independent variational symmetries
modulo gauge symmetries. 
\end{cor}

\subsection{Computation of variational symmetries and Noether conservation laws}

Whenever a locally variational PDE system \eqref{ELpde} is regular, 
the determining condition \eqref{varsymmdeteq} for finding variational symmetries 
$\hat\X=P(t,x,u,\p u,\ldots,\p^r u)\p_u$ 
can be converted into a linear system of equations for $P(t,x,u,\p u,\ldots,\p^r u)$ 
by the following steps. 

On the solution space $\Esp$ of the PDE system, 
the Fr\'echet derivative adjoint operator $\delta_G^*|_\Esp$ vanishes.
Thus, the determining condition \eqref{varsymmdeteq} implies
$(\delta_P G^\t)|_\Esp =0$ 
which coincides with the determining equation \eqref{modernsymmdeteq} 
for an infinitesimal symmetry of the PDE system. 
This shows that $P$ is the characteristic function of an infinitesimal symmetry. 
From \lemref{hadamard}, it then follows that $P$ satisfies the relation 
\begin{equation}\label{symmdeteqoffE}
\delta_P G^\t = R_P(G^\t)
\end{equation}
for some linear differential operator 
\begin{equation}\label{RPop}
R_P = R_P^{(0)} + R_P^{(1)}\cdot D + R_P^{(2)}\cdot D^2 +\cdots + R_P^{(r)}\cdot D^r
\end{equation}
whose coefficients are non-singular on $\Esp$, 
as the PDE system is assumed to be regular, 
where $r$ is the differential order of $P$. 
Note that if the PDE system satisfies a differential identity \eqref{diffid} 
then $R_P$ is determined by $P$ only up to $\chi\D^\t$ 
where $\chi$ is an arbitrary differential function 
and $\D$ is the linear differential operator defining the identity. 
Substitution of the relation \eqref{symmdeteqoffE} 
into the determining condition \eqref{varsymmdeteq} 
yields 
\begin{equation}\label{varsymmoffE}
0=R_P(G^t)+ \delta_G^* P^\t . 
\end{equation}
Note that $\delta_G^* P^\t$ can be expressed in an operator form 
\begin{equation}
\delta_G^* P^\t = E_u(P)G^\t -E_u^{(1)}(P)\cdot(DG^\t) + \cdots + E_u^{(r)}(P)\cdot(-D)^rG^\t
\end{equation}
using the relation \eqref{adjfrechet-euler-rel}. 
Consequently, 
when the PDEs $G=(G^1,\ldots,G^M)$ are expressed in a solved form 
\eqref{pde-solvedform}--\eqref{pde-leadingder} for a set of leading derivatives,
equation \eqref{varsymmoffE} can be split with respect to 
these leading derivatives and their differential consequences. 
This yields a linear system of equations
\begin{equation}\label{helmholtzsymmeq}
0= R_P^{(k)} +(-1)^k E_{u}^{(k)}(P),
\quad 
k=0,1,\ldots,r . 
\end{equation}
Note that these equations are similar in structure to the Helmholtz conditions \eqref{helmholtz}.

Hence, the following result has been established. 

\begin{thm}\label{varsymmdetsys}
The determining equation \eqref{varsymmdeteq} for variational symmetries 
$\hat\X=P(t,x,u,\p u,\ldots,\p^r u)\p_u$ 
of any locally variational, regular PDE system \eqref{ELpde} 
is equivalent to a linear system of equations consisting of 
the determining condition \eqref{modernsymmdeteq} for $\hat\X$ 
to be an infinitesimal symmetry of the PDE system, 
and Helmholtz-type conditions \eqref{helmholtzsymmeq} for $\hat\X$ 
to leave any Lagrangian of the PDE system invariant modulo a total divergence. 
This linear determining system \eqref{modernsymmdeteq}, \eqref{helmholtzsymmeq} is formulated
entirely in terms of the symmetry characteristic function $P$ and the PDE expressions $G=(G^1,\ldots,G^M)$, 
without explicit use of a Lagrangian. 
\end{thm}

It is important to emphasize that 
the determining system \eqref{modernsymmdeteq}, \eqref{helmholtzsymmeq} 
can be solved computationally by the same standard procedure \cite{Olv,1stbook,2ndbook}
that is used to solve the standard determining equation \eqref{symmdeteq} for symmetries.

\section{Main results}
\label{mainresults}

For any regular PDE system \eqref{pde}, whether or not it has a variational principle, 
all local conservation laws have a characteristic form given by multipliers,
as shown by the general correspondence stated in \thmref{correspondence}. 
In the case of regular PDE systems that are locally variational, 
the modern form of Noether's theorem given by \thmref{modernNoether} 
shows that multipliers for local conservation laws 
are the same as characteristic functions for variational symmetries. 
These symmetries satisfy a determining equation \eqref{varsymmdeteq} 
which can be split into an equivalent determining system 
for the symmetry characteristic functions,
without explicit use of a Lagrangian, as shown in \thmref{varsymmdetsys}. 
A similar determining system can be derived for multipliers, 
by splitting the multiplier determining equation \eqref{multrdeteq} in the same way. 

On the solution space $\Esp$ of a given regular PDE system \eqref{pde},
the Fr\'echet derivative adjoint operator $\delta_G^*|_\Esp$ vanishes.
Thus, the multiplier determining equation \eqref{multrdeteq} implies
\begin{equation}\label{adjsymmdeteq}
(\delta_Q^* G)|_\Esp =0
\end{equation}
which is the adjoint of the symmetry determining equation \eqref{modernsymmdeteq},
and its solutions $Q(t,x,u,\p u,\ldots,\p^r u)$ 
are called {\em adjoint-symmetries} \cite{AncBlu97,AncBlu02a,AncBlu02b} 
(or sometimes {\em cosymmetries}). 
Then $Q$ satisfies the identity 
\begin{equation}\label{adjsymmdeteqoffE}
\delta_Q^* G = \delta_{Q^\t}^* G^\t = R_{Q^\t}(G^\t)
\end{equation}
from \lemref{hadamard}, 
where 
\begin{equation}\label{RQop}
R_{Q^\t} = R_{Q^\t}^{(0)} + R_{Q^\t}^{(1)}\cdot D + R_{Q^\t}^{(2)}\cdot D^2 +\cdots + R_{Q^\t}^{(r)}\cdot D^r
\end{equation}
is some linear differential operator whose coefficients are non-singular on $\Esp$, 
and $r$ is the differential order of $Q$. 
Note that if the PDE system satisfies a differential identity \eqref{diffid} 
then $R_{Q^\t}$ is determined by $Q$ only up to $\chi\D^\t$ 
where $\chi$ is an arbitrary differential function 
and $\D$ is the linear differential operator defining the identity. 
The determining equation \eqref{multrdeteq} now becomes
\begin{equation}\label{multroffE}
0=R_{Q^\t}(G^\t)+ \delta_G^* Q . 
\end{equation}
From the relation \eqref{adjfrechet-euler-rel}, 
note that $\delta_G^* Q$ can be expressed in an operator form 
\begin{equation}
\delta_G^* Q = E_u(Q^\t)G^\t -E_u^{(1)}(Q^\t)\cdot(DG^\t) + \cdots + E_u^{(r)}(Q^\t)\cdot(-D)^rG^\t . 
\end{equation}
Consequently, 
when the PDEs $G=(G^1,\ldots,G^M)$ are expressed in a solved form 
\eqref{pde-solvedform}--\eqref{pde-leadingder} in terms of a set of leading derivatives,
equation \eqref{multroffE} can be split with respect to 
these leading derivatives and their differential consequences. 
This yields a linear system of equations
\begin{equation}\label{helmholtzmultreq}
0= R_{Q^\t}^{(k)} +(-1)^k E_{u}^{(k)}(Q^\t),
\quad 
k=0,1,\ldots,r
\end{equation}
which is similar in form to the Helmholtz conditions \eqref{helmholtz}.

Thus, the following result has been established. 

\begin{thm}\label{multrdetsys}
The determining equation \eqref{multrdeteq} for conservation law multipliers 
of any regular PDE system \eqref{pde} 
is equivalent to the linear system of equations \eqref{adjsymmdeteq}, \eqref{helmholtzmultreq}. 
In particular, multipliers are adjoint-symmetries \eqref{adjsymmdeteq} 
satisfying Helmholtz-type conditions \eqref{helmholtzmultreq},
where these conditions are necessary and sufficient for an adjoint-symmetry 
$Q(t,x,u,\p u,\ldots,\p^r u)$ to have the variational form \eqref{QfromTX} 
derived from a conserved current $\Phi=(T,X^i)$. 
\end{thm}

A comparison of the determining systems formulated in \thmref{multrdetsys} and \thmref{varsymmdetsys} 
shows how the correspondence between the local conservation laws and the multipliers for regular PDE systems 
is related to the Noether correspondence between the local conservation laws and the variational symmetries for locally variational, regular PDE systems. 

\begin{cor}\label{adjnoether}
When a regular PDE system is locally variational \eqref{ELpde}, 
the adjoint-symmetry determining equation \eqref{adjsymmdeteq} 
is the same as the symmetry determining equation \eqref{modernsymmdeteq} 
and the Helmholtz-type conditions \eqref{helmholtzmultreq} 
under which an adjoint-symmetry is a multiplier 
are equivalent to the variational conditions \eqref{helmholtzsymmeq} 
under which a symmetry is a variational symmetry. 
\end{cor}

Thus, \thmrefs{correspondence}{multrdetsys} provide a direct generalization of 
the modern form of Noether's theorem given by \thmrefs{modernNoether}{varsymmdetsys},
in which the role of symmetries in the derivation of local conservation laws for variational PDE systems 
is replaced by adjoint-symmetries in the derivation of local conservation laws for non-variational PDE systems.

\subsection{Computation of multipliers and conserved currents}

For any given regular PDE system, 
all of its non-trivial local conservation laws (up to equivalence) 
can be obtained by the following three steps. \\
Step 1: Solve the determining system \eqref{adjsymmdeteq}, \eqref{helmholtzmultreq} 
to obtain all multipliers. \\
Step 2: Find all linearly independent equivalence classes of non-trivial multipliers. \\
Step 3: Construct the conserved current determined by a representative multiplier 
in each equivalence class. 

The multiplier determining system \eqref{adjsymmdeteq}, \eqref{helmholtzmultreq}
can be solved computationally by the same standard procedure \cite{Olv,1stbook,2ndbook}
that is used to solve the determining equation \eqref{modernsymmdeteq} for symmetries.
Moreover, for multipliers of a given differential order $r$, 
the multiplier determining system is, in general, 
more overdetermined than is the symmetry determining equation
for infinitesimal symmetries of the same differential order $r$. 
Consequently, 
the computation of multipliers is typically easier than the computation of symmetries. 

As an alternative to solving the whole multiplier determining system together, 
only the adjoint-symmetry determining equation can be solved first,
and the Helmholtz-type conditions \eqref{helmholtzmultreq} then can be checked for 
each adjoint-symmetry to obtain all multipliers. 

In practice, it can be computationally hard to obtain the complete solution to 
the multiplier determining system (or the adjoint-symmetry determining equation)
because this will involve going to an arbitrarily high differential order 
for the dependence of the multiplier (or the adjoint-symmetry) on the derivatives of the dependent variables in the PDE system. 
Moreover, for computations using computer algebra, 
this differential order must be specified in advance. 
The same issue arises when symmetries are being sought, 
but often these obstacles are set aside by looking for just Lie symmetries, 
or higher symmetries of a special form. 

A similar approach can be used for multipliers, 
by looking just for all low-order conservation laws
or by looking just for higher order conservation laws with a special form or with a particular differential order.
In physical applications, 
there is often a specific class of conserved densities that is of interest. 
The form for multipliers corresponding to a given class of conserved densities 
can be derived directly by balancing derivatives on both sides of the characteristic equation,
as shown in the running examples in \secref{loworder}. 

For each non-trivial multiplier, 
the construction of a corresponding non-trivial conserved current 
can be carried out by several different methods. 

First, the homotopy integral formula \eqref{TfromQ}--\eqref{XfromQ} can be applied. 
An advantage of this formula compared to the standard linear-homotopy formula in the literature \cite{Olv,AncBlu02a,AncBlu02b}
is that the homotopy curve can be adapted to the structure of the expressions 
for the multiplier $Q$ and the PDE system $G$, 
which allows avoiding integration singularities. 

Second, the characteristic equation \eqref{chareqn} can be converted into 
a linear system of determining equations for the conserved density $\til T$ and the flux $\til X$. 
The determining equations are derived in a straightforward way
starting from the expression for the multiplier $Q$, 
similarly to the derivation of the form for low-order conservation laws 
explained in \secref{loworder}. 
This method is computationally advantageous as it can be implemented in the same way as 
setting up and solving the determining system for multipliers \cite{Wol02b,2ndbook}.

Third, if a given PDE system possesses a scaling symmetry 
then an algebraic formula that yields a scaling multiple of the conserved current 
$\til\Phi|_\Esp=(\til T,\til X)|_\Esp$ is available \cite{Anc03},
where the scaling multiple is simply the scaling weight of the corresponding conserved integral. 
The formula can be derived by applying the scaling relation \eqref{inveulerlagr-scaling}--\eqref{inveulerlagr-scalingcurrent}
directly to the function $f=GQ$. 
This gives 
\begin{align}
&\begin{aligned}
T= \omega\til T|_\Esp & = \Big( 
P\sum_{l=1}^{k}(-D)^{l-1}\cdot\Big(\parder{G}{(\p^{l-1}\p_t u)}Q\Big)
+(D P)\cdot \Big(\sum_{l=2}^{k}(-D)^{l-2}\cdot\Big(\parder{G}{(\p^{l-1}\p_t u)}Q\Big)\Big)
\\&\qquad
+\cdots 
+(D^{k-1} P)\cdot\Big(\parder{G}{(\p^{k-1}\p_t u)}Q\Big) 
\Big)\Big|_\Esp , 
\end{aligned}
\label{scalTfromQ}\\
&\begin{aligned}
X= \omega\til X|_\Esp & = \Big(
P\sum_{l=1}^{k}(-D)^{l-1}\cdot\Big(\parder{G}{(\p^{l-1}\p_x u)}Q\Big)
+(D P)\cdot\Big(\sum_{l=2}^{k}(-D)^{l-2}\cdot\Big(\parder{G}{(\p^{l-1}\p_x u)}Q\Big)\Big)
\\&\qquad
+\cdots 
+(D^{k-1}P)\cdot\Big(\parder{G}{(\p^{k-1}\p_x u)}Q\Big) 
\Big)\Big|_\Esp , 
\end{aligned}
\label{scalXfromQ}
\end{align}
modulo a locally trivial current $\Phi_\triv = (D_x\Theta,-D_t\Theta+D_x\cdot\Gamma)$, 
where
\begin{equation}\label{scalgenerator}
P= \eta -u_t\tau - u_x\cdot\xi,
\quad
\tau = at,
\quad
\xi =(b_{(1)}x^1,\ldots,b_{(n)}x^n),
\quad
\eta = (c_{(1)} u^1,\ldots,c_{(m)} u^m) 
\end{equation}
are the characteristic functions in the generator of the scaling symmetry \eqref{scaling}. 
Here 
\begin{equation}\label{scalweight}
\omega= s +D_t\tau +D_x\cdot\xi = s +a + \sum_{i=1}^{n} b_{(i)}
\end{equation}
is a scaling factor, 
with $s$ being the scaling weight of the function $GQ$. 
Note, as seen from the characteristic equation \eqref{chareqn}, 
$\omega$ is equal to the scaling weight of the conserved integral 
$\int_\Omega \til T|_\Esp dV$, 
as defined on any given spatial domain $\Omega\subseteq\Rnum^n$. 

This algebraic formula \eqref{scalTfromQ}--\eqref{scalgenerator}
has the advantage that it does not require any integrations. 
However, it assumes that the scaling multiple $\omega$ is non-zero,
which means that it can be used only for constructing conserved currents 
whose corresponding conserved integral has a non-zero scaling weight, $\omega\neq 0$.

A more general algebraic construction formula can be derived by utilizing dimensional analysis,
which is applicable to PDE systems without a scaling symmetry. 
Any given PDE system arising in physical applications will be scaling homogeneous 
under dimensional scaling transformations that act by 
rescaling the fundamental physical units of all variables and all parametric constants \cite{Olv,1stbook}
(whether or not the PDE system admits a scaling symmetry). 
In particular, these dimensional scaling transformations will comprise 
independent rescalings of length, time, mass, charge, and so on. 
For each dimensional scaling transformation, 
a scaling formula will arise for $T$ and $X$, 
generalizing the algebraic formula \eqref{scalTfromQ}--\eqref{scalgenerator}
in a way that involves the dependence of $Q$ and $G$ 
on all of the dimensionful parametric constants appearing in their expressions. 
If a conserved integral represents a dimensionful physical quantity, 
then the scaling multiple in the resulting formula will be non-zero. 

A derivation of this general construction formula will be given elsewhere \cite{AncHer}. 
Here, it will be illustrated in a running example. 

{\bf Running \Ex{(1)}}
All low-order conservation laws will now be derived 
for the gKdV equation \eqref{gkdv-eqn}. 
As shown previously, 
low-order conserved currents correspond to low-order multipliers,
which have the general form $Q(t,x,u,u_x,u_{xx})$. 
Multipliers are adjoint-symmetries that satisfy Helmholtz-type conditions.
To set up the determining system for multipliers, 
first note  $\delta_Q^* G =-(D_tQ +D_x^3Q +u^pD_x Q)$,
where $G=u_t+u^pu_x+u_{xxx}$.
Hence the adjoint-symmetry determining equation for $Q$ is given by 
\begin{equation*}
(D_tQ +D_x^3Q +u^pD_x Q)|_\Esp =0 . 
\end{equation*}
Next look at the terms that contain the leading derivative $u_t$ and its $x$-derivatives
in this equation. 
This yields 
\begin{equation*}
-D_tQ +D_x^3Q +u^pD_x Q = -\parder{Q}{u} -\parder{Q}{u_x}D_x G -\parder{Q}{u_{xx}}D_x^2 G 
= R_Q(G)
\end{equation*}
holding off of the gKdV solution space, 
where the components of the operator $R_Q$ are given by 
\begin{equation*}
R_Q^{(0)} =-\parder{Q}{u},
\quad
R_Q^{(x)} = -\parder{Q}{u_x},
\quad
R_Q^{(x,x)} = -\parder{Q}{u_{xx}} . 
\end{equation*}
Then the Helmholtz-type equations on $Q$ consist of 
\begin{equation*}
\begin{aligned}
0= & R_Q^{(0)} + E_{u}(Q) = -D_x\parder{Q}{u_x} + D_x^2\parder{Q}{u_{xx}} , 
\\
0= & R_Q^{(x)} -E_{u}^{(x)}(Q) = -2\parder{Q}{u_x} + 2D_x\parder{Q}{u_{xx}} , 
\\
0= & R_Q^{(x,x)} + E_{u}^{(x,x)}(Q^\t) = 0 , 
\end{aligned}
\end{equation*}
which reduce to a single equation
\begin{equation*}
D_x\parder{Q}{u_{xx}} -\parder{Q}{u_x} =0 . 
\end{equation*}
This Helmholtz-type equation and the adjoint-symmetry equation can be split 
with respect to all derivatives of $u$ which do not appear in $Q$, 
with $u_t$ eliminated through the gKdV equation.
This gives, after some simplifications, 
a linear overdetermined system of 8 equations:
\begin{gather*}
\parder{Q}{u_x} =0,
\quad
\parder{^2Q}{u_{xx}^2} =0,
\quad
\parder{{}^2Q}{x\p u_{xx}}=0,
\quad
\parder{{}^2Q}{u\p u_{xx}} =0,
\quad
\parder{{}^3Q}{x\p u^2} =0,
\\
\\
\parder{{}^3Q}{u^3} -p(p-1)u^{p-2}\parder{Q}{u_{xx}} =0,
\quad
\parder{{}^3Q}{x^2\p u} + u_{xx}\parder{{}^2Q}{u^2} - pu^{p-1}u_{xx}\parder{Q}{u_{xx}} =0,
\\
\parder{Q}{t} +u^p\parder{Q}{x} +\parder{{}^3Q}{x^3} +3u_{xx}\parder{{}^2Q}{x\p u} =0 . 
\end{gather*}
These equations can be solved for $Q$, with $p$ treated as an unknown, to get 
\begin{equation*}
Q=c_1 +c_2 u +c_3(u_{xx} + \tfrac{1}{p+1}u^{p+1}) +c_4 (x-tu) +c_5 (t(3u_{xx} + u^3) -xu)
\end{equation*}
with $c_4=0$ if $p\neq 1$, and $c_5=0$ if $p\neq 2$.  
Hence, 5 low-order multipliers are obtained, 
\begin{gather*}
Q_1=1,
\quad
Q_2 = u, 
\quad
Q_3 = u_{xx} + \tfrac{1}{p+1}u^{p+1}, 
\quad 
p>0 , 
\\
Q_4 = x-tu,
\quad 
p=1 , 
\\
Q_5 =t(3u_{xx} + u^3) -xu, 
\quad 
p=2 . 
\end{gather*}

The corresponding low-order conserved currents will now be derived 
using the three different construction methods. 
First is the homotopy integral method. 
The simplest choice for the homotopy is $u_{(\lambda)}=\lambda u$ 
since the gKdV equation is a homogeneous PDE, $G|_{u=0}=0$. 
Hence the homotopy integral is simply given by 
\begin{align*}
\til T & = \int_0^1 u \parder{(GQ)}{u_t}\Big|_{u=u_{(\lambda)}}\;d\lambda 
\\
& = \int_0^1 u \big( 
c_1 +c_4 x +(c_2 -c_4 t -c_5 x)u\lambda +(c_3+c_5 3t)u_{xx}\lambda 
+ c_5 tu^3 \lambda^3 + c_3 \tfrac{1}{p+1}u^{p+1} \lambda^{p+1}  
\big)\;d\lambda 
\\
& = (c_1 +c_4 x)u +\tfrac{1}{2}(c_2 -c_4 t -c_5 x)u^2 
+\tfrac{1}{2}(c_3 +c_5 3t) uu_{xx} + c_5 \tfrac{1}{4}tu^4 
+ c_3 \tfrac{1}{(p+1)(p+2)}u^{p+2} 
\end{align*}
and 
\begin{align*}
\til X & 
= \int_0^1 \Big( 
u \Big( \parder{(GQ)}{u_x}\Big|_{u=u_{(\lambda)}} -D_x\parder{(GQ)}{u_{xx}}\Big|_{u=u_{(\lambda)}} +D_x^2\parder{(GQ)}{u_{xxx}}\Big|_{u=u_{(\lambda)}} \Big)
\\&\qquad
+u_x \Big( \parder{(GQ)}{u_{xx}}\Big|_{u=u_{(\lambda)}} -D_x\parder{(GQ)}{u_{xxx}}\Big|_{u=u_{(\lambda)}} \Big)
+u_{xx} \parder{(GQ)}{u_{xxx}}\Big|_{u=u_{(\lambda)}} 
\Big)d\lambda
\\
& = \int_0^1 \Big( 
u \big( (c_4xu-2c_5 u_x-(c_3+c_5 3t)u_{tx} -(c_4t+c_5 x-c_2)u_{xx})\lambda 
\\&\qquad
-c_4tu^2\lambda^2
+c_5(3u^2u_{xx}-xu^3)\lambda^3 
+c_5 tu^5\lambda^5
+c_1 u^p\lambda^p
+(c_2u^{p+1} +c_3u^pu_{xx})\lambda^{p+1}
\\&\qquad
+c_3\tfrac{1}{p+1} u^{2p+1}\lambda^{2p+1} \big)
+u_x \big( -c_4 +(c_5 u+(c_3+c_5 3t)u_t +(c_4 t+c_5 x-c_2)u_x)\lambda \big)
\\&\qquad
+u_{xx} \big( c_1 +c_4 x +(c_2 -c_4 t -c_5 x)u +(c_3+c_5 3t)u_{xx}) \lambda 
+ c_5 tu^3 \lambda^3 + c_3 \tfrac{1}{p+1}u^{p+1} \lambda^{p+1}  \big) 
\Big)d\lambda
\end{align*}
which is easiest to evaluate when separated into the non-overlapping cases 
$p=1$ with $c_5=c_1=c_2=c_3=0$, $p=2$ with $c_4=c_1=c_2=c_3=0$, 
and $p>0$ with $c_4=c_5=0$. 
This yields the 5 low-order conserved currents 
\begin{align*}
& 
\til T_1 = u ,
\quad
\til X_1 = \tfrac{1}{p+1}u^{p+1} +u_{xx}
\\
& 
\til T_2 = \tfrac{1}{2} u^2 ,
\quad
\til X_2 = \tfrac{1}{p+2}u^{p+2} +uu_{xx} - \tfrac{1}{2} u_x^2
\quad
\\
&
\til T_3 = \tfrac{1}{2} uu_{xx} + \tfrac{1}{(p+1)(p+2)}u^{p+2} ,
\quad
\til X_3 = \tfrac{1}{2(p+1)^2}u^{2p+2} +\tfrac{1}{p+1}u^{p+1} u_{xx} + \tfrac{1}{2}(u_{xx}^2 +u_tu_x)-uu_{tx}
\\
&
\til T_4 = xu -\tfrac{1}{2} tu^2 ,
\quad
\til X_4 = t(\tfrac{1}{2}u_x^2-uu_{xx} -\tfrac{1}{3}u^3) +x(u_{xx}+\tfrac{1}{2}u^2)-u_x
\quad
p=1
\\
&\begin{aligned}
\til T_5 = \tfrac{1}{2}(3t uu_{xx} -xu^2) +\tfrac{1}{4}tu^4 ,
\quad
\til X_5 & = t(\tfrac{3}{2}(u_{xx}^2+u_tu_x)+u^3u_{xx} -\tfrac{3}{2}uu_{tx}+\tfrac{1}{6}u^6) 
\\&\qquad
+x(\tfrac{1}{2}u_x^2-uu_{xx}-\tfrac{1}{4}u^4)-\tfrac{1}{2}uu_x,
\quad
p=2
\end{aligned}
\end{align*}
whose respective multipliers are $Q_1,\ldots,Q_5$. 
Each of these conserved currents is in characteristic form, 
namely $D_t\til T_i +D_x\til X_i = Q_iG$. 

Second is the integration method using the characteristic equation 
$D_t\til T +D_x\til X = GQ$,
where 
\begin{equation*}
GQ=\big( c_1 +c_2 u +c_3(u_{xx} + \tfrac{1}{p+1}u^{p+1}) +c_4 (x-tu) +c_5 (t(3u_{xx} + u^3) -xu) \big)(u_t+u^pu_x+u_{xxx})
\end{equation*}
with $c_4=0$ if $p\neq 1$, and $c_5=0$ if $p\neq 2$.  
There are three steps in this method. 
First, as shown previously 
from balancing derivatives on both sides of the characteristic equation,
the general form for all low-order conserved conserved currents 
$\til\Phi=(\til T,\til X)$ is found to be given by 
\begin{equation*}
\til\Phi|_\Esp =(\til T(t,x,u,u_x),\til X(t,x,u,u_t,u_x,u_{xx})) . 
\end{equation*}
Second, 
the characteristic equation can then be split with respect to $u_{tx}$ and $u_{xxx}$,
which yields a linear overdetermined system of three equations:
\begin{gather*}
\parder{\til T}{u_x} + \parder{\til X}{u_t} =0,
\\
\parder{\til X}{u_{xx}} 
= c_3\tfrac{1}{p+1}u^{p+1} +c_1+c_4x +(c_2-c_4t-c_5x)u+(c_3+c_53t)u_{xx} +c_5tu^3,
\\
\begin{aligned}
\parder{\til T}{t} +\parder{\til X}{x} +u_t\parder{\til T}{u} +u_x\parder{\til X}{u} +u_{xx}\parder{\til X}{u_x} 
& 
= (u_t +u^pu_x) \big( c_3\tfrac{1}{p+1}u^{p+1} +c_1+c_4x 
\\&\qquad
+(c_2-c_4t-c_5x)u+(c_3+c_53t)u_{xx} +c_5tu^3 \big) .
\end{aligned}
\end{gather*}
These equations can be integrated directly. 
It is simplest to consider separately the non-overlapping cases 
$p=1$ with $c_5=c_1=c_2=c_3=0$, $p=2$ with $c_4=c_1=c_2=c_3=0$, 
and $p>0$ with $c_4=c_5=0$. 
The first case is found to reproduce $\til T_4$ and $\til X_4$;
the second case yields 
$\til T_5 -D_x\til\Theta_5$ and $\til X_5+D_t\til\Theta_5$
where $\til\Theta_5=\tfrac{3}{2}tuu_x$.
Similarly, the third case with $c_3=0$ is found to reproduce 
$\til T_1$, $\til T_2$, $\til X_1$, $\til X_2$,
and with $c_3\neq 0$ it yields
$\til T_3 -D_x\til\Theta_3$ and $\til X_3+D_t\til\Theta_3$
where $\til\Theta_3=\tfrac{1}{2}uu_x$.
Thus, the resulting conserved currents 
agree with those obtained from the homotopy integral, up to locally trivial currents. 
In particular, each of these currents is in characteristic form. 

Third is the scaling symmetry method. 
The gKdV equation possesses a scaling symmetry 
\begin{equation*}
t\rightarrow \lambda^3 t,
\quad
x\rightarrow \lambda x,
\quad
u\rightarrow \lambda^{-2/p} u,
\quad
\lambda \neq 0
\end{equation*}
with the characteristic $P=-(2/p)u - 3t u_t -x u_x$. 
Note the multipliers $Q_1,\ldots,Q_5$ are each homogeneous under the scaling symmetry,
with respective scaling weights $q_1=0$, $q_2=-2/p$, $q_3=-2-2/p$, $q_4=1$, $q_5=0$. 
Hence the corresponding scaling factors \eqref{scalweight} are given by 
$\omega_1=1-2/p$, $\omega_2=1-4/p$, $\omega_3=-1-4/p$, $\omega_4=0$, $\omega_5=0$,
where $s_i=q_i +c-a$, $a=3$, $b=1$, $c=-2/p$. 
Then the scaling symmetry formula is given by 
\begin{align*}
&
\begin{aligned}
T_i= \omega_i\til T_i|_\Esp = \Big(P \parder{G}{u_t}Q_i\Big)\Big|_\Esp, 
\end{aligned}
\\
&
\begin{aligned}
X_i= \omega_i\til X_i|_\Esp & = \Big(
P\Big(\parder{G}{u_x}Q_i -D_x\Big(\parder{G}{u_{xx}}Q_i\Big) + D_x^2\Big(\parder{G}{u_{xxx}}Q_i\Big)\Big)
\\&\qquad
+D_x P \Big( \parder{G}{u_{xx}}Q_i -D_x\Big(\parder{G}{u_{xxx}}Q_i\Big) \Big)
+D_x^2 P \Big( \parder{G}{u_{xxx}}Q_i \Big) 
\Big)\Big|_\Esp ,
\end{aligned}
\end{align*}
modulo a locally trivial current. 
For $i=1,2,3$,
this yields the conserved density expressions
\begin{align*}
& 
T_1= -(\tfrac{2}{p}u +3t u_t +x u_x)|_\Esp 
= (1-2/p)\til T_1|_\Esp +D_x\Theta_1, 
\quad
\Theta_1 = 3t\til X_1 -x\til T_1,
\\
&
T_2= -((\tfrac{2}{p}u +3t u_t +x u_x) u)|_\Esp = (1-4/p)\til T_2|_\Esp +D_x\Theta_2, 
\quad
\Theta_2 = 3t\til X_2 -x\til T_2, 
\\
&
\begin{aligned}
& T_3= -(\tfrac{2}{p}u +3t u_t +x u_x) (u_{xx} + \tfrac{1}{p+1}u^{p+1})|_\Esp 
= (-1-4/p)\til T_3|_\Esp +D_x\Theta_3 ,
\\&\qquad
\Theta_3 = \tfrac{1}{2}(1+4/p)uu_x +3t(\til X_3 +D_t\til\Theta_3)-x(\til T_3 -D_x\til\Theta_3) . 
\end{aligned}
\end{align*}
Note their scaling factors are non-zero when $p\neq2$, $p\neq4$, and $p\neq -4$, respectively.
When $p=2$, $T_1$ reduces to a locally trivial conserved density $D_x\Theta_1$, 
and when $p=4$, $T_2$ reduces to a locally trivial conserved density $D_x\Theta_2$. 
Likewise, when $p=-4$, $T_3$ reduces to a locally trivial conserved density $D_x\Theta_3$. 
The expressions given by the scaling symmetry formula for $i=4,5$
yield 
\begin{gather*}
T_4 = -(2u +3t u_t +x u_x)(x-tu)|_\Esp =D_x\Theta_4,
\\
\Theta_4 = (x-tu)(t(3u_{xx} +u^2)-xu) +\tfrac{3}{2}(tu_x-1)^2, 
\end{gather*}
and 
\begin{gather*}
T_5 = -(u +3t u_t +x u_x)(3t(u_{xx} + u^3) -xu)|_\Esp =D_x\Theta_5,
\\
\Theta_5 =\tfrac{1}{2}(t(3u_{xx}+u^3)-xu)^2. 
\end{gather*}
These cases for $p>0$ in which the scaling symmetry formula 
yields locally trivial currents 
are called the critical powers for the corresponding conserved currents. 
To obtain the conserved currents for a critical power, 
it is necessary to use the more general dimensional scaling formula. 

Several steps are needed to set up the dimensional scaling formula. 

The first step is to introduce dimensionful constants into the gKdV equation
so that it is homogeneous under separate dimensional scalings of 
$t$ $[\text{time}]$, $x$ $[\text{length}]$, and $u$ $[\text{mass}]$. 
Thus, let 
\begin{equation*}
\til G = u_t +\mu u^pu_x +\nu u_{xxx},
\quad
\mu,\nu = \const
\end{equation*}
where $\mu$ has dimensions of 
$[\text{time}]^{-1}[\text{length}][\text{mass}]^{-p}$,
and $\nu$ has dimensions of 
$[\text{time}]^{-1}[\text{length}]^{3}$. 
Note $\til G=G$ will be the gKdV equation 
when these constants have the numerical values $\mu=1$ and $\nu=1$. 

The next step is to insert factors of $\mu$ and $\nu$ 
into the expressions for the low-order multipliers so that $Q_1,\ldots,Q_5$
are each dimensionally homogeneous:
\begin{gather*}
Q_1=1,
\quad
Q_2 = u, 
\quad
Q_3 = \nu u_{xx} + \tfrac{1}{p+1}\mu u^{p+1}, 
\quad 
p>0, 
\\
Q_4 = x-\mu tu,
\quad 
p=1, 
\\
Q_5 =t(3\nu u_{xx} + \mu u^3) -xu, 
\quad 
p=2 . 
\end{gather*}

The main step consists of generalizing the scaling relation \eqref{inveulerlagr-scaling}--\eqref{inveulerlagr-scalingcurrent}
so that it applies to dimensional scaling transformations. 
These transformations are given by 
\begin{align*}
& t \rightarrow \lambda t,
\quad
\mu \rightarrow \lambda^{-1}\mu,
\quad
\nu \rightarrow \lambda^{-1}\nu;
\\
& x \rightarrow \lambda x,
\quad
\mu \rightarrow \lambda\mu,
\quad
\nu \rightarrow \lambda^3\nu;
\\
& u \rightarrow \lambda u,
\quad
\mu \rightarrow \lambda^{-p}\mu,
\quad
\nu \rightarrow \nu;
\end{align*}
as determined by the dimensions of $\mu$ and $\nu$. 
Since the scaling relation \eqref{inveulerlagr-scaling}--\eqref{inveulerlagr-scalingcurrent}
only holds for variables in jet space, 
the constants $\mu$ and $\nu$ now must be treated as variables 
by introducing the equations
\begin{equation*}
\til G^{(\mu)} = (\mu_t,\mu_x)=0,
\quad
\til G^{(\nu)} = (\nu_t,\nu_x)=0 . 
\end{equation*}
Then the augmented PDE system 
\begin{equation*}
\til G =0,
\quad 
\til G^{(\mu)} = 0,
\quad
\til G^{(\nu)} = 0
\end{equation*}
will admit each of the three scaling transformations as symmetries
formulated in the augmented jet space 
$\til J=(t,x,u,\mu,\nu,u_t,u_x,\mu_t,\mu_x,\nu_t,\nu_x,\ldots)$. 
Note that the characteristic equation for conserved currents 
will have additional multiplier terms
\begin{equation*}
D_t\til T +D_x\til X = 
\til GQ + \til G^{(\mu)}\til Q_{(\mu)} +\til G^{(\nu)}\til Q_{(\nu)}
\end{equation*}
for some expressions 
$\til Q_{(\mu)} =(\til Q_{(\mu)}^t,\til Q_{(\mu)}^x)^\t$
and 
$\til Q_{(\nu)} =(\til Q_{(\nu)}^t,\til Q_{(\nu)}^x)^\t$,
where $Q$ is unchanged. 
These expressions can be found in a straightforward way 
by setting up and solving the multiplier determining system,
with $Q=Q_i$ being the previously derived low-order multipliers 
for the gKdV equation. 
Since $\mu$ and $\nu$ appear linearly in each $Q_i$ 
as well as in the PDE expression $\til G$, 
note $\til Q_{(\mu)}$ and $\til Q_{(\nu)}$
can have at most linear dependence on these variables 
and cannot contain any derivatives of these variables.
Also, since $Q_i$ depends on $u,u_x,u_{xx},u_{xxx}$,
and $\til G$ depends on $u,u_t,u_x,u_{xxx}$, 
note $\til Q_{(\mu)}$ and $\til Q_{(\nu)}$ can depend on only 
$u,u_t,u_x,u_{tx},u_{xx}$ in addition to $t,x$ and $\mu,\nu$:
\begin{equation*}
\til Q_{(\mu)}(t,x,u,\mu,\nu,u_t,u_x,u_{tx},u_{xx}), 
\quad
\til Q_{(\nu)}(t,x,u,\mu,\nu,u_t,u_x,u_{tx},u_{xx}) . 
\end{equation*}
The multiplier determining system is then given by 
\begin{gather*}
E_u(\til GQ_i + \til G^{(\mu)}\til Q_{(\mu)} +\til G^{(\nu)}\til Q_{(\nu)})=0,
\\
E_{(\mu)}(\til GQ_i + \til G^{(\mu)}\til Q_{(\mu)} +\til G^{(\nu)}\til Q_{(\nu)})=0,
\quad
E_{(\nu)}(\til GQ_i + \til G^{(\mu)}\til Q_{(\mu)} +\til G^{(\nu)}\til Q_{(\nu)})=0
\end{gather*}
for $i=1,\ldots,5$. 
This system splits with respect to all derivatives of $u,\mu,\nu$ which do not appear in $\til Q_{(\mu)}$ and $\til Q_{(\nu)}$. 
Integration of the resulting equations yields 
\begin{gather*}
\til Q_{1(\mu)}^t = 0,
\quad
\til Q_{1(\mu)}^x = \tfrac{1}{p+1}u^{p+1},
\quad
\til Q_{1(\nu)}^t = 0,
\quad
\til Q_{1(\nu)}^x = u_{xx}, 
\\
\til Q_{2(\mu)}^t = 0,
\quad
\til Q_{2(\mu)}^x = \tfrac{1}{p+2}u^{p+2},
\quad
\til Q_{2(\nu)}^t = 0,
\quad
\til Q_{2(\nu)}^x = uu_{xx}-\tfrac{1}{2}u_x^2, 
\\
\begin{aligned}
& \til Q_{3(\mu)}^t =\tfrac{1}{(p+1)(p+2)}u^{p+2}, 
\quad
\til Q_{3(\mu)}^x = \tfrac{1}{p+1}\nu u^{p+1} +\tfrac{1}{(p+1)(p+2)}\mu u^{2p+2},
\\
& \til Q_{3(\nu)}^t = -\tfrac{1}{2}u_x^2,
\quad
\til Q_{3(\nu)}^x = \nu u_{xx}^2 +u_tu_x +\mu u^{p+1}u_{xx}, 
\end{aligned}
\\
\begin{aligned}
& \til Q_{4(\mu)}^t =-\tfrac{1}{2}tu^2,
\quad
\til Q_{4(\mu)}^x = t\nu(\tfrac{1}{2} u_x^2 -uu_{xx}) +\tfrac{1}{2}x u^2-\tfrac{2}{3}\mu u^3,
\\
& \til Q_{4(\nu)}^t = 0, 
\quad
\til Q_{4(\nu)}^x = t\mu(\tfrac{1}{2} u_x^2 -uu_{xx}) +xu_{xx} -u_x, 
\end{aligned}
\\
\begin{aligned}
& \til Q_{5(\mu)}^t =\tfrac{1}{4}tu^2,
\quad
\til Q_{5(\mu)}^x = t(\nu u^3u_{xx} + \tfrac{1}{3}\mu u^6) -\tfrac{1}{4}x u^4,
\\
& \til Q_{5(\nu)}^t = -\tfrac{3}{2}tu_x^2,
\quad
\til Q_{5(\nu)}^x = t(\mu u^3u_{xx} +3\nu u_{xx}^2+3u_tu_x) +x(\tfrac{1}{2} u_x^2 -uu_{xx}) +uu_x . 
\end{aligned}
\end{gather*}
The scaling relation \eqref{inveulerlagr-scaling}--\eqref{inveulerlagr-scalingcurrent} 
can now be applied to the function 
$f_i=\til G Q_i + \til G^{(\mu)}\til Q_{i(\mu)} +\til G^{(\nu)}\til Q_{i(\nu)}$
in the augmented jet space 
$\til J=(t,x,u,\mu,\nu,u_t,u_x,\mu_t,\mu_x,\nu_t,\nu_x,\ldots)$
by using an infinitesimal scaling symmetry 
given by one of the scaling transformation generators
\begin{gather*}
\hat\X_{\text{time}}  = -(\mu+t\mu_t)\p_\mu -(\nu+t\nu_t)\p_\nu -t u_t\p_u , 
\\
\hat\X_{\text{length}} = (\mu-x\mu_x)\p_\mu +(3\nu-x\nu_x)\p_\nu -x u_x\p_u , 
\\
\hat\X_{\text{mass}} = -p\mu\p_\mu +u\p_u . 
\end{gather*}
Let $P$, $P^{(\mu)}$, $P^{(\nu)}$ denote the characteristic functions 
in the selected scaling transformation generator $\hat\X$. 
Then this yields the dimensional scaling formula 
\begin{align*}
& T_i=\omega_i\til T_i|_\Esp = \Big(
P \parder{\til G}{u_t}Q_i + P^{(\mu)}\til Q^t_{i(\mu)} + P^{(\nu)}\til Q^t_{i(\nu)} 
\Big)\Big|_\Esp , 
\\
&\begin{aligned}
X_i =\omega_i\til X_i|_\Esp & = \Big(
P\Big(\parder{\til G}{u_x}Q_i -D_x\Big(\parder{\til G}{u_{xx}}Q_i\Big) + D_x^2\Big(\parder{\til G}{u_{xxx}}Q_i\Big)\Big)
\\&\qquad
+D_x P \Big( \parder{\til G}{u_{xx}}Q_i -D_x\Big(\parder{\til G}{u_{xxx}}Q_i\Big) \Big)
+D_x^2 P \Big( \parder{\til G}{u_{xxx}}Q_i \Big) 
\\&\qquad
+ P^{(\mu)}\til Q^x_{i(\mu)} + P^{(\nu)}\til Q^x_{i(\nu)} 
\Big)\Big|_\Esp , 
\end{aligned}
\end{align*}
modulo a locally trivial current,
where 
\begin{equation}
\omega_i= q_i +s +D_t\tau+D_x\xi
\end{equation}
is a scaling factor defined in terms of 
the scaling weights $q_i,s$ of $Q_i,\til G$ 
and the divergence factor $D_t\tau+D_x\xi$ 
arising from the selected dimensional scaling transformation. 
In particular, for each $i=1,\ldots,5$, 
there will be some (possibly combined) transformation 
such that the scaling factor $w_i$ is non-zero,
as seen from Tables~\ref{scalproperties} and~\ref{Phiweights}. 

\begin{table}[ht!]
\begin{center}
\begin{tabular}{|c|c|c|c|c|c|c|c|c|c|c|}
\hline
\hfill  
& $P$ & $P^{(\mu)}$ & $P^{(\nu)}$ & $D_t\tau+D_x\xi$ 
& $s$ & $q_1$ & $q_2$ & $q_3$ & $q_4$ & $q_5$
\\
\hline
time
& $-t u_t$
& $-(\mu+t\mu_t)$
& $-(\nu+t\nu_t)$
& $1$
& $-1$
& $0$
& $0$
& $-1$
& $0$
& $0$
\\
\hline
length
& $-x u_x$
& $\mu-x\mu_x$
& $3\nu-x\nu_x$
& $1$
& $0$
& $0$
& $0$
& $1$
& $1$
& $1$
\\
\hline
mass
& $u$
& $-p\mu$
& $0$
& $0$
& $1$
& $0$
& $1$
& $1$
& $0$
& $1$
\\
\hline
\end{tabular}
\end{center}
\caption{Properties of dimensional scaling transformations for the gKdV equation and its low-order multipliers}
\label{scalproperties}
\end{table}

\begin{table}[ht!]
\begin{center}
\begin{tabular}{|c|c|c|c|c|c|}
\hline
\hfill  & $\omega_1$ & $\omega_2$ & $\omega_3$ & $\omega_4$ & $\omega_5$
\\
\hline
time
& $0$
& $0$
& $-1$
& $0$
& $0$
\\
\hline
length
& $1$
& $1$
& $2$
& $2$
& $2$
\\
\hline
mass
& $1$
& $2$
& $2$
& $1$
& $2$
\\
\hline
\end{tabular}
\end{center}
\caption{Dimensional scaling weights for low-order conserved currents of the gKdV equation}
\label{Phiweights}
\end{table}

The dimensional scaling formula will now be used to obtain 
the conserved currents $\Phi_i=(T_i,X_i)|_\Esp$ that were missed previously 
by the scaling symmetry formula.
These cases are: $i=4,5$; and, $i=1,2$ when $p$ is a critical power. 
From the form of the dimensional scaling generators, 
the mass scaling transformation is simplest choice to use. 
Then the formula becomes 
\begin{align*}
T_i = \omega_i\til T_i & = ( u Q_i -p\til Q^t_{i(\mu)} )|_{\mu=\nu=1} , 
\\
X_i = \omega_i\til X_i & = ( u^{p+1} Q_i + uD_x^2Q_i -u_x D_xQ_i +u_{xx} Q_i -p\til Q^x_{i(\mu)} )|_{\mu=\nu=1}, 
\end{align*}
modulo a locally trivial current.
This mass scaling formula yields the conserved density and flux expressions
\begin{align*}
T_1 = \til T_1, 
\quad
X_1 = \til X_1, 
\\
T_2 = 2\til T_2, 
\quad
X_2 = 2\til X_2, 
\end{align*}
which hold for all powers $p>0$ 
(including the critical powers $p=2$ and $p=4$, respectively),
and also
\begin{align*}
T_4 = \til T_4, 
\quad
X_4=\til X_4, 
\\
T_5 = 2\til T_5|, 
\quad
X_5 = 2\til X_5 . 
\end{align*}

\section{Concluding remarks}
\label{remarks}

The main results presented in \secref{mainresults} 
provide a broad generalization of Noether's theorem in modern form using multipliers,
yielding a general method which is applicable to all typical PDE systems 
arising in physical applications. 
In this generalization, 
the problem of finding all conservation laws for a given PDE system 
becomes an adjoint version of the problem of finding all infinitesimal symmetries of the PDE system. 

For any given variational PDE system, 
conservation laws arise from variational symmetries,
which are infinitesimal symmetries that satisfy variational conditions 
corresponding to invariance of any variational principle for the PDE system. 
Noether's theorem shows that the characteristic functions in a variational symmetry 
are precisely the component functions in a multiplier. 
For any given non-variational PDE system, 
the role of symmetries in the derivation of conservation laws is replaced by adjoint-symmetries, 
and the variational conditions under which an infinitesimal symmetry is a variational symmetry
are replaced by Helmholtz-type conditions under which an adjoint-symmetry is a multiplier. 
Also, 
the role of a Lagrangian in constructing a conserved integral from a variational symmetry
is replaced by several different constructions:
an explicit integral formula, an explicit algebraic scaling formula,
and a system of determining equations, 
all of which use only a multiplier and the given PDE system itself. 

Most importantly, 
the completeness of this general method in finding all conservation laws for a given PDE system 
is established by working with the system expressed in a solved-form for a set of leading derivatives without restricting it to have a generalized Cauchy-Kovalevskaya form. 
This means that the method applies equally well to PDE systems that possess differential identities. 

As a consequence, 
there is no need to use special methods or anstazes for determining the conservation laws of any given PDE system, 
just as there is no necessity to use special methods or ansatzes for finding its symmetries.

The formulation of the general method as a generalization of Noether's theorem 
rests on the adjoint relationship between variational symmetries and multipliers, 
which originates from the algebraic relationship between symmetries and adjoint-symmetries. 
An interesting question is whether this algebraic relationship has a geometrical interpretation. 

As will be shown in more detail elsewhere \cite{Anc17b}, 
adjoint-symmetries indeed can be given a simple geometrical meaning. 
In the case of PDE systems comprised of dynamical evolution equations, 
$G=\p_t u -g(t,x,u,\p_x u,\p_x^2 u, \ldots, \p_x^N u)=0$, 
an adjoint-symmetry defines a 1-form (or covector field) $Q\mathbf{d}u$ that is invariant under the dynamical flow on $u(t,x)$, 
similarly to how a symmetry $P\p_u$ defines an invariant vector field. 
This geometrical statement essentially relies on the number of dependent variables 
being the same as the number of equations in the PDE system. 
For general PDE systems $G=0$, 
it seems necessary to use the well-known procedure \cite{Olv} of embedding the PDE system 
into a larger, variational system defined by a Lagrangian $L=G v^\t$ 
where $v$ denotes additional dependent variables 
which are paired with the equations $G=0$ in the given PDE system. 
In this setting, 
an adjoint-symmetry defines a symmetry vector field $Q\p_v$ of the enlarged system, 
$G=0$ and $G'{}^*(v)=0$,
where $G'$ is the Fr\'echet derivative of $G$, and $G'{}^*$ is its adjoint. 
Then, it is straightforward to show that an adjoint-symmetry is a multiplier 
precisely when $Q\p_v$ is a variational symmetry.

\section*{Acknowledgements}

S.C. Anco is supported by an NSERC research grant. 
The referees are thanked for valuable comments which have improved this work.

\end{document}